\documentclass[twocolumn,aps,pra,10pt]{revtex4-2}
\usepackage{amssymb}
\usepackage{amsfonts}
\usepackage{amsmath}
\usepackage{amsxtra}
\usepackage{amscd}
\usepackage{amsthm}
\usepackage{graphicx}
\usepackage{epsf}
\usepackage{bbold}
\usepackage{xcolor}
\usepackage{mathalfa}
\usepackage{eucal}
\usepackage{bm}
\usepackage{array}

\begin{document}

\title{Solitons in binary compounds with stacked two-dimensional honeycomb lattices}

\author{James H. Muten}
\author{Louise H. Frankland}
\author{Edward McCann}
\email{ed.mccann@lancaster.ac.uk}
\affiliation{Department of Physics, Lancaster University, Lancaster, LA1 4YB, United Kingdom}

\begin{abstract}
We model the electronic properties of thin films of binary compounds with stacked layers where each layer is a two-dimensional honeycomb lattice with two atoms per unit cell.
The two atoms per cell are assigned different onsite energies in order to consider six different stacking orders: ABC, ABA, AA, ABC$^{\prime}$, ABA$^{\prime}$, and AA$^{\prime}$.
Using a minimal tight-binding model with nearest-neighbor hopping, we consider whether a fault in the texture of onsite energies in the vertical, stacking direction supports localized states, and we find localized states within the bulk band gap for ABC, ABA, and AA$^{\prime}$ stacking.
Depending on the stacking type, parameter values, and whether the soliton is atomically sharp or a smooth texture, there are a range of different band structures including soliton bands that are either isolated or that hybridize with other states, such as surface states, and soliton bands that are either dispersive or flat, the latter yielding narrow features in the density of states.
We discuss the relevance of our results to specific materials including graphene, hexagonal boron nitride and other binary compounds.
\end{abstract}

\maketitle

\setcounter{MaxMatrixCols}{20}

\section{Introduction}

\begin{figure*}[t]
\includegraphics[scale=0.32]{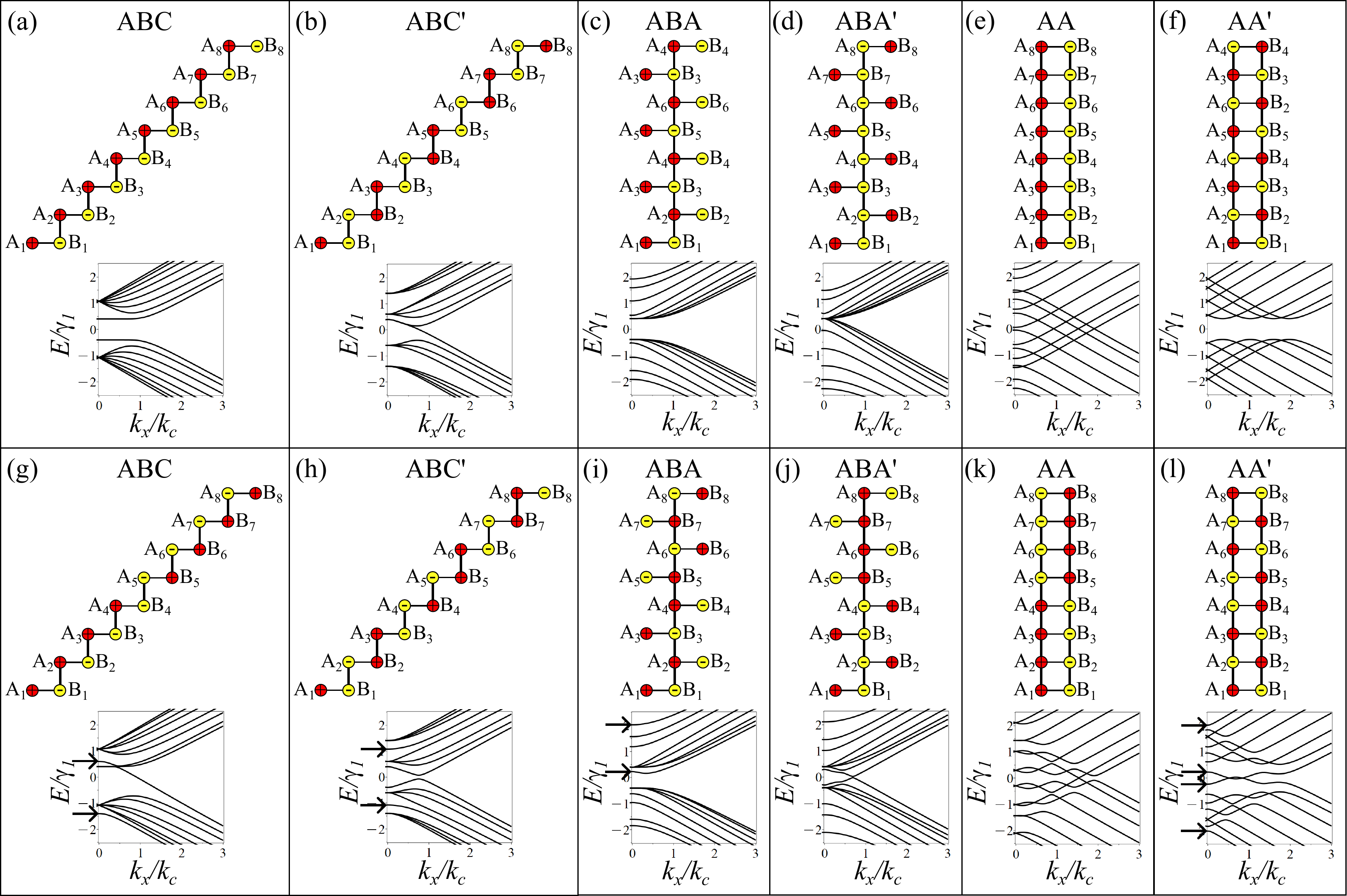}
\caption{Summary of the main results for the six different stacking types.
The first row (a)-(f) shows the faultless systems with a schematic side view of the stacking for $N=8$ layers with red (yellow) circles indicating A (B) atoms, horizontal (vertical) solid lines indicating nearest-neighbor intralayer (interlayer) hopping. The bottom of each panel shows the corresponding band structure near low energy as a function of wave vector component $k_x$ (with $k_y=0$) measured from the center of the $K_{+}$ valley, where $k_c = \gamma_1 / (\hbar v)$.
The second row (g)-(l) shows the lattices for $N=8$ layers with an atomically-sharp fault in the onsite energies occurring between the middle two layers and the bottom of each panel shows the corresponding band structures.
Horizontal arrows show the energies at $k_x=0$ of the states localized on the soliton.
In all plots, parameter values are $\gamma_0 = 3.16\,$eV, $\gamma_1 = 0.381\,$eV~\cite{kuzmenko09}, $a = 2.46$\AA~\cite{saito98},
and the magnitude of the alternating onsite energies is $u = U/\gamma_1 = 0.4$.
}\label{fig:summary}
\end{figure*}

Low-energy flat bands in thin films of rhombohedral graphite have attracted recent attention, due to advances in fabrication~\cite{pierucci15,henni16,henck18,latychevskaia19,yang19,geisenhof19,bouhafs21,shi20,kerelsky21,hagymasi22} and experimental observations of correlated states in rhombohedral trilayer graphene~\cite{zhou21a,zhou21b} and in thin films with several layers~\cite{shi20,kerelsky21,hagymasi22}.
The low-energy flat bands arise from surface states, and their existence may be related to the edge states of the Su-Schrieffer-Heeger (SSH) model~\cite{su79,asboth16,cayssol21,mccann23} by dimensional reduction~\cite{ryu10,xiao11,heikkila11}, considering the in-plane wave vector to be a fixed parameter.

Recently, stacking faults (in the vertical, out-of-plane direction) have been considered for thin films of rhombohedral graphite~\cite{taut14,slizovskiy19,garciaruiz19,shi20,muten21,garciaruiz23}. As with solitons in the SSH model~\cite{jackiw76,su79,asboth16}, stacking faults in thin graphitic films support localized states, manifested as bands in quasi-2D. Such bands tend to appear in pairs at about the same energy because the sequence of intra- and interlayer bonding either side of the fault is fixed, i.e., the texture of bonding strengths away from the fault cannot be changed by the fault. Hence, such faults are effectively coupled soliton-antisoliton pairs~\cite{muten21}.

In this paper, we investigate the possible existence of a single, isolated band localized on a stacking fault in quasi-2D materials consisting of layers of atoms on a honeycomb lattice. We consider materials with two different chemical elements such as hexagonal boron nitride~\cite{pacile08,naclerio22},
using a minimal tight-binding model to provide a generic description, modeling stacking faults for different stacking orders.
Specifically, for each layer, we consider two nonequivalent atoms per unit cell, labeled A and B, each with a single orbital which is isotropic within the plane, e.g. a $p_z$ orbital for sp$^2$ hybridization in graphene or hexagonal boron nitride. The two atoms have different onsite energies $\epsilon_A = U$ and $\epsilon_B = - U$ such that $U = (\epsilon_A - \epsilon_B)/2$ and $\epsilon_A + \epsilon_B = 0$.
We consider naturally occurring stackings, i.e., those whereby every atom is either directly above or below another atom, or above or below the center of a hexagon.
There are six of them~\cite{ribeiro11,constantinescu13,kim13,gilbert19}, as shown in the top row of Fig.~\ref{fig:summary}, consisting of rhombohedral stacking (ABC), Bernal stacking (ABA), and AA stacking. For each of these, we also include a primed version (ABC$^{\prime}$, ABA$^{\prime}$, and AA$^{\prime}$) whereby every other layer has the sign of the onsite energy $U$ reversed~\cite{ribeiro11,gilbert19}.

For rhombohedral stacking (ABC) with alternating onsite energies, dimensional reduction~\cite{ryu10}, by considering the in-plane wave vector as a fixed parameter, relates the Hamiltonian to that of the Rice-Mele model~\cite{ricemele82}, which is a generalization of the SSH model with alternating onsite energies as well as alternating hopping in one dimension. This is generally not topological, but, at certain wave vectors ($k_x = k_c$ in Fig.~\ref{fig:summary}), the intra- and interlayer hoppings are effectively equal, and ABC stacking is related to the charge density wave (CDW) model~\cite{kivelson83,brzezicki20,cayssol21,fuchs21,allen22,mccann23}
which is a one-dimensional $\mathbb{Z}_2$ topological insulator~\cite{shiozaki15} with constant hoppings and alternating onsite energies. Stacking faults for ABC stacking are analogous to solitons in the CDW model~\cite{kivelson83,brzezicki20,allen22}, albeit only at a certain wave vector, and our aim is to explore this analogy.

We focus on electronic properties near the corner of the first Brillouin zone (K point) and Fig.~\ref{fig:summary} summarizes our results. The top row shows the six faultless systems with their electronic band structures. Three of them (ABC, ABA, and AA$^{\prime}$ stacking) exhibit band gaps, and these are the configurations which support localized states on a stacking fault that lie within the band gap.
A sharp stacking fault consists of an inversion of the signs of the onsite energy $U$ between two layers without any change in the intra- or interlayer hopping, as shown in the bottom row of Fig.~\ref{fig:summary}; we refer to this as a soliton because it creates a change in texture of the onsite energies either side of the fault.
We find a range of different behaviors, depending on the stacking; arrows in Fig.~\ref{fig:summary} (bottom row) show the energies of states localized on the soliton exactly at the K point.
For ABC stacking, there are two localized states. One of them is within the valence band, but another generally lies within the bulk band gap. Depending on parameter values, it may hybridize with surface states, as shown in Fig.~\ref{fig:summary}(g).
For ABC$^{\prime}$, there are also two localized states, but they don't lie near zero energy; the two states within the band gap in Fig.~\ref{fig:summary}(h) are surface states.
For ABA stacking, there are two localized states. One of them is within the conduction band, but another generally lies within the bulk band gap, Fig.~\ref{fig:summary}(i), and there are no surface states to hybridize with.
As this band eventually moves into the conduction band, we refer to the corresponding texture as an antisoliton whereas the band shown in Fig.~\ref{fig:summary}(g) for ABC stacking moves into the valence band, and its texture is called a soliton. In general, both of these stackings (ABC and ABA) can support either a soliton or an antisoliton, depending on the position of the fault.

For both ABA$^{\prime}$ and AA stacking, there are no states localized on the fault, although the presence of the fault does have an impact on the band structure, Fig.~\ref{fig:summary}(j) and Fig.~\ref{fig:summary}(k), respectively.
For AA$^{\prime}$, a single fault supports four localized states: Two of them are in the bulk bands, but two of them lie within the bulk band gap and, depending on parameter values, they may hybridize with each other as shown in Fig.~\ref{fig:summary}(l).

In the following, we describe the cases shown in Fig.~\ref{fig:summary}, focusing on the systems with a single, isolated soliton band, and modeling their parameter dependence.
We describe the methodology used in Section~\ref{s:methodology}.
We describe rhombohedral stacking (ABC and ABC$^{\prime}$) in Section~\ref{s:rhomb}, Bernal stacking (ABA and ABA$^{\prime}$) in Section~\ref{s:bernal}, and AA stacking (AA and AA$^{\prime}$) in Section~\ref{s:aa}.
The properties of smooth solitons with a finite width are described in Section~\ref{s:smooth}, and defects which consist of a reversal of the signs of the onsite energies on a single layer only~\cite{yin11} are described in Section~\ref{s:single}. Section~\ref{s:disorder} briefly describes the robustness of soliton features in the density of states to the presence of interlayer disorder. In Section~\ref{s:transport}, we describe ballistic, coherent electronic transport for ABC stacking, comparing the energy dependence of conductivity for systems with and without a soliton.
Section~\ref{s:tight} describes the influence of additional tight-binding parameters beyond the minimal model, which only has nearest-neighbor intra- and interlayer coupling. Relevance to particular materials and the stability of structural defects is discussed in Section~\ref{s:materials}.

\section{Methodology}\label{s:methodology}

We use a minimal tight-binding model with nearest-neighbor intralayer hopping $\gamma_0$ and nearest-neighbor interlayer hopping $\gamma_1$.
Assuming translational invariance within each layer, we Fourier transform to reciprocal space where ${\mathbf{q}} = (q_x,q_y)$ is the in-plane wave vector measured with respect to the center of the Brillouin zone (the $\Gamma$ point).
In a basis of a single orbital on each atomic site $( A_1 , B_1 , A_2 , B_2 , \ldots , A_N , B_N)$, the Hamiltonian is
\begin{eqnarray}
H = \begin{pmatrix}
D_1 & V_1 & 0 & 0 & 0 & \hdots \\
V_1^{\dagger} & D_2 & V_2 & 0 & 0 & \hdots \\
0 & V_2^{\dagger} & D_3 & V_1 & 0 & \hdots \\
0 & 0 & V_1^{\dagger} & D_4 & V_2 & \hdots \\
\vdots & \vdots & \vdots & \vdots & \vdots & \ddots
\end{pmatrix} , \label{h1}
\end{eqnarray}
written in terms of $2 \times 2$ blocks.
Intralayer blocks are
\begin{eqnarray}
D_i = \begin{pmatrix}
U_i & - \gamma_0 f({\bf q}) \\
- \gamma_0 f^{\ast}\! ({\bf q}) & - U_i
\end{pmatrix} , \label{di}
\end{eqnarray}
where $\pm U_i$ are onsite energies and intralayer hopping is described~\cite{mccann13} by
\begin{eqnarray}
f({\bf q}) = e^{iq_y a / \sqrt{3}} + 2 e^{-iq_y a / (2\sqrt{3})}
\cos (q_x a / 2) , \label{fq}
\end{eqnarray}
where $a$ is the in-plane lattice constant.
For faultless systems with non-primed stackings, e.g. Fig.~\ref{fig:summary}(a), (c), (e), then $U_i = U$ for all $i$, whereas, for faultless systems with primed stackings, e.g. Fig.~\ref{fig:summary}(b), (d), (f), then $U_i = U$ for odd $i$ and $U_i = -U$ for even $i$.

The form of the interlayer block $V_j$ in the Hamiltonian~(\ref{h1}) depends on the stacking type:
\begin{eqnarray}
{\mathrm{ABC}}\,{\mathrm{(Rhombohedral)}}\!:\qquad 
V_1 = V_2 &=& \begin{pmatrix}
0 & 0 \\
\gamma_1 & 0
\end{pmatrix} \! , \label{vabc} \\
{\mathrm{ABA}}\,{\mathrm{(Bernal)}}\!:\qquad 
V_1 = V_2^{\dagger} &=& \begin{pmatrix}
0 & 0 \\
\gamma_1 & 0
\end{pmatrix} \! , \label{vaba} \\
{\mathrm{AA}}\!:\qquad 
V_1 = V_2 &=& \begin{pmatrix}
\gamma_1 & 0 \\
0 & \gamma_1
\end{pmatrix} \! , \label{vaa}
\end{eqnarray}
without any dependence on whether the stacking is primed or not.

A system of $N$ layers with an atomically-sharp fault in the onsite energies is labelled $(m,n)$ where $m$ is the number of layers below the fault, $n$ is the number of layers above the fault, and $N = m + n$.
Such a fault is modeled by reversing the signs of $U_i$ on the layers above the fault, as shown in Fig.~\ref{fig:summary}(g)-(l).
Note that the two sites, $A_1$ and $B_1$, on the first layer are always considered to have fixed energies of $\epsilon_{A1} = U$ and $\epsilon_{B1} = -U$. If one were to reverse this convention, there would be the same outcomes with electron-hole inversion of the band structures so that, for example, solitons would become antisolitons. 

For the band structure plots we numerically diagonalize the Hamiltonian~(\ref{h1}).
We use tight-binding parameters experimentally measured~\cite{kuzmenko09} for Bernal-stacked bilayer graphene $\gamma_0 = 3.16\,$eV and $\gamma_1 = 0.381\,$eV, and lattice constant $a = 2.46$\AA~\cite{saito98}.
Band structures are calculated in the vicinity of valley $K_{+}$ with wave vector ${\mathbf{K_{+}}} = ( 4 \pi / (3a) , 0 )$ by shifting the wave vector as ${\mathbf{q}} = {\mathbf{K_{+}}} + {\mathbf{k}}$, with ${\mathbf{k}} = (k_x,k_y)$.
Specifically, we determine the energy eigenvalues $E_n$ on a square grid of points centered on $K_{+}$.
The band structures have valley degeneracy (under a suitable rotation) and they are approximately isotropic around each valley at the energies we consider, so we plot them as a function of component $k_x$ (with $k_y=0$), normalized by the characteristic wave vector $k_c = \gamma_1 / \hbar v$~\cite{snyman07} where $v = \sqrt{3} a \gamma_0 / (2\hbar)$ is the velocity related to intralayer hopping. By plotting $k_x$ as a function of $k_c$ and energy $E$ as a function of $\gamma_1$, there are only two free parameters in our model, namely the ratios $\gamma_0 / \gamma_1$ and $U/\gamma_1$.
The density of states $g (E)$ per unit energy per unit area ($L^2$)  is determined numerically by approximation using a Lorentzian with a finite width~$\delta$,
\begin{eqnarray}
g (E) = \frac{1}{\pi L^2} \sum_n \frac{\delta}{(E - E_n)^2 + \delta^2} . \label{dos}
\end{eqnarray}
States localized on solitons may or may not appear in the bulk band gap, but we focus on those that do.
Such states are extended states in the in-plane direction, but are localized on the soliton in the out-of-plane ($z$) direction. We identify them by observing the form of the state in the atomic basis of the Hamiltonian~(\ref{h1}); an example is shown in Fig.~\ref{figABC2}(a).

For analytic calculations, we expand the function $f({\bf q})$, Eq.~(\ref{fq}), for $|{\mathbf{k}}| \ll k_c$, as $f({\mathbf{K_{\xi}}} + {\bf k}) \approx - \sqrt{3} a (\xi k_x - i k_y)/2$ so that the intralayer hopping matrix element for layer $i$ is $H_{A_iB_i} \approx \hbar v (\xi k_x - i k_y)$ where $\xi = \pm 1$ is a valley index for the valley centers at wave vectors ${\mathbf{K_{\xi}}} = \xi ( 4 \pi / (3a) , 0 )$.
With this approximation, the phase of $H_{A_iB_i}$ may be gauged away (using a diagonal unitary transformation) so that the intralayer matrix element may be replaced by $H_{A_iB_i} \approx \hbar v k$ where
$k = | \mathbf{k} | = (k_x^2 + k_y^2)^{1/2}$, and the intralayer block is
\begin{eqnarray}
D_i \approx \begin{pmatrix}
U_i & \hbar v k \\
\hbar v k & - U_i
\end{pmatrix} . \label{di2}
\end{eqnarray}

In the interpretation of numerical results, we use dimensional reduction~\cite{ryu10} whereby we consider the in-plane wave vector $\mathbf{k}$ to be a fixed parameter, typically either $k=0$ or $k=k_c$. Then, we can relate the model of stacked two-dimensional layers to a one-dimensional tight-binding model.
In particular, we use analogy with the Rice-Mele model~\cite{ricemele82} which is a one-dimensional tight-binding model with two orbitals per unit cell, alternating nearest-neighbor hopping, and alternating onsite energies. It has two phases with chiral symmetry: The Su-Schrieffer-Heeger (SSH) model~\cite{su79,ssh80,heeger88} with alternating nearest-neighbor hopping but constant onsite energies, and the charge density wave (CDW) model~\cite{kivelson83,brzezicki20,cayssol21,fuchs21,allen22,mccann23} with alternating onsite energies but constant nearest-neighbor hopping.
In position space with open boundary conditions, the CDW model for a system with $J$ orbitals may be written in the atomic basis as
\begin{eqnarray}
H_{\mathrm{CDW}} = \begin{pmatrix}
U & t & 0 & 0 & 0 & \hdots \\
t & -U & t & 0 & 0 & \hdots \\
0 & t & U & t & 0 & \hdots \\
0 & 0 & t & -U & t & \hdots \\
\vdots & \vdots & \vdots & \vdots & \vdots & \ddots
\end{pmatrix} , \label{hcdw}
\end{eqnarray}
where $\pm U$ is the alternating onsite energy and $t$ is the constant nearest-neighbor hopping parameter.
This is a two band insulator with a band gap of $2U$.
For a soliton in the center of the system, i.e. with onsite energies of $\ldots, +U, -U, +U, -U, -U, +U, -U, +U, \ldots$ it was recently shown~\cite{allen22} that a
state is localised on the soliton with an energy $E_{\mathrm{sol}}$ that isn't exactly at zero energy, but within the band gap for $0 < t/U < J/2$, i.e.,
\begin{eqnarray}
-U < E_{\mathrm{sol}} < U \qquad \mathrm{for} \qquad 0 < t/U < J/2 , \label{cdw1}
\end{eqnarray}
where $J$ is the total number of orbitals. For $U/t \gg 1$, perturbation theory to second order in $t$ shows~\cite{allen22} that the energy of this level $E_{\mathrm{sol}}$ is
\begin{eqnarray}
E_{\mathrm{sol}} = - U + t - \frac{t^2}{2U} \qquad \mathrm{for} \qquad U/t \gg 1 . \label{cdw2}
\end{eqnarray}
This energy $E_{\mathrm{sol}}$ is not at zero because the chiral symmetry of the CDW model is nonsymmorphic~\cite{shiozaki15,zhao16,brzezicki20,han20,allen22,mccann23} meaning it involves a translation by half a unit cell, and the ends of the system and   a soliton of finite width break the chiral symmetry.
The nonsymmorphic chiral symmetry may be written as $T_{a/2}S_z$~\cite{allen22,mccann23} where
\begin{eqnarray}
T_{a/2}S_z = \begin{pmatrix}
0 & -1 & 0 & 0 & \hdots & 0 & 0 \\
0 & 0 & 1 & 0 & \hdots & 0 & 0 \\
0 & 0 & 0 & -1 & \hdots & 0 & 0 \\
0 & 0 & 0 & 0 & \hdots & 0 & 0 \\
\vdots & \vdots & \vdots & \vdots & \vdots & \vdots & \vdots \\
0 & 0 & 0 & 0 & \hdots & 0 & -1 \\
1 & 0 & 0 & 0 & \hdots & 0 & 0
\end{pmatrix} . \label{tatsz}
\end{eqnarray}
Here $T_{a/2}$ describes translation by half a unit cell,
\begin{eqnarray}
T_{a/2} = \begin{pmatrix}
0 & 1 & 0 & 0 & \hdots & 0 & 0 \\
0 & 0 & 1 & 0 & \hdots & 0 & 0 \\
0 & 0 & 0 & 1 & \hdots & 0 & 0 \\
0 & 0 & 0 & 0 & \hdots & 0 & 0 \\
\vdots & \vdots & \vdots & \vdots & \vdots & \vdots & \vdots \\
0 & 0 & 0 & 0 & \hdots & 0 & 1 \\
1 & 0 & 0 & 0 & \hdots & 0 & 0
\end{pmatrix} . \label{tat}
\end{eqnarray}
and $S_z$ describes sublattice chiral symmetry for the SSH model, $S_z = \mathrm{diag} (1,-1,1,-1,1,-1,\ldots )$.

The expectation value of $T_{a/2}S_z$ is a generalization of electric polarization,
\begin{eqnarray}
p_y = \langle \psi | T_{a/2}S_z | \psi \rangle , \label{py1}
\end{eqnarray}
where subscript `$y$' is used because operator $T_{a/2}S_z$ is represented as $\sigma_y$ in reciprocal space.
In one-dimensional topological insulators with chiral symmetry, the polarization related to a chiral operator takes values $\pm 1$ for zero-energy topological states~\cite{asboth16}. However, since the nonsymmorphic chiral symmetry is broken in a finite system, the corresponding polarization takes values with magnitude less than one.
For $U/t \gg 1$, perturbation theory~\cite{allen22} gives
\begin{eqnarray}
p_y^{\mathrm{sol}} = \frac{(1 + \tau )^2}{2 (1 + \tau^2)} , \qquad
\tau = \frac{t}{2U} ,\label{py2}
\end{eqnarray}
for a soliton state, so that $p_y^{\mathrm{sol}} = 1/2$ for $t/U = 0$ and $p_y^{\mathrm{sol}} > 1/2$ for $t/U > 0$.
For an antisoliton, $p_y^{\mathrm{asol}} = - p_y^{\mathrm{sol}}$.

\section{Rhombohedral stacking}\label{s:rhomb}

\subsection{ABC}\label{s:rabc}

\begin{figure*}[t]
\includegraphics[scale=0.4]{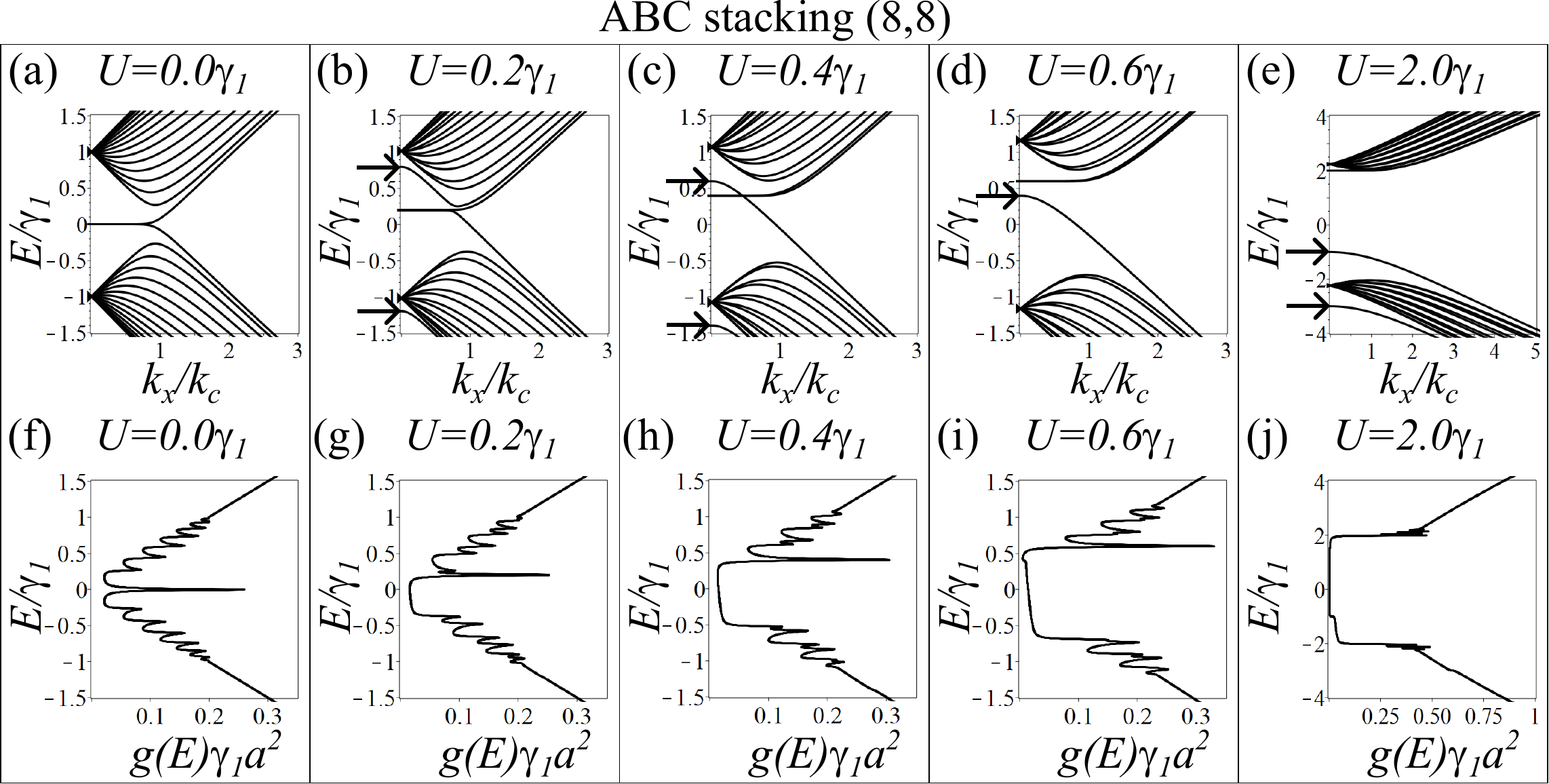}
\caption{ABC stacking with $N=16$ layers and a sharp soliton at the center $(8,8)$. The top row shows band structures, the bottom row shows the corresponding density of states $g(E)$ for different values of the magnitude of the alternating onsite energies as (a), (f) $U/\gamma_1 = 0.0$, (b), (g) $U/\gamma_1 = 0.2$, (c), (h) $U/\gamma_1 = 0.4$, (d), (i) $U/\gamma_1 = 0.6$, and (e), (j) $U/\gamma_1 = 2.0$.
In the band structure plots, horizontal arrows show the energies at $k_x=0$ of the states localized on the soliton.
Note that the axis scales for (e), (j) differ from the rest. In all plots, parameter values are $\gamma_0 = 3.16\,$eV, $\gamma_1 = 0.381\,$eV~\cite{kuzmenko09}, $a = 2.46$\AA~\cite{saito98}. For the band structures, $k_y=0$, and, for the density of states, $\delta = 0.01 \gamma_1$.
}\label{figABC1}
\end{figure*}

\subsubsection{Faultless system}

For bulk, faultless ABC stacking, the unit cell consists of two atoms $A_i$, $B_i$ on the $i$th layer, and the Hamiltonian may be written as
\begin{eqnarray}
H_{\mathrm{ABC}} = \begin{pmatrix}
U & \hbar v k + \gamma_1 e^{-i q_z d} \\
\hbar v k + \gamma_1 e^{i q_z d} & - U
\end{pmatrix} , \label{habc}
\end{eqnarray}
using the approximation for the intralayer hopping~(\ref{di2}) with $k<k_c$, where $q_z$ is the $z$ component of the wave vector and $d$ is the layer separation.
The two corresponding bulk bands are
\begin{eqnarray*}
E_{\mathrm{ABC}} = \pm \sqrt{U^2 + \gamma_1^2 + (\hbar vk)^2 + 2 \hbar v k \gamma_1 \cos q_z d} ,
\end{eqnarray*}
which have a band gap of $2U$.
If we consider the in-plane wave vector $k$ to be a fixed parameter, then the Hamiltonian~(\ref{habc}) is that of the Rice-Mele model~\cite{ricemele82}.

For $U = 0$,  the system is rhombohedral graphite which has a bulk energy gap and two flat bands at zero energy arising from surface states (localised on sites $A_1$ and $B_N$), as shown in Fig.~\ref{figABC1}(a) and (f). The flat bands give rise to a narrow peak in the density of states at zero energy.
This band structure may be understood by dimensional reduction~\cite{ryu10,heikkila11,xiao11} as being related to the SSH model, whereby, on treating the in-plane wave vector ${\mathbf{k}}$ as a fixed parameter, the Hamiltonian of rhombohedral graphite is the same as that of the SSH model. The point in the band structure where $k_x$ is equal to the characteristic wave vector $k_c = \gamma_1 / (\hbar v)$ is equivalent to the metallic phase of the SSH model, and, thus, $k_c$ defines the extent of the flat bands in ${\mathbf{k}}$ space.

\subsubsection{Single soliton}

For finite $U$, we consider a sharp fault consisting of two adjacent sites with negative onsite energies as shown in Fig.~\ref{fig:summary}(g).
Band structures and density of states for $N=16$ layers and a sharp fault occurring between the middle two layers, denoted $(8,8)$, are shown in Fig.~\ref{figABC1}.
There are two bands localized on the fault, as indicated by the arrows in the band structure plots of Fig.~\ref{figABC1}.
Both of these bands move into the valence band for $k_x \gg k_c$, hence we refer to this fault as a soliton~\cite{conventionnote}.
We focus on the higher energy band of the two because it is near zero energy.
For $U \leq \gamma_1/2$, this soliton band hybridizes with the flat bands arising from the surface states, Fig.~\ref{figABC1}(b) and  Fig.~\ref{figABC1}(c).
Note that the soliton changes the texture of onsite energies, flipping the sign of the onsite energy on the top surface (site $B_N$). This can be seen in Fig.~\ref{fig:summary}(g) where the soliton (between sites $B_4$ and $A_5$) changes the order of the onsite energies for higher layers (with layer index $j \geq 5$) so that the final orbital, $B_8$, has the same onsite energy as the first, $A_1$, in contrast to the faultless case, Fig.~\ref{fig:summary}(a). As a result, both surface states, localized on $A_1$ and $B_8$, are at positive energy and move into the conduction band for $k_x \gg k_c$.

For $U > \gamma_1/2$, Fig.~\ref{figABC1}(d) and Fig.~\ref{figABC1}(e), the soliton band and the surface bands are always separate, although the soliton band remains within the band gap. Thus, for $U > \gamma_1/2$, there is a band gap, but it is smaller than the bulk band gap of $2U$, Fig.~\ref{figABC1}(j).
The numerical plots may be understood by examining the form of the Hamiltonian~(\ref{h1}) at $k = 0$ because the intralayer hopping $-\gamma_0 f({\mathbf{q}})$ is zero there, and the Hamiltonian~(\ref{h1}) becomes block diagonal.
The two surface states, on $A_1$ and $B_N$, are completely disconnected at $k = 0$ and they give two degenerate states at energy $E = +U$.
There are $N-2$ dimers with onsite energies $\pm U$ which each give two states of energies $E = \pm \sqrt{U^2 + \gamma_1^2}$.
Finally, the soliton is a dimer of sites $B_{m}$ and $A_{m+1}$ with onsite energies both at $-U$, yielding  two states with energies $E = - U \pm \gamma_1$. The soliton level is the highest energy of the pair,
\begin{eqnarray}
E_{\mathrm{sol}} = - U + \gamma_1 \quad \mathrm{for} \, k = 0 . \label{abcesol1}
\end{eqnarray}
Given that the soliton band begins at energy $- U + \gamma_1$ for $k = 0$, and joins the valence band for $k \gg k_c$, there will be a part of this band within the bulk band gap for all non-zero values of $U$.
In the regime $U \gg \gamma_1$ and $U \gg \hbar v k$, it is possible to estimate the energy of the soliton level using perturbation theory in the hopping strength. Using the analytic approximation~(\ref{di2}) for the intralayer hopping matrix element gives
\begin{eqnarray}
\!\!\! \!\!\! E_{\mathrm{sol}}(\mathbf{k}) \approx - U + \gamma_1 - \frac{(\hbar v k)^2}{2U} \,\, \mathrm{for} \,\, \{ \gamma_1 , \hbar v k \} \ll U . \label{abcesol3}
\end{eqnarray}
Note that expressions~(\ref{abcesol1},\ref{abcesol3}) for the energy $E_{\mathrm{sol}}$ of the band localized on the soliton are independent of the position of the soliton and the number of layers $N$ in the system.

Soliton states are extended states in the in-plane direction, but are localized in the out-of-plane ($z$) direction.
At $k = 0$, the soliton state is localized on the two dimer sites $B_{m}$ and $A_{m+1}$, i.e. it is given by $\psi_{\mathrm{sol}} = (\psi_{B_{m}} + \psi_{A_{m+1}})/\sqrt{2}$ where $\psi_j$ denotes the atomic orbital on site $j$.
For $U \approx \gamma_1/2$ and $k \ll k_c$ we derive an effective $3 \times 3$ Hamiltonian describing the hybridization of the soliton state, $\psi_{\mathrm{sol}} = (\psi_{B_{m}} + \psi_{A_{m+1}})/\sqrt{2}$, with the two surface states $\psi_{A_{1}}$ and $\psi_{B_{N}}$.
This is done using the linear expansion $f({\mathbf{K_{\xi}}} + {\bf k}) \approx - \sqrt{3} a (\xi k_x - i k_y)/2$, eliminating the other orbitals~\cite{mccann06,mccann13} and performing an expansion for $k/k_c \ll 1$ and $|U/\gamma_1 - 1/2| \ll 1$.
In basis $\psi_{A_1} , (\psi_{B_{m}} + \psi_{A_{m+1}})/\sqrt{2} , \psi_{B_N}$, the effective Hamiltonian is
\begin{widetext}
\begin{eqnarray}
H_{\mathrm{sol}}^{(m,n)} = \gamma_1 \!
\begin{pmatrix}
\frac{1}{2} + \Delta + \frac{2}{9} ( \frac{4}{5} )^{2(m-1)} (\kappa \kappa^{\dagger})^{m} &
\frac{1}{\sqrt{2}} ( - \frac{4}{5} )^{m-1} (\kappa^{\dagger})^m &
- \frac{2}{9} ( - \frac{4}{5} )^{m+n-2} (\kappa^{\dagger})^{m+n} \\
\frac{1}{\sqrt{2}} ( - \frac{4}{5} )^{m-1} \kappa^m &
\frac{1}{2} - \Delta - \frac{4}{5} c_{m,n} \kappa \kappa^{\dagger}
& \frac{1}{\sqrt{2}} ( - \frac{4}{5} )^{n-1} (\kappa^{\dagger})^n  \\
- \frac{2}{9} ( - \frac{4}{5} )^{m+n-2} \kappa^{m+n} &
\frac{1}{\sqrt{2}} ( - \frac{4}{5} )^{n-1} \kappa^n  &
\frac{1}{2} + \Delta + \frac{2}{9} ( \frac{4}{5} )^{2(n-1)} (\kappa \kappa^{\dagger})^{n}
\end{pmatrix} \! , \label{hsol}
\end{eqnarray}
\end{widetext}
where $\Delta$ is a dimensionless contribution to the onsite energies, $U/\gamma_1 = 1/2 + \Delta$ with $|\Delta| \ll 1$, $\kappa$ is a dimensionless complex wave vector $\kappa = (\xi k_x + i k_y)/k_c$, $\kappa^{\dagger} = (\xi k_x - i k_y)/k_c$ with $|\kappa| \ll 1$, and $c_{m,n}$ is a numerical factor,
\begin{eqnarray}
c_{m,n} = \begin{cases}
1 &\mbox{if } n = m , \\
1/2 & \mbox{if } n \neq m .
\end{cases}
\end{eqnarray}
For each matrix element in the effective Hamiltonian~(\ref{hsol}), we retain the leading order term in $|\kappa|$.

An antisoliton $(m,n)$ has the same structure as a soliton $(m,n)$ but with the signs of all of the onsite energies reversed. At $k = 0$, the antisoliton is a dimer of sites $B_{m}$ and $A_{m+1}$ with onsite energies both at $U$, yielding  two states with energies $E = U \pm \gamma_1$.
The antisoliton state is the lowest state of the pair,
\begin{eqnarray}
E_{\mathrm{asol}} = U - \gamma_1 \quad \mathrm{for} \, k = 0 , \label{abcesol2}
\end{eqnarray}
and it is localized on the two dimer sites $B_{m}$ and $A_{m+1}$ as $\psi_{\mathrm{asol}} = (\psi_{B_{m}} - \psi_{A_{m+1}})/\sqrt{2}$.
The effective Hamiltonian in the basis $\psi_{A_1} , (\psi_{B_m} - \psi_{A_{m+1}})/\sqrt{2} , \psi_{B_N}$, is
\begin{widetext}
\begin{eqnarray}
H_{\mathrm{asol}}^{(m,n)} = \gamma_1
\begin{pmatrix}
- \frac{1}{2} - \Delta - \frac{2}{9} ( \frac{4}{5} )^{2(m-1)} (\kappa \kappa^{\dagger})^{m} &
\frac{1}{\sqrt{2}} ( - \frac{4}{5} )^{m-1} (\kappa^{\dagger})^m &
- \frac{2}{9} ( - \frac{4}{5} )^{m+n-2} (\kappa^{\dagger})^{m+n} \\
\frac{1}{\sqrt{2}} ( - \frac{4}{5} )^{m-1} \kappa^m &
- \frac{1}{2} + \Delta + \frac{4}{5} c_{m,n} \kappa \kappa^{\dagger}
& - \frac{1}{\sqrt{2}} ( - \frac{4}{5} )^{n-1} (\kappa^{\dagger})^n  \\
- \frac{2}{9} ( - \frac{4}{5} )^{m+n-2} \kappa^{m+n} &
- \frac{1}{\sqrt{2}} ( - \frac{4}{5} )^{n-1} \kappa^n  &
- \frac{1}{2} - \Delta - \frac{2}{9} ( \frac{4}{5} )^{2(n-1)} (\kappa \kappa^{\dagger})^{n}
\end{pmatrix} ,
\end{eqnarray}
\end{widetext}
where $U/\gamma_1 = 1/2 + \Delta$ with $|\Delta| \ll 1$, and $|\kappa| \ll 1$.

\begin{figure}[t]
\includegraphics[scale=0.5]{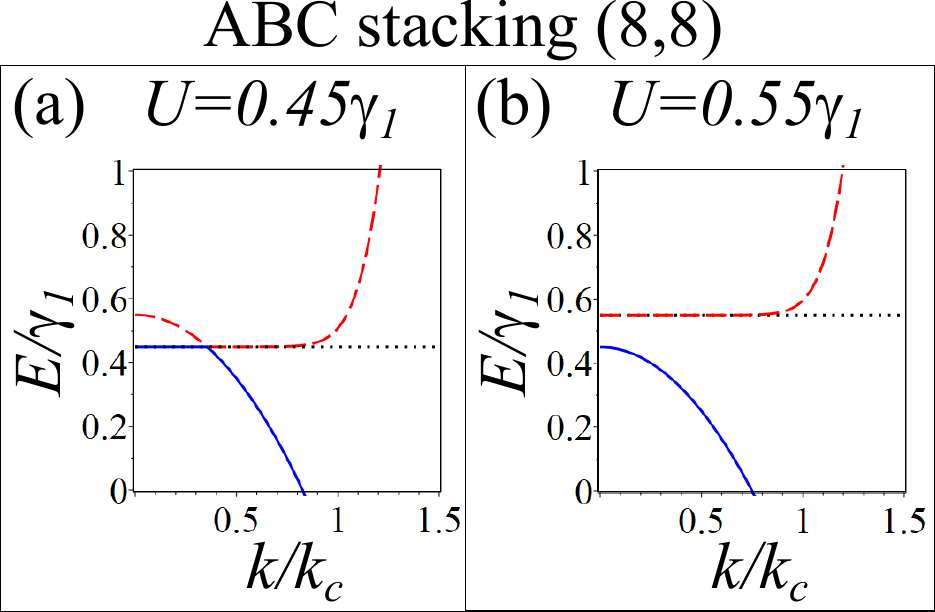}
\caption{ABC stacking with $N=16$ layers and a sharp soliton at the center $(8,8)$. Plots of the analytic band energies of bands hybridized between the soliton and surface states for (a) $U/\gamma_1 = 0.45$ and (b) $U/\gamma_1 = 0.55$. The black dotted line is the flat band $E=U$, the red dashed line is the solution corresponding to the `$+$' sign in Eq.~(\ref{abchyb}), and the blue solid line corresponds to the `$-$' sign.
}\label{figABC3}
\end{figure}

To analyze the hybridization of the soliton band with the surface states, we simplify the soliton effective Hamiltonian~(\ref{hsol}) by considering a soliton at the center of a system with an even number of layers, $m=n=N/2$, and we consider a large number of layers $N \gg 1$ to neglect terms of order $(k/k_c)^{N}$,
\begin{eqnarray}
H_{\mathrm{sol}}^{(m,m)} \approx \gamma_1 \!
\begin{pmatrix}
\frac{1}{2} + \Delta &
\beta^{\dagger} &
0 \\
\beta &
\frac{1}{2} - \Delta - \alpha
& \beta^{\dagger}  \\
0 &
\beta  &
\frac{1}{2} + \Delta
\end{pmatrix} \! , \label{hsol2}
\end{eqnarray}
where $\alpha =  (4/5) |\kappa|^2$ and $\beta = (1/\sqrt{2}) ( - 4/5 )^{m-1} \kappa^m$.
The eigenvalues of the three states are $E/\gamma_1 = 1/2 + \Delta$ and $E/\gamma_1 = 1/2 - \alpha/2 \pm \sqrt{(\Delta + \alpha/2)^2 + 2|\beta|^2}$ or, in terms of physical parameters, there is a flat band $E=U$ and two hybridized bands,
\begin{eqnarray}
\frac{E}{\gamma_1} &=& \frac{1}{2} - \frac{2}{5} \bigg( \frac{k}{k_c} \bigg)^2 \label{abchyb} \\
&& \pm \sqrt{\bigg[ \frac{U}{\gamma_1} - \frac{1}{2} + \frac{2}{5}\bigg( \frac{k}{k_c} \bigg)^2 \bigg]^2 + \bigg( \frac{4}{5} \bigg)^{N-2}\bigg( \frac{k}{k_c} \bigg)^{N}} , \nonumber
\end{eqnarray}
for $U \approx \gamma_1/2$, $k \ll k_c$, and layer number $N \gg 1$.
By considering the eigenstates of the effective Hamiltonian~(\ref{hsol2}), it is possible to see that the flat band $E=U$ has zero weight on the soliton state, i.e., it is a linear combination of surface states only and, hence, it is dispersionless because of the approximation neglecting terms of order $(k/k_c)^{N}$.
The analytic solutions for the flat band $E=U$ and the hybridized bands Eq.~(\ref{abchyb}) are plotted in Fig.~\ref{figABC3} for energy values near $U = \gamma_1/2$.

\begin{figure}[t]
\includegraphics[scale=0.5]{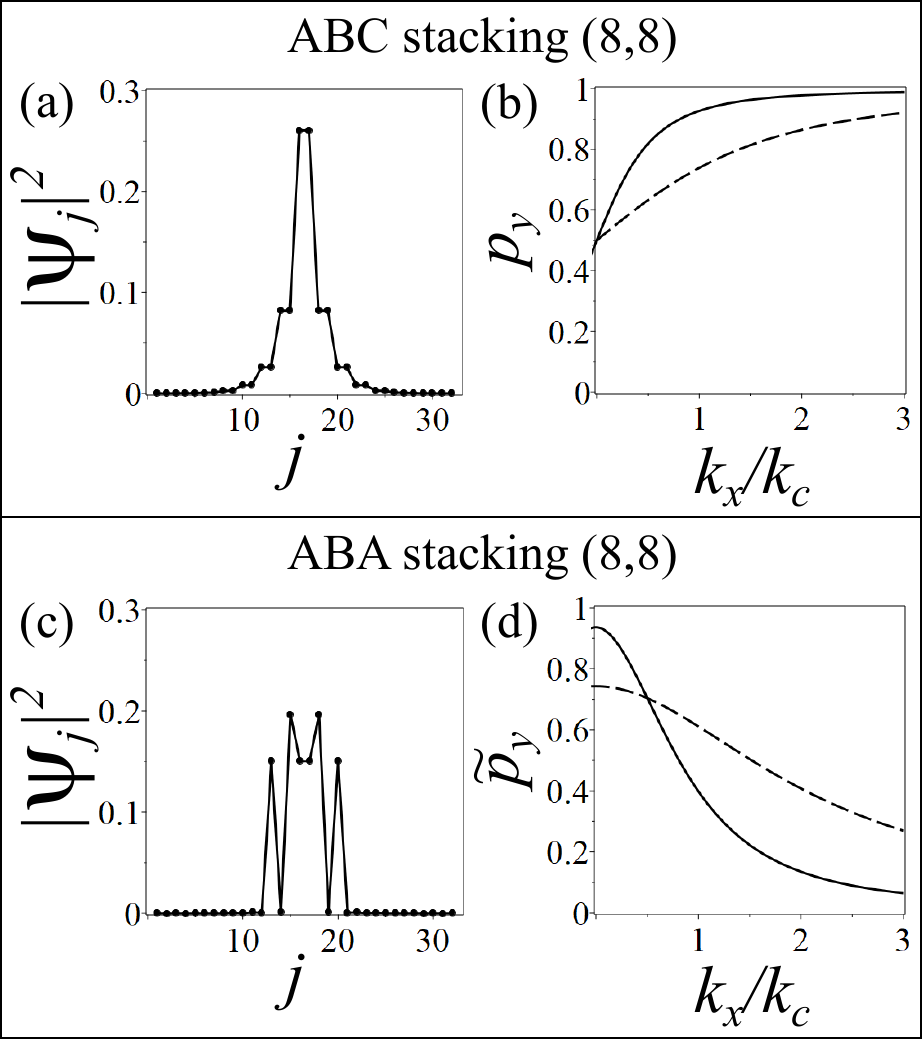}
\caption{A sharp soliton at the center of the system $(8,8)$, $N=16$ layers, for ABC stacking (top) and ABA stacking (bottom).
(a) The probability density $|\psi_j|^2$ per site $j = 1, 2, \ldots , 32$ for the energy level localized on the soliton at $k_x = k_c$, $k_y = 0$ and $U/\gamma_1 = 0.6$. (b) The polarization $p_y = \langle \psi | T_{a/2}S_z | \psi \rangle$, Eq.~(\ref{py1}), of the soliton state as a function of $k_x$ with $k_y = 0$ for $U/\gamma_1 = 0.6$ (solid) and $U/\gamma_1 = 1.8$ (dashed).
(c) The probability density $|\psi_j|^2$ per site $j = 1, 2, \ldots , 32$ for the energy level localized on the soliton at $k_x = k_c$, $k_y = 0$ and $U/\gamma_1 = 0.6$. (d) The polarization $\tilde{p}_y = \langle \psi | \tilde{T}_{a/2}S_z | \psi \rangle$, Eq.~(\ref{altpy1}), of the soliton state as a function of $k_x$ with $k_y = 0$ for $U/\gamma_1 = 0.6$ (solid) and $U/\gamma_1 = 1.8$ (dashed).
In all plots, $\gamma_0 = 3.16\,$eV, $\gamma_1 = 0.381\,$eV~\cite{kuzmenko09}, $a = 2.46$\AA~\cite{saito98}.
}\label{figABC2}
\end{figure}

For $k=k_c$, the Hamiltonian~(\ref{habc}) is approximately equal to that of the CDW model (this equivalence is only approximate because $k / k_c$ is not small when $k = k_c$) with the replacement $t \equiv \gamma_1$.
For $k = 0$, the soliton state is localized on the two adjacent dimer sites as $\psi_{\mathrm{sol}} = (\psi_{B_{m}} + \psi_{A_{m+1}})/\sqrt{2}$. 
For $k = k_c$, it remains localized on the soliton, but has a broader extent in position space as shown in Fig.~\ref{figABC2}(a) for $U/\gamma_1 = 0.6$.
Fig.~\ref{figABC2}(b) shows the polarization $p_y = \langle \psi | T_{a/2}S_z | \psi \rangle$, Eq.~(\ref{py1}), as a function of $k_x$ (with $k_y = 0$) for $U/\gamma_1 = 0.6$ (solid line) and $U/\gamma_1 = 1.8$ (dashed line).
For $U / \gamma_1 \agt 1$, the value of $p_y$ at $k_x = k_c$ is in good agreement with the analytic expectation~(\ref{py2}) with $\tau = \gamma_1 / (2U)$ (dashed line), and the value of $p_y$ increases as $U/\gamma_1$ decreases (solid line). The polarization $p_y$ never reaches unity because the nonsymmorphic chiral symmetry is broken by the ends of the system and the finite width of the soliton.

\subsubsection{Soliton-antisoliton pair}

A soliton-antisoliton pair is denoted $(\ell,m,n)$ meaning a soliton (with consecutive onsite energies of $-U,-U$) after $\ell$ layers and an antisoliton (with consecutive onsite energies of $+U,+U$) after $\ell + m$ layers, with $N = \ell + m + n$. For example, $(2,2,1)$ would indicate onsite energies of $+U,-U,+U,-U,-U,+U,-U,+U,+U,-U$.

\begin{figure*}[t]
\includegraphics[scale=0.38]{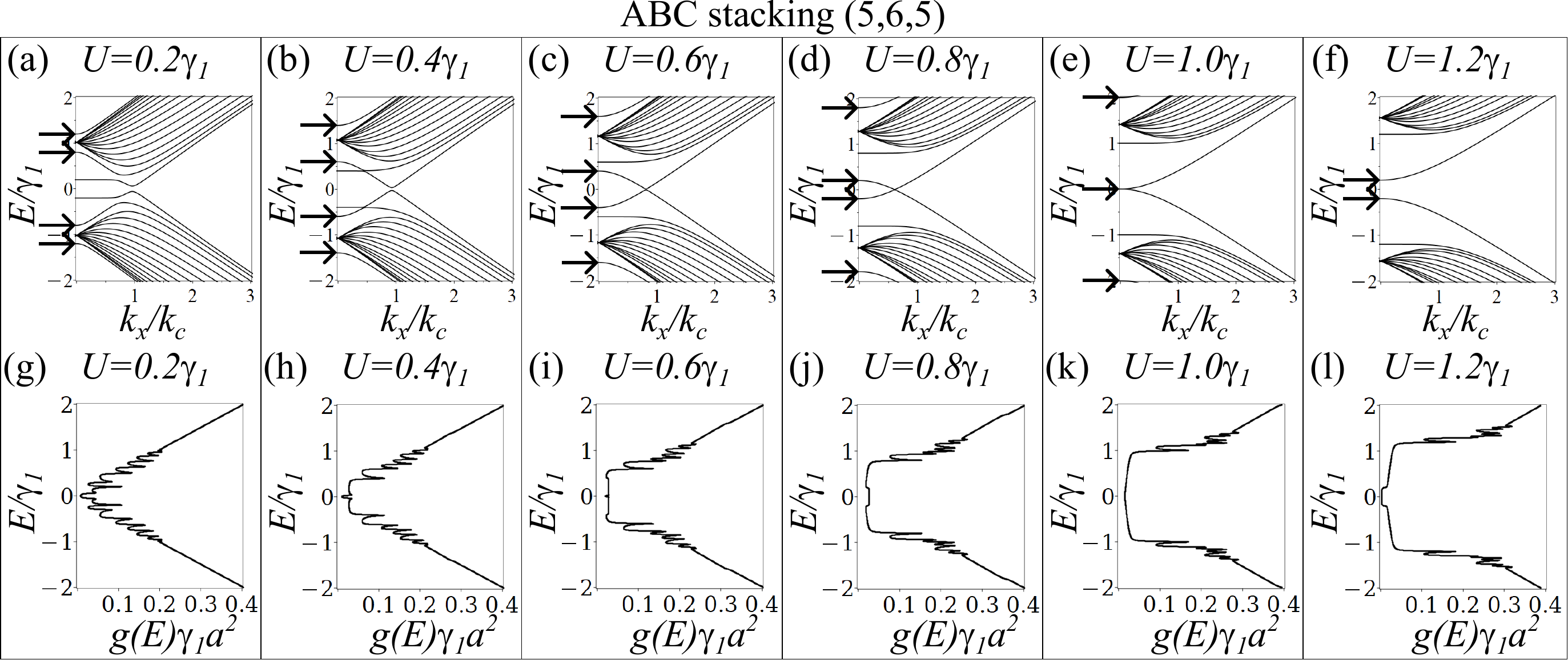}
\caption{ABC stacking with $N=16$ layers and a soliton-antisoliton pair $(5,6,5)$. The top row shows band structures, the bottom row shows the corresponding density of states $g(E)$ for different values of the magnitude of the alternating onsite energies as (a), (g) $U/\gamma_1 = 0.2$, (b), (h) $U/\gamma_1 = 0.4$, (c), (i) $U/\gamma_1 = 0.6$, (d), (j) $U/\gamma_1 = 0.8$, (e), (k) $U/\gamma_1 = 1.0$, and (f), (l) $U/\gamma_1 = 1.2$.
In the band structure plots, horizontal arrows show the energies at $k_x=0$ of the states localized on the soliton or antisoliton.
In all plots, parameter values are $\gamma_0 = 3.16\,$eV, $\gamma_1 = 0.381\,$eV~\cite{kuzmenko09}, $a = 2.46$\AA~\cite{saito98}. For the band structures, $k_y=0$, and, for the density of states, $\delta = 0.01 \gamma_1$.
}\label{figABC4}
\end{figure*}

The band structure and density of states of a soliton-antisoliton pair $(5,6,5)$ with $N=16$ layers is shown in Fig.~\ref{figABC4} for different values of $U$. There are two localized states associated with the soliton, two with the antisoliton, and two surface states. The two surface states, one soliton state and one antisoliton state generally lie within the bulk band gap. 
At $k = 0$, the soliton has energy $E_{\mathrm{sol}} = - U + \gamma_1$, the antisoliton $E_{\mathrm{asol}} = U - \gamma_1$, and the surface states have energies $\pm U$.

For $U \leq \gamma_1/2$, the soliton and antisoliton state hybridize with the two surface states, Fig.~\ref{figABC4}(b) and Fig.~\ref{figABC4}(h). For $\gamma_1 > U > \gamma_1/2$, the soliton and antisoliton are closer to zero energy than the surface states, and they tend to hybridize together with a tiny anticrossing, Fig.~\ref{figABC4}(c) and Fig.~\ref{figABC4}(d). For $U > \gamma_1$, the soliton and antisoliton states separate, leaving an overall band gap $E_{\mathrm{g}} = 2 (U - \gamma_1)$, Fig.~\ref{figABC4}(f) and Fig.~\ref{figABC4}(l).

For $U \approx \gamma_1/2$, the soliton state will strongly hybridize with the bottom surface state (on $A_1$) at energy $+\gamma_1/2$, and the antisoliton state will strongly hybridize with the top surface state (on $B_N$) at energy $-\gamma_1/2$.
We derive an effective $4 \times 4$ Hamiltonian in the basis
$\psi_{A_1} , (\psi_{B_{\ell}} + \psi_{A_{\ell+1}})/\sqrt{2} , (\psi_{B_{\ell +m}} - \psi_{A_{\ell +m+1}})/\sqrt{2} , \psi_{B_N}$ as
\begin{widetext}
\begin{eqnarray*}
\!\!\!\!\!\! \!\!\! H_{\mathrm{sol-asol}}^{(\ell,m,n)}
= \gamma_1 \!\!
\begin{pmatrix}
\frac{1}{2} + \Delta + \frac{2}{9} (\frac{4}{5})^{2(\ell-1)} (\kappa \kappa^{\dagger})^{\ell} &
\frac{1}{\sqrt{2}} (-\frac{4}{5})^{\ell-1} (\kappa^{\dagger})^{\ell} &
- \frac{1}{3\sqrt{2}} (-\frac{4}{5})^{\ell + m -2} (\kappa^{\dagger})^{\ell + m} &
0  \\
\frac{1}{\sqrt{2}} (-\frac{4}{5})^{\ell-1} (\kappa)^{\ell} &
\frac{1}{2} - \Delta - f_{\ell,m} \kappa \kappa^{\dagger} &
\frac{1}{2} (-\frac{4}{5})^{m-1} (\kappa^{\dagger})^{m} &
-\frac{1}{3\sqrt{2}} (-\frac{4}{5})^{n + m -2} (\kappa^{\dagger})^{n + m} \\
- \frac{1}{3\sqrt{2}} (-\frac{4}{5})^{\ell + m -2} \kappa^{\ell + m} &
\frac{1}{2} (-\frac{4}{5})^{m-1} \kappa^{m} &
-\frac{1}{2} + \Delta + f_{n,m} \kappa \kappa^{\dagger} &
-\frac{1}{\sqrt{2}} (- \frac{4}{5})^{n-1} (\kappa^{\dagger})^{n} \\
0  &
-\frac{1}{3\sqrt{2}} (-\frac{4}{5})^{n + m -2} \kappa^{n + m} &
-\frac{1}{\sqrt{2}} (-\frac{4}{5})^{n-1} \kappa^{n} &
-\frac{1}{2} - \Delta - \frac{2}{9} (\frac{4}{5})^{2(n-1)} (\kappa \kappa^{\dagger})^{n}
\end{pmatrix} \!\! ,
\end{eqnarray*}
\end{widetext}
where $U/\gamma_1 = 1/2 + \Delta$ with $|\Delta| \ll 1$, and $\kappa = (\xi k_x + i k_y)/k_c$ with $|\kappa| \ll 1$.
Here we neglect the contributions $(H_{\mathrm{sol-asol}}^{(\ell,m,n)})_{14} = (H_{\mathrm{sol-asol}}^{(\ell,m,n)})_{41}^{\ast} \sim {\cal O} ( \gamma_1 |\kappa|^{N+2} )$, and $f_{n,m}$ is a numerical factor,
\begin{eqnarray}
f_{n,m} =
\begin{cases}
2/9 &\mbox{for } n=1\,\,\mathrm{and}\,\,m=1 \\
2/5 &\mbox{for } n=1\,\,\mathrm{and}\,\,m>1 \\
28/45 &\mbox{for } n>1\,\,\mathrm{and}\,\,m=1 \\
4/5 &\mbox{for } n>1\,\,\mathrm{and}\,\,m>1
\end{cases}
\end{eqnarray}

For $U \approx \gamma_1$, the soliton and antisoliton state are both near zero energy at $k = 0$ and they hybridize for $k \neq 0$ (the surface bands are distant, at energies $\pm U$).
We use a two-component basis $(\psi_{B_{\ell}} + \psi_{A_{\ell+1}})/\sqrt{2} , (\psi_{B_{\ell +m}} - \psi_{A_{\ell +m+1}})/\sqrt{2}$ to write
\begin{eqnarray}
H_{\mathrm{sol-asol}}^{(\ell,m,n)}
= \gamma_1
\begin{pmatrix}
-\delta - g_{m,\ell} \kappa \kappa^{\dagger} &
-(-\kappa^{\dagger}/2)^{m} \\
-(-\kappa /2)^{m} &
\delta + g_{m,n} \kappa \kappa^{\dagger}
\end{pmatrix} , \label{h2}
\end{eqnarray}
where $U/\gamma_1 = 1 + \delta$ with $|\delta| \ll 1$, $|\kappa| \ll 1$, and $g_{m,\ell}$ is a numerical factor,
\begin{eqnarray}
g_{m,\ell} =
\begin{cases}
5/8 &\mbox{for } m=1\,\,\mathrm{and}\,\,\ell=1 \\
3/8 &\mbox{for } m=1\,\,\mathrm{and}\,\,\ell>1 \\
3/4 &\mbox{for } m>1\,\,\mathrm{and}\,\,\ell=1 \\
1/2 &\mbox{for } m>1\,\,\mathrm{and}\,\,\ell>1
\end{cases}
\end{eqnarray}
Hamiltonian~(\ref{h2}) has eigenvalues given by
\begin{eqnarray}
\frac{E}{\gamma_1} &=& \frac{1}{2} ( g_{m,n} - g_{m,\ell} ) \bigg( \frac{k}{k_c} \bigg)^2 \\
&& \!\!\!\!\!\! \pm \sqrt{ \bigg[ \frac{U}{\gamma_1} - 1 + \frac{1}{2} ( g_{m,n} + g_{m,\ell} ) \bigg( \frac{k}{k_c} \bigg)^2 \bigg]^2
+ \bigg( \frac{k}{2k_c} \bigg)^{2m} } , \nonumber
\end{eqnarray}
describing hybridization of the soliton and antisoliton states for $U \approx \gamma_1$ and $k \ll k_c$.

\subsection{ABC$^{\prime}$}

\subsubsection{Faultless system}

For bulk, faultless ABC$^{\prime}$ stacking, the unit cell consists of four atoms such as $A_1$, $B_1$, $A_2$, $B_2$ (in Fig.~\ref{fig:summary}(b)), and the Hamiltonian may be written as
\begin{eqnarray}
H_{\mathrm{ABC}^{\prime}} = \begin{pmatrix}
U & \hbar v k & 0 & \gamma_1 e^{-i q_z d} \\
\hbar v k & - U & \gamma_1 e^{i q_z d}& 0 \\
0 & \gamma_1 e^{-i q_z d} & -U & \hbar v k \\
\gamma_1 e^{i q_z d} & 0 & \hbar v k & U
\end{pmatrix} \!\! ,
\end{eqnarray}
using the approximation for the intralayer hopping~(\ref{di2}) with $k<k_c$, where $q_z$ is the $z$ component of the wave vector and $d$ is the layer separation (the length of the unit cell in the $z$ direction is $2d$).
The four corresponding bulk bands are given by
\begin{eqnarray*}
(E_{\mathrm{ABC}^{\prime}})^2 &=& U^2 + \gamma_1^2 + (\hbar vk)^2 \\
&& \qquad \pm \, 2 \gamma_1 \sqrt{ U^2 + (\hbar vk)^2 \cos^2 q_z d} .
\end{eqnarray*}

\begin{figure*}[t]
\includegraphics[scale=0.34]{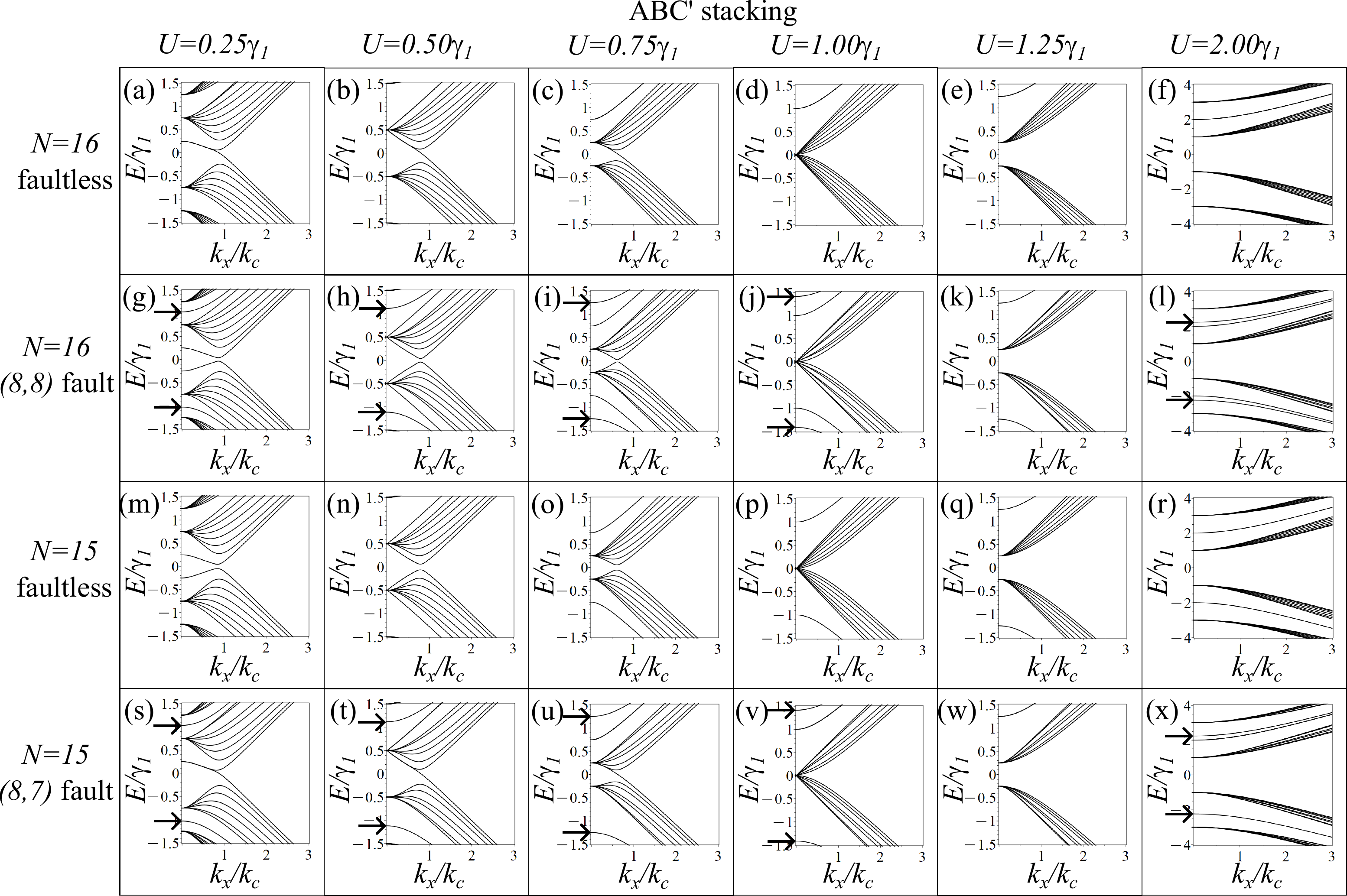}
\caption{Low-energy band structures for ABC$^{\prime}$ stacking. Each column shows a different value of the onsite energies $U$, each row shows a different system. The top row shows a faultless system with $N=16$ layers, whereas the second row shows $N=16$ layers with a sharp fault at its center $(8,8)$.
The third row shows a faultless system with $N=15$ layers, and the final row shows $N=15$ layers with a sharp fault near its center $(8,7)$.
Horizontal arrows show the energies at $k_x=0$ of the states localized on the soliton.
Note that the axis scales for the final column (with $U = 2.00\gamma_1$) differ from the rest.
In all plots, parameter values are $\gamma_0 = 3.16\,$eV, $\gamma_1 = 0.381\,$eV~\cite{kuzmenko09}, $a = 2.46$\AA~\cite{saito98}.
}\label{fig:ABCp1}
\end{figure*}

For a thin film of finite thickness, we begin by considering the faultless system for an even number of layers $N$, Fig.~\ref{fig:ABCp1} (top row).
At $k = 0$, the intralayer hopping is zero, and the system simplifies as a collection of separate dimers plus two isolated surface states.
Thus, the $N$ bands (for $N \geq 4$) are at five distinct energies at $k = 0$ with $E = + U$ (twice) arising from the surface states, and from the dimers: $E = + U + \gamma_1$ ($N/2 - 1$ times), $E = + U - \gamma_1$ ($N/2 - 1$ times), $E = - U + \gamma_1$ ($N/2$ times), and $E = - U - \gamma_1$ ($N/2$ times).
The surface bands appear at lower energy ($E = + U$) than the bulk bands for $0 \leq U < \gamma_1/2$, and the bulk bands touch at $k = 0$ for $U = \gamma_1$.
Considering the bands for all $k$ values, the numerical plots in Fig.~\ref{fig:ABCp1} show that there is always a band within the nominal band gap for $U < \gamma_1$, whereas there is a bandgap for $U > \gamma_1$.

\subsubsection{Single soliton}

For even $N$ and a single fault, Fig.~\ref{fig:ABCp1} (second row), we consider $k = 0$ again. In this case, the $N$ bands (for $N \geq 4$) are at eight distinct energies at $k = 0$ with $E = + U$ and $E= - U$ levels arising from the surface states; the presence of the fault switches the texture of onsite energies, causing the energy of the top surface to flip sign as compared to the faultless case. For dimers at $k = 0$, the energies are $E = + U + \gamma_1$, $E = + U - \gamma_1$, $E = - U + \gamma_1$, and $E = - U - \gamma_1$, each of these occurring $N/2 - 1$ times. In addition, there is the dimer consisting of the stacking fault giving energies $E = \pm \sqrt{U^2 + \gamma_1^2}$.
For nonzero $k$, for $0 < U < \gamma_1$, there is a small anticrossing between two low-energy bands, then, for $U = \gamma_1$, the bands touch, and, for $U > \gamma_1$, there is a band gap $E_{{\mathrm{g}}} = 2 (U - \gamma_1)$.
However, the states localized on the soliton are not near low energy, as indicated by the arrows in Fig.~\ref{fig:ABCp1} (second row). They are at energies $E = \pm \sqrt{U^2 + \gamma_1^2}$ for $k = 0$, and then move further away from zero for nonzero $k$.

The band structure for ABC$^{\prime}$ stacking depends on whether there is an even or odd number of layers, even for a faultless system. The band structure for a faultless system with an odd number of layers $N$ is shown in Fig.~\ref{fig:ABCp1} (third row). For $N \geq 3$, at $k = 0$,  the $N$ bands are at six distinct energies with $E = + U$ and $E= - U$ levels arising from the surface states, and from the dimers: $E = + U + \gamma_1$, $E = + U - \gamma_1$, $E = - U + \gamma_1$, and $E = - U - \gamma_1$, each of these occurring $(N-1)/2$ times.
As with the faultless even $N$ case, the surface bands appear at lower energy ($E = \pm U$) than the bulk bands for $0 \leq U < \gamma_1/2$, and the bulk bands touch at $k = 0$ and $U = \gamma_1$.
Nevertheless, the band structure is more similar to that of the even $N$ case with a fault. This can be seen by considering the bands for nonzero $k$ values: For $0 < U < \gamma_1$, there is a small anticrossing between two low-energy bands, then, for $U = \gamma_1$, the bands touch, and, for $U > \gamma_1$, there is a band gap $E_{{\mathrm{g}}} = 2 (U - \gamma_1)$.

The band structure for an odd number of layers $N$ and a fault near the center is shown in Fig.~\ref{fig:ABCp1} (bottom row). In this case, the $N$ bands (for $N \geq 5$) are at seven distinct energies at $k = 0$ with $E = + U$ (twice) arising from the surface states, and from the dimers: $E = + U + \gamma_1$ ($(N-3)/2$ times), $E = + U - \gamma_1$ ($(N-3)/2$ times), $E = - U + \gamma_1$ ($(N-1)/2$ times), and $E = - U - \gamma_1$ ($(N-1)/2$ times). In addition, there is the dimer consisting of the stacking fault giving energies $E = \pm \sqrt{U^2 + \gamma_1^2}$.
Again, the surface bands appear at lower energy ($E = \pm U$) than the bulk bands for $0 \leq U < \gamma_1/2$, and the bulk bands touch at $k = 0$ and $U = \gamma_1$.
Near low energy, the band structure is similar to that of the faultless even $N$ case. Considering the bands for nonzero $k$ values, there is always a band within the nominal band gap for $U < \gamma_1$, whereas the bandgap, for $U > \gamma_1$, is given by $E_{{\mathrm{g}}} = 2 (U - \gamma_1)$.
As with the even $N$ case, the states localized on the soliton are not near low energy, as indicated by the arrows in Fig.~\ref{fig:ABCp1} (bottom row). They are at energies $E = \pm \sqrt{U^2 + \gamma_1^2}$ for $k= 0$, and then move further away from zero for nonzero $k$.

\section{Bernal stacking}\label{s:bernal}

\subsection{ABA}\label{s:baba}

\subsubsection{Faultless system}

\begin{figure*}[t]
\includegraphics[scale=0.4]{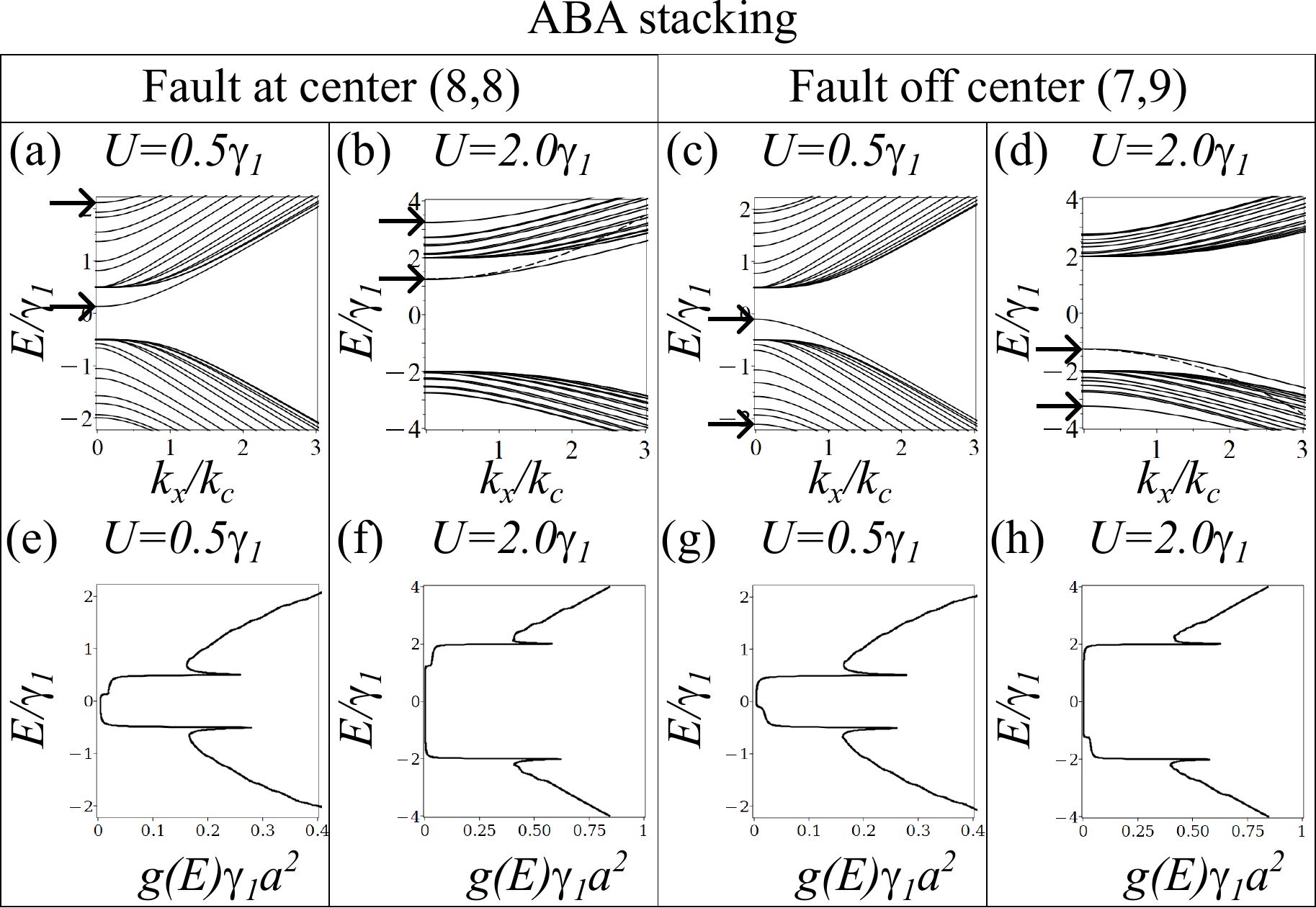}
\caption{Low-energy band structures for ABA stacking with $N=16$ layers and a sharp soliton near the center.
The top row shows band structures, the bottom row shows the corresponding density of states $g(E)$, with the left side showing a soliton at the center $(8,8)$ after an even number of layers, and the right side shows a soliton near the center $(7,9)$ after an odd number of layers.
There are different values of the magnitude of the alternating onsite energies as (a), (c), (e), (g) $U/\gamma_1 = 0.5$, (b), (d), (f), (h) $U/\gamma_1 = 2.0$, with different scales on the axes.
Horizontal arrows in the top row show the energies at $k_x=0$ of the states localized on the soliton.
The dashed line in (b) shows the perturbative approximation for the antisoliton energy $E_{\mathrm{asol}}(\mathbf{k})$ and that in (d) shows the soliton energy $E_{\mathrm{sol}}(\mathbf{k})$ where $E_{\mathrm{asol}}(\mathbf{k}) = - E_{\mathrm{sol}}(\mathbf{k})$ and $E_{\mathrm{sol}}(\mathbf{k})$ is given in Eq.~(\ref{abaesol1}).
In all plots, parameter values are $\gamma_0 = 3.16\,$eV, $\gamma_1 = 0.381\,$eV~\cite{kuzmenko09}, $a = 2.46$\AA~\cite{saito98}. For the band structures, $k_y=0$, and, for the density of states, $\delta = 0.01 \gamma_1$.
}\label{figABA1}
\end{figure*}

For bulk, faultless ABA stacking, the unit cell consists of four atoms such as $A_1$, $B_1$, $A_2$, $B_2$ (in Fig.~\ref{fig:summary}(c)), and the Hamiltonian may be written as
\begin{eqnarray}
H_{\mathrm{ABA}} = \begin{pmatrix}
U & \hbar v k & 0 & 0 \\
\hbar v k & - U & 2\gamma_1 \!\cos (q_z d) & 0 \\
0 & 2\gamma_1 \!\cos (q_z d) & U & \hbar v k \\
0 & 0 & \hbar v k & -U
\end{pmatrix} \!\! , \label{haba}
\end{eqnarray}
using the approximation for the intralayer hopping~(\ref{di2}) with $k<k_c$, where $q_z$ is the $z$ component of the wave vector and $d$ is the layer separation (the length of the unit cell in the $z$ direction is $2d$).
The four corresponding bulk bands are given by
\begin{eqnarray*}
(E_{\mathrm{ABA}})^2 &=& U^2 + \Big( \sqrt{ \gamma_1^2 \cos^2 \!q_z d + (\hbar vk)^2} \pm \gamma_1 \cos q_z d \Big)^2 ,
\end{eqnarray*}
which shows that the band gap is $E_{\mathrm{g}} = 2U$.
The two conduction bands are degenerate at the Brillouin zone edge ($q_z d = \pm \pi / 2$) because $\cos q_z d = 0$ there giving a block diagonal Hamiltonian~(\ref{haba}), likewise the two valence bands.

For a thin film of finite thickness, a faultless system is gapped with bandgap $E_{\mathrm{g}} \approx 2U$, Fig.~\ref{fig:summary}(c).
This can be understood by considering $k = 0$ where the intralayer hopping is zero and the system separates into isolated atoms with energies $E = \pm U$ plus a $N$-mer which is a single chain of $N$ atoms ($N=8$ in Fig.~\ref{fig:summary}(c)). This chain corresponds to the CDW model~\cite{kivelson83,brzezicki20,cayssol21,fuchs21,allen22,mccann23} which is a one-dimensional chain with constant hopping $\gamma_1$ and alternating onsite energies $\pm U$, having a band gap of $2U$, as described in the methodology, Section~\ref{s:methodology}.

\subsubsection{Single soliton}

For a system with a single fault, there are two bands localized on the fault, one of which generally lies within the bulk band gap, Fig.~\ref{fig:summary}(i) and Fig.~\ref{figABA1}. The nature of this band depends on whether the fault occurs after an even or odd number of layers~\cite{conventionnote} as shown in Fig.~\ref{figABA1}.
When the fault occurs after an even number of layers, Fig.~\ref{fig:summary}(i), Fig.~\ref{figABA1}(a), and Fig.~\ref{figABA1}(b), the fault consists of two consecutive onsite energies of $+U$ (on atoms $A_4$ and $B_5$ in Fig.~\ref{fig:summary}(i)) and we refer to this as an antisoliton; the energy band moves into the conduction band for $k \gg k_c$.
However, when the fault occurs after an odd number of layers, Fig.~\ref{figABA1}(c) and Fig.~\ref{figABA1}(d), the fault consists of two consecutive onsite energies of $-U$, a soliton texture, and the energy band moves into the valence band for $k \gg k_c$.

We consider $k = 0$ where the intralayer hopping is zero. Once again, the system separates into isolated atoms with energies $E = \pm U$ plus a $N$-mer  ($N=8$ in Fig.~\ref{fig:summary}(i)). But now the $N$-mer, which corresponds to the CDW model~\cite{kivelson83,brzezicki20,cayssol21,fuchs21,allen22,mccann23}, has a soliton or antisoliton in it~\cite{brzezicki20,allen22}.
It was shown~\cite{allen22} that the energy of the state localized on such a fault lies within the bulk band gap as long as $0 < \gamma_1 / U < N$, when the fault lies at the center of the system of $N$ layers.

For large $U$ ($U \gg \gamma_1$ and $U \gg \hbar v k$), it is possible to estimate the energy of the soliton level using perturbation theory in the hopping strength. Using the analytic approximation~(\ref{di2}) for the intralayer hopping matrix element gives
\begin{eqnarray}
E_{\mathrm{sol}}(\mathbf{k}) \approx - U + \gamma_1 - \frac{\gamma_1^2}{2U} - \frac{(\hbar v k)^2}{2U} , \label{abaesol1}
\end{eqnarray}
for $\{ \gamma_1 , \hbar v k \} \ll U$, assuming the soliton is not at the surface of the system. For an antisoliton, $E_{\mathrm{asol}}(\mathbf{k}) = - E_{\mathrm{sol}}(\mathbf{k})$. The approximation~(\ref{abaesol1}) is shown as the dashed line in  Fig.~\ref{figABA1}(b) and Fig.~\ref{figABA1}(d).

\begin{figure*}[t]
\includegraphics[scale=0.38]{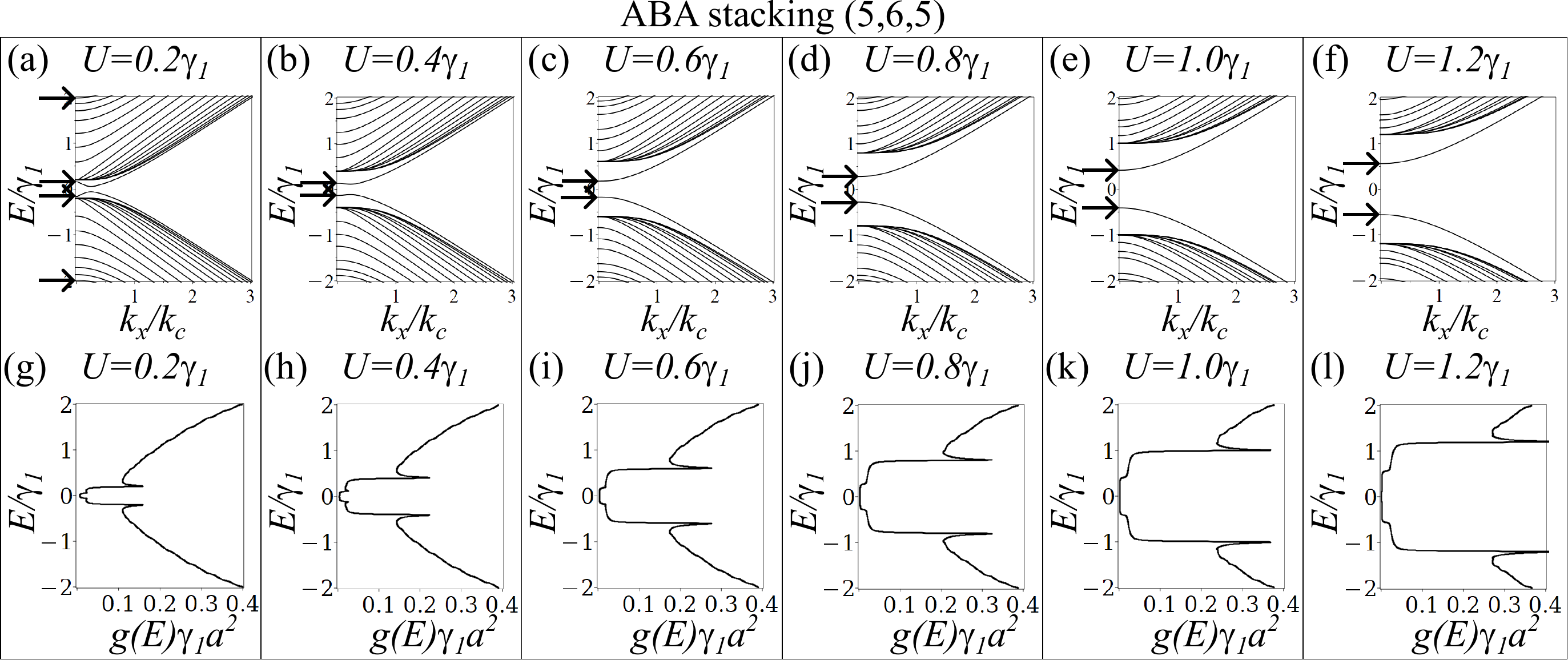}
\caption{ABA stacking with $N=16$ layers and a soliton-antisoliton pair $(5,6,5)$. The top row shows band structures, the bottom row shows the corresponding density of states $g(E)$ for different values of the magnitude of the alternating onsite energies as (a), (g) $U/\gamma_1 = 0.2$, (b), (h) $U/\gamma_1 = 0.4$, (c), (i) $U/\gamma_1 = 0.6$, (d), (j) $U/\gamma_1 = 0.8$, (e), (k) $U/\gamma_1 = 1.0$, and (f), (l) $U/\gamma_1 = 1.2$.
In the band structure plots, horizontal arrows show the energies at $k_x=0$ of the states localized on the soliton or antisoliton.
In all plots, parameter values are $\gamma_0 = 3.16\,$eV, $\gamma_1 = 0.381\,$eV~\cite{kuzmenko09}, $a = 2.46$\AA~\cite{saito98}. For the band structures, $k_y=0$, and, for the density of states, $\delta = 0.01 \gamma_1$.
}\label{figABA2}
\end{figure*}

Soliton states are extended states in the in-plane direction, but are localized in the out-of-plane ($z$) direction.
For $k = 0$, the soliton state is localized on the soliton~\cite{brzezicki20,allen22}, and, for $k_x = k_c$ (and $k_y = 0$), the state is still localized near the soliton as shown in Fig.~\ref{figABC2}(c).
Since the system is equivalent to the CDW model at $k = 0$, the soliton state is polarized, with maximum polarization at $k = 0$, Fig.~\ref{figABC2}(d).
In this case, the equivalence with the CDW model only applies to the $N$-mer (the atomic sites connected by interlayer hopping $\gamma_1$), excluding the isolated non-$N$-mer atoms. Therefore, we modify the definition of the translation operator $T_{a/2}$, Eq.~(\ref{tat}), in order to describe translation between the atoms on the $N$-mer only,
\begin{eqnarray}
\tilde{T}_{a/2} = \begin{pmatrix}
0 & 0 & 0 & 1 & 0 & 0 & 0 & 0 & \hdots \\
0 & 0 & 1 & 0 & 0 & 0 & 0 & 0 & \hdots \\
0 & 0 & 0 & 0 & 0 & 1 & 0 & 0 & \hdots \\
0 & 0 & 0 & 0 & 1 & 0 & 0 & 0 & \hdots \\
0 & 0 & 0 & 0 & 0 & 0 & 0 & 1 & \hdots \\
0 & 0 & 0 & 0 & 0 & 0 & 1 & 0 & \hdots \\
\vdots & \vdots & \vdots & \vdots & \vdots & \vdots & \vdots & \vdots & \ddots \\
0 & 1 & 0 & 0 & 0 & 0 & 0 & 0 & \hdots \\
1 & 0 & 0 & 0 & 0 & 0 & 0 & 0 & \hdots
\end{pmatrix} , \label{alttat}
\end{eqnarray}
which is shown for an even number of layers $N$ (for $N \geq 4$).
The translation $\tilde{T}_{a/2}$ for an odd number of layers is the same with the replacements $(\tilde{T}_{a/2})_{N,1} = (\tilde{T}_{a/2})_{N-1,2} = 0$ and $(\tilde{T}_{a/2})_{N-1,1} = (\tilde{T}_{a/2})_{N,2} = 1$.
Hence, for an even number of layers,
\begin{eqnarray}
\tilde{T}_{a/2}S_z = \begin{pmatrix}
0 & 0 & 0 & -1 & 0 & 0 & 0 & 0 & \hdots \\
0 & 0 & 1 & 0 & 0 & 0 & 0 & 0 & \hdots \\
0 & 0 & 0 & 0 & 0 & -1 & 0 & 0 & \hdots \\
0 & 0 & 0 & 0 & 1 & 0 & 0 & 0 & \hdots \\
0 & 0 & 0 & 0 & 0 & 0 & 0 & -1 & \hdots \\
0 & 0 & 0 & 0 & 0 & 0 & 1 & 0 & \hdots \\
\vdots & \vdots & \vdots & \vdots & \vdots & \vdots & \vdots & \vdots & \ddots \\
0 & -1 & 0 & 0 & 0 & 0 & 0 & 0 & \hdots \\
1 & 0 & 0 & 0 & 0 & 0 & 0 & 0 & \hdots
\end{pmatrix} , \label{alttatsz}
\end{eqnarray}
using $S_z = \mathrm{diag} (1,-1,1,-1,1,-1,\ldots )$. Thus, we define the modified polarization as
\begin{eqnarray}
\tilde{p}_y = \langle \psi | \tilde{T}_{a/2}S_z | \psi \rangle . \label{altpy1}
\end{eqnarray}
Fig.~\ref{figABC2}(d) shows the polarization $\tilde{p}_y$, Eq.~(\ref{altpy1}), as a function of $k_x$ (with $k_y = 0$) for $U/\gamma_1 = 0.6$ (solid line) and $U/\gamma_1 = 1.8$ (dashed line).
The polarization is a maximum at $k = 0$, and, at $k = 0$, it is larger for smaller $U/\gamma_1$ (solid line).
Note that the values of $\tilde{p}_y$ at $k = 0$ in Fig.~\ref{figABC2}(d) are approximately the same as the values of $p_y$ at $k_x = k_c$ in Fig.~\ref{figABC2}(b) because, at these points, both systems are equivalent to the CDW model.
The polarization $\tilde{p}_y$ never reaches unity because the nonsymmorphic chiral symmetry is broken by the ends of the system and the finite width of the soliton.

\subsubsection{Soliton-antisoliton pair}

For ABA stacking, the band structure and density of states of a soliton-antisoliton pair $(5,6,5)$ with $N=16$ layers is shown in Fig.~\ref{figABA2} for different values of $U$. There are two localized states associated with the soliton and two with the antisoliton, of which one soliton state and one antisoliton lie within the bulk bandgap of $E_{\mathrm{g}} = 2U$. For small values of $U/\gamma_1$, the low-energy soliton and antisoliton states hybridize with an anticrossing, Fig.~\ref{figABA2}(a) and Fig.~\ref{figABA2}(g). For larger values of $U/\gamma_1$, the soliton and antisoliton bands separate. For large $U$ ($U \gg \gamma_1$ and $U \gg \hbar v k$), we can use the perturbation expression~(\ref{abaesol1}) for the energy of a soliton to estimate the separation of these bands as $2U - 2\gamma_1 + \gamma_1^2/U$.
This separation appears as an extra feature within the bulk band gap in the density of states, Fig.~\ref{figABA2} (bottom row).

\begin{figure*}[t]
\includegraphics[scale=0.34]{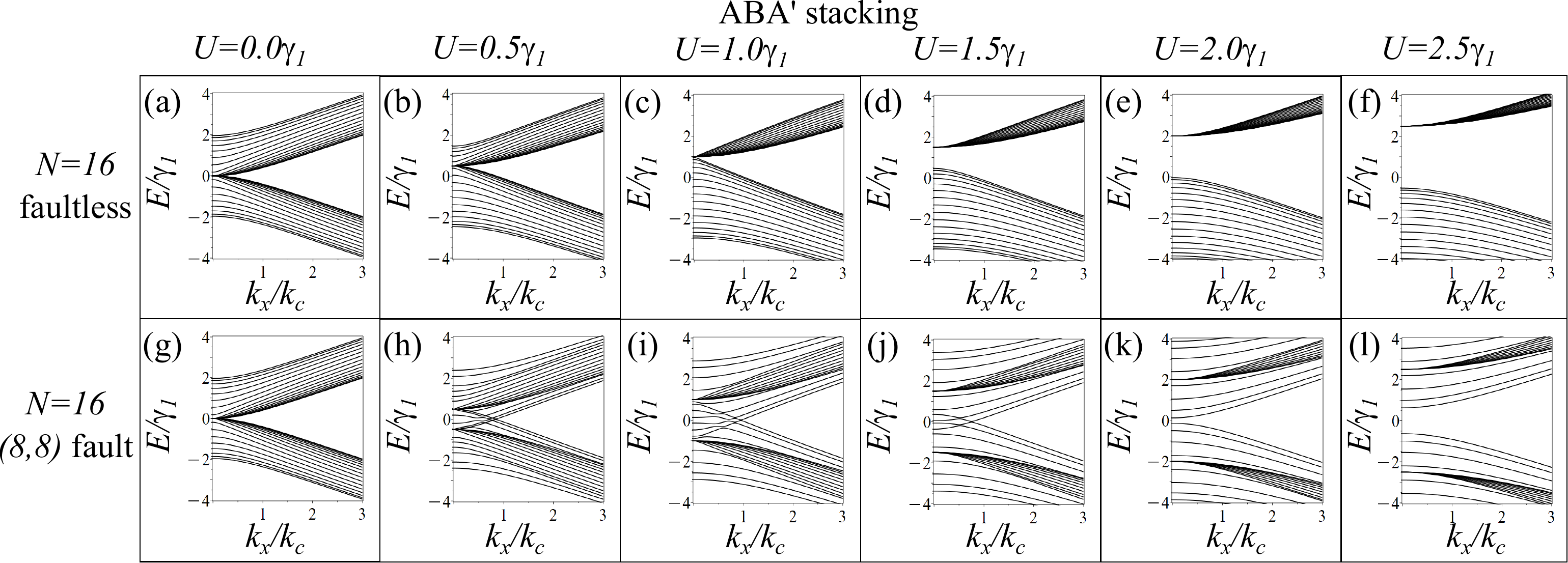}
\caption{Low-energy band structures for ABA$^{\prime}$ stacking. Each column shows a different value of the onsite energies $U$. The top row shows a faultless system with $N=16$ layers, whereas the second row shows $N=16$ layers with a sharp fault at its center $(8,8)$.
In all plots, parameter values are $\gamma_0 = 3.16\,$eV, $\gamma_1 = 0.381\,$eV~\cite{kuzmenko09}, $a = 2.46$\AA~\cite{saito98}.
}\label{fig:ABAp1}
\end{figure*}

\subsection{ABA$^{\prime}$}

\subsubsection{Faultless system}

For bulk, faultless ABA$^{\prime}$ stacking, using the approximation for the intralayer hopping~(\ref{di2}) with $k<k_c$, we can use a unit cell of two atoms, either $A_i$, $B_i$ on an odd layer or $B_i$, $A_i$ on an even layer, to give the Hamiltonian
\begin{eqnarray}
H_{\mathrm{ABA}^{\prime}} = \begin{pmatrix}
U & \hbar v k \\
\hbar v k  & - U + 2\gamma_1 \cos q_z d
\end{pmatrix} , \label{habap}
\end{eqnarray}
where $q_z$ is the $z$ component of the wave vector and $d$ is the layer separation.
The two corresponding bulk bands are
\begin{eqnarray*}
E_{\mathrm{ABA}^{\prime}} = \gamma_1 \cos q_z d \pm \sqrt{(\hbar vk)^2 + ( U - \gamma_1 \cos q_z d)^2} .
\end{eqnarray*}
This shows that there is no band gap for $U \leq \gamma_1$, but that the band gap is $E_{\mathrm{g}} = 2(U-\gamma_1)$ for $U > \gamma_1$.

For a thin film of finite thickness, we begin by considering the faultless system, Fig.~\ref{fig:ABAp1} (top row).
At $k = 0$, the system separates into isolated atoms with energy $+U$~\cite{conventionnote} and a $N$-mer monoatomic chain with onsite energies $-U$ and nearest neighbor hopping $\gamma_1$, Fig.~\ref{fig:summary}(d). Such a chain has a bulk band given by $E(q_z) = - U + 2 \gamma_1 \cos (q_z d)$ where $d$ is the layer separation and $q_z$ is the $z$ component of the wave vector; this band is centered on energy $-U$ and it has a bandwidth of $4\gamma_1$. Thus, this band overlaps with the isolated atoms of energy $+U$ for $U \leq \gamma_1$.
Hence, Fig.~\ref{fig:ABAp1}(c) for $U=\gamma_1$ shows the bands due to the isolated atoms of energy $+U$ just touching the lower bands due to the $N$-mer chain whereas Fig.~\ref{fig:ABAp1}(d) for $U > \gamma_1$ shows that they are separated.

\subsubsection{Single soliton}

Band structures for ABA$^{\prime}$ stacking with a sharp fault at its center are shown in Fig.~\ref{fig:ABAp1} (bottom row).
At $k = 0$, the system separates into isolated atoms with energies $\pm U$ plus the $N$-mer chain. Now the $N$-mer chain contains a fault, so it consists of a monoatomic chain with onsite energies $-U$ and nearest neighbor hopping $\gamma_1$ connected to a monoatomic chain with onsite energies $+U$ and nearest neighbor hopping $\gamma_1$.
The former has a bulk band $E(q_z) = - U + 2 \gamma_1 \cos (q_z d)$, the latter has a bulk band $E(q_z) = U + 2 \gamma_1 \cos (q_z d)$, i.e., the system consists of one part with a band centered on energy $+U$ and of bandwidth $4\gamma_1$, connected to another part with a band centered on energy $-U$ and of bandwidth $4\gamma_1$. For $U \leq 2 \gamma_1$, the two bands overlap leading, in a system of a finite number of layers, to overlapping bands with anticrossings at low energy as in Fig.~\ref{fig:ABAp1}(h) to Fig.~\ref{fig:ABAp1}(j). For $U > 2 \gamma_1$, the two bands do not overlap, giving an overall band gap as in Fig.~\ref{fig:ABAp1}(l).

Although the presence of a fault for ABA$^{\prime}$ stacking creates changes in the band structure, Fig.~\ref{fig:ABAp1}, there are no states localized on the fault. This can be understood by considering the system at $k = 0$: The fault divides two sections of monoatomic chains with different onsite energies ($+U$ or $-U$) and, thus, the two sections are not degenerate, nor do they have a Dirac-like low-energy continuum description~\cite{jackiw76}.

\section{AA stacking}\label{s:aa}

\subsection{AA}

\subsubsection{Faultless system}

\begin{figure*}[t]
\includegraphics[scale=0.34]{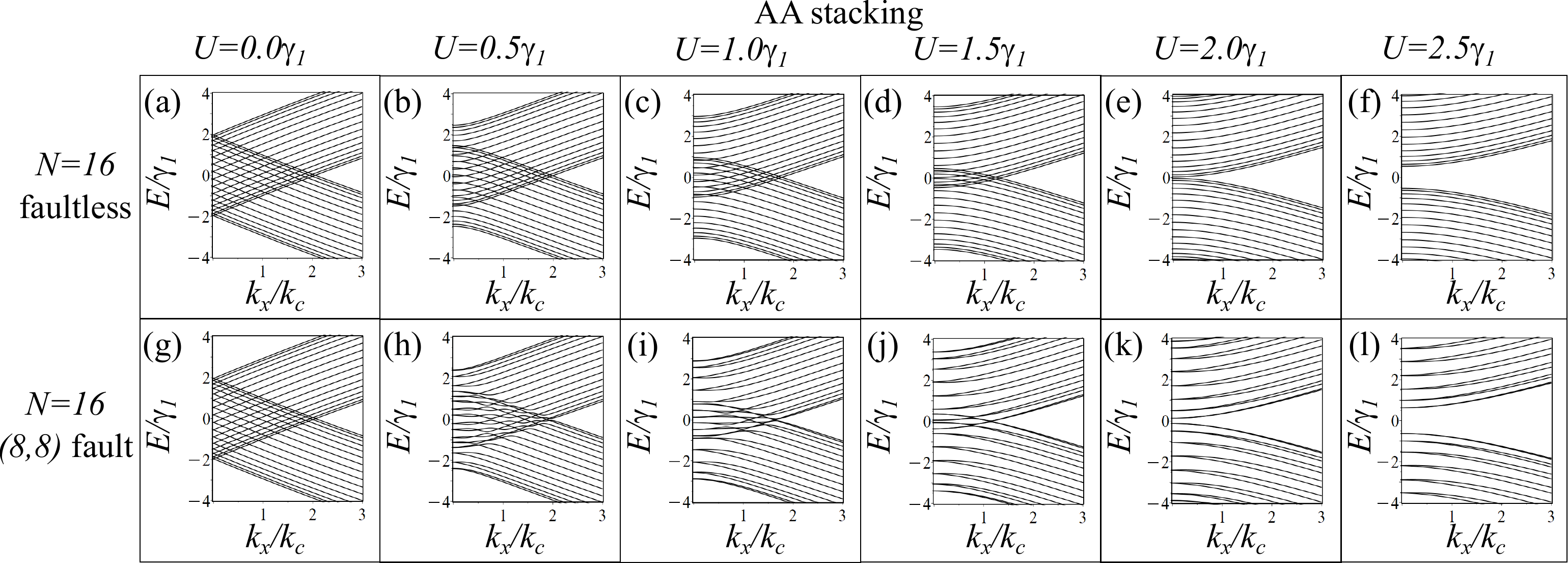}
\caption{Low-energy band structures for AA stacking. Each column shows a different value of the onsite energies $U$. The top row shows a faultless system with $N=16$ layers, whereas the second row shows $N=16$ layers with a sharp fault at its center $(8,8)$.
In all plots, parameter values are $\gamma_0 = 3.16\,$eV, $\gamma_1 = 0.381\,$eV~\cite{kuzmenko09}, $a = 2.46$\AA~\cite{saito98}.
}\label{fig:AA1}
\end{figure*}

For bulk, faultless AA stacking, the unit cell consists of two atoms $A_i$, $B_i$ on the $i$th layer, and the Hamiltonian may be written as
\begin{eqnarray}
H_{\mathrm{AA}} = \begin{pmatrix}
U + 2 \gamma_1 \cos q_z d & \hbar v k \\
\hbar v k & - U + 2 \gamma_1 \cos q_z d
\end{pmatrix} , \label{haa}
\end{eqnarray}
using the approximation for the intralayer hopping~(\ref{di2}) with $k<k_c$, where $q_z$ is the $z$ component of the wave vector and $d$ is the layer separation.
The two corresponding bulk bands are
\begin{eqnarray*}
E_{\mathrm{AA}} = 2 \gamma_1 \cos q_z d \pm \sqrt{U^2 + (\hbar vk)^2} .
\end{eqnarray*}
At $k = 0$, the bands are centered on energy $+U$ or $-U$ and each of them has a bandwidth of $4 \gamma_1$.
This shows that there is no band gap for $U \leq 2\gamma_1$, but that the band gap is $E_{\mathrm{g}} = 2(U-2\gamma_1)$ for $U > \gamma_1$.

For a thin film of finite thickness, band structures for the faultless system are plotted in Fig.~\ref{fig:AA1} (top row).
The two bands overlap for $U \leq 2\gamma_1$ as shown in Fig.~\ref{fig:AA1} (top row).
It is also noteable that there are no anticrossings when the bands overlap. This is because the system has reflection symmetry (swap $A_1$ with $A_N$, $B_1$ with $B_N$, etc.) which may be used to block diagonalize the position space Hamiltonian~(\ref{h1}) into two separate blocks with either even or odd parity eigenstates.

\subsubsection{Single soliton}

In the presence of an atomically-sharp fault, there are different ladders either side of the fault, Fig.~\ref{fig:summary}(k), resulting in band structures as plotted in Fig.~\ref{fig:AA1} (bottom row). As with the faultless case, there are two bands overall, centered on energy $+U$ or $-U$ and each of them with a bandwidth of $4 \gamma_1$.
Thus, the two bands overlap for $U \leq 2\gamma_1$ as shown in Fig.~\ref{fig:AA1} (bottom row). Unlike the faultless case, however, there are some anticrossings (and crossings) of the bands in the region when the bands overlap.
The situation shown in Fig.~\ref{fig:AA1} (bottom row) is somewhat special because the fault is at the center so that the system has inversion symmetry (swap $A_1$ with $B_N$, $B_1$ with $A_N$, etc.) and this may be used to block diagonalize the position space Hamiltonian~(\ref{h1}) into two separate blocks with either even or odd parity eigenstates. Hence there are level crossings and all bands are doubly degenerate at $k=0$. However, within each even or odd parity block, anticrossings can arise giving the band structure in  Fig.~\ref{fig:AA1}(h) and  Fig.~\ref{fig:AA1}(i).

As with ABA$^{\prime}$ stacking, although the presence of a fault for AA stacking creates changes in the band structure, Fig.~\ref{fig:AA1}, there are no states localized on the fault.
The term describing hopping along the ladder, $2 \gamma_1 \cos (q_z d)$, appears on the main diagonal in the ladder Hamiltonian~(\ref{haa}) and it breaks chiral symmetry.
The trivial nature of the fault may be further understood by considering the limit $k=0$ when the system separates into monoatomic chains which do not have a Dirac-like low-energy continuum description~\cite{jackiw76}.

\begin{figure*}[t]
\includegraphics[scale=0.4]{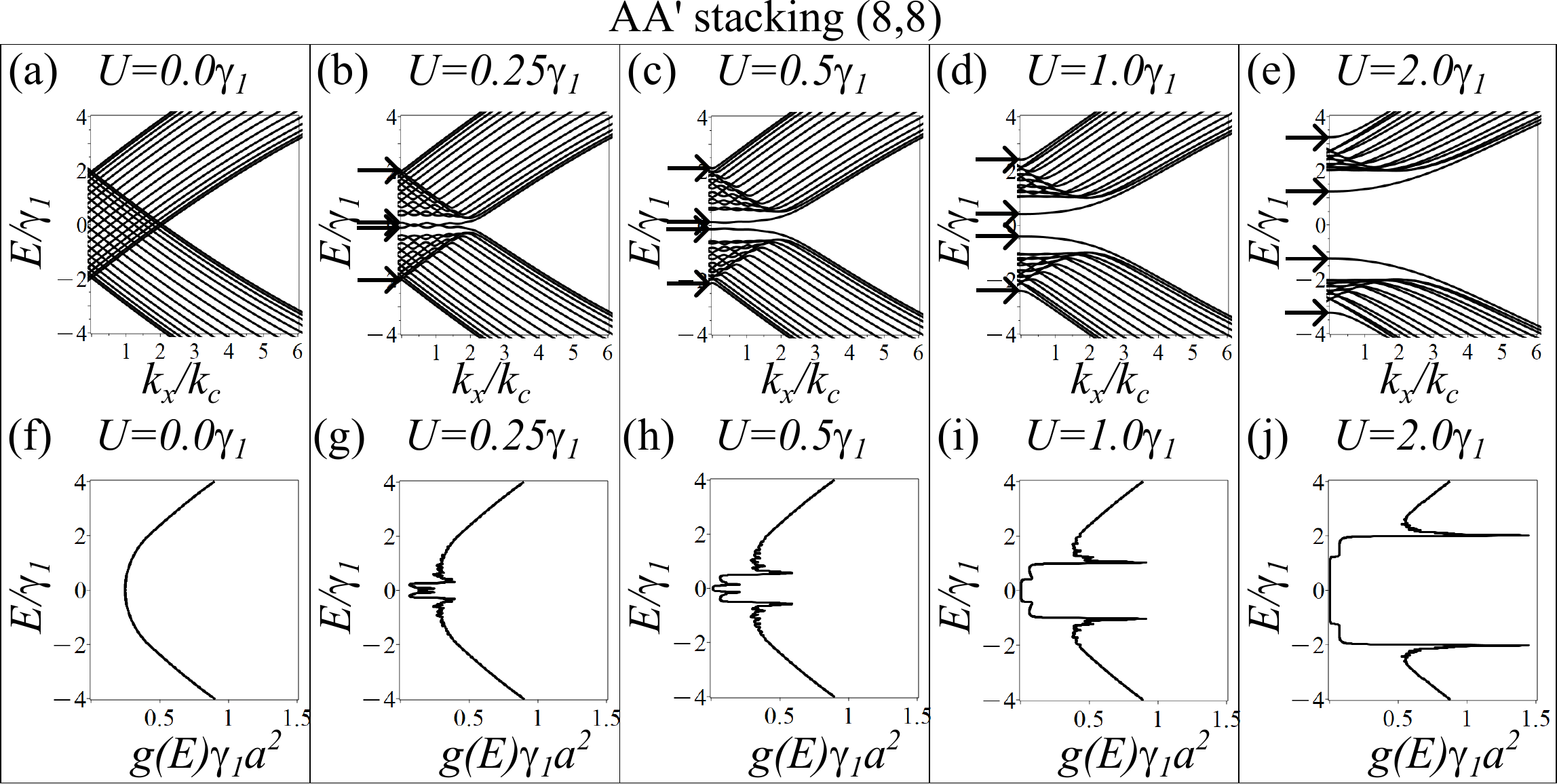}
\caption{AA$^{\prime}$ stacking with $N=16$ layers and a sharp soliton at the center $(8,8)$. The top row shows band structures, the bottom row shows the corresponding density of states $g(E)$ for different values of the magnitude of the alternating onsite energies as (a), (f) $U/\gamma_1 = 0.0$, (b), (g) $U/\gamma_1 = 0.25$, (c), (h) $U/\gamma_1 = 0.5$, (d), (i) $U/\gamma_1 = 1.0$, and (e), (j) $U/\gamma_1 = 2.0$.
In the band structure plots, horizontal arrows show the energies at $k_x=0$ of the states localized on the soliton.
In all plots, parameter values are $\gamma_0 = 3.16\,$eV, $\gamma_1 = 0.381\,$eV~\cite{kuzmenko09}, $a = 2.46$\AA~\cite{saito98}. For the band structures, $k_y=0$, and, for the density of states, $\delta = 0.01 \gamma_1$.
}\label{figAAp1}
\end{figure*}

\subsection{AA$^{\prime}$}\label{s:aap}

\subsubsection{Faultless system}

For bulk, faultless AA$^{\prime}$ stacking, using the approximation for the intralayer hopping~(\ref{di2}) with $k<k_c$, we can use a unit cell of two atoms, either $A_i$, $B_i$ on an odd layer or $B_i$, $A_i$ on an even layer, to give the Hamiltonian
\begin{eqnarray}
H_{\mathrm{AA}^{\prime}} = \begin{pmatrix}
U & \hbar v k + 2\gamma_1 \cos q_z d \\
\hbar v k + 2\gamma_1 \cos q_z d & - U
\end{pmatrix} \!\! , \label{haap}
\end{eqnarray}
where $q_z$ is the $z$ component of the wave vector and $d$ is the layer separation.
The two corresponding bulk bands are
\begin{eqnarray*}
E_{\mathrm{AA}^{\prime}} = \pm \sqrt{U^2 + ( \hbar vk + 2\gamma_1 \cos q_z d)^2} ,
\end{eqnarray*}
which have a band gap $E_{\mathrm{g}} = 2U$.

For a thin film of finite thickness, there is indeed a band gap of $E_{\mathrm{g}} \approx 2U$ in the numerically-derived band structure, Fig.~\ref{fig:summary}(f). Note that the system at $k=0$ consists of two separated CDW chains and, for $k \neq 0$, the chains are coupled by intralayer hopping to form a ladder.
By dimensional reduction, treating $k$ as a fixed parameter, this ladder is equivalent to a one-dimensional model in class CI~\cite{mccann23} which is a topologically trivial insulator.

\subsubsection{Single soliton}\label{s:aapss}

In the presence of a single atomically-sharp fault, there are four states localized on the fault, Fig.~\ref{fig:summary}(l) and Fig.~\ref{figAAp1}. Of these, one is at the top of the conduction band, one is at the bottom of the valence band, and the other two generally lie within the band gap. At low $U$ values, they can hybridize together as in Fig.~\ref{figAAp1}(b) and Fig.~\ref{figAAp1}(g).

At $k=0$, the system consists of two separated CDW chains, each of them has either a soliton or an antisoliton. This is why there are two low-energy states. The analogy with the CDW model means that, at $k=0$, we can say that the energies of the two low-energy states are within the band gap for $0 < \gamma_1 / U < N$ for a fault at the center of a system with $N$ layers. For $U / \gamma_1 \gg 1$, these energies are
\begin{eqnarray}
E_{\mathrm{sol}/\mathrm{asol}} \approx \pm \bigg( - U + \gamma_1 - \frac{\gamma_1^2}{2U} \bigg) , \label{esolaap}
\end{eqnarray}
at $k=0$, as indicated by the two middle arrows in Fig.~\ref{figAAp1}(e).
The solitons give rise to a step-like nonzero density of states within the bulk band gap. For $U \gg \gamma_1$, the nonzero density of states due to the solitons is visible at energies $-U$ to $-(U - \gamma_1 + \gamma_1^2/(2U))$ and at $(U - \gamma_1 + \gamma_1^2/(2U))$ to $U$, Eq.~(\ref{esolaap}), see Fig.~\ref{figAAp1}(j).
In addition to the soliton feature, there are also sharp peaks in the density of states at the bulk band edges, $E = \pm U$.

For small values of $U/\gamma_1$, there are crossings between levels, including the two low-energy levels, Fig.~\ref{figAAp1}(b), whereby the energies appear to oscillate as a function of $k$. Using spatial inversion symmetry, it is possible to block diagonalize the Hamiltonian~(\ref{h1}) into a block with even parity states and a block with odd parity states. The odd parity block has eigenstates which are the electron-hole reflection ($E \rightarrow - E$) of the even parity eigenstates. Each individual block describes a system similar to AA$^{\prime}$ stacking with a spectrum having anticrossings at $U \neq 0$ and a single soliton level.
For small $U/\gamma_1$, the soliton level crosses zero energy, at which point its electron-hole partner (from the opposite parity block) crosses zero energy in the opposite direction. Oscillations are caused by anticrossings with levels from the same block. The anticrossings, and resulting oscillations, are reduced as $U/\gamma_1$ increases because the other levels (in the bulk conduction and valence bands) move further away in energy and the soliton-antisoliton pair move away from zero energy and separate.
This picture holds in the presence of spatial inversion symmetry. When the soliton is off-center, the band structure is similar, but the level crossings between the levels with even or odd parity are replaced by anticrossings.

\section{Smooth solitons}\label{s:smooth}

\begin{figure*}[t]
\includegraphics[scale=0.375]{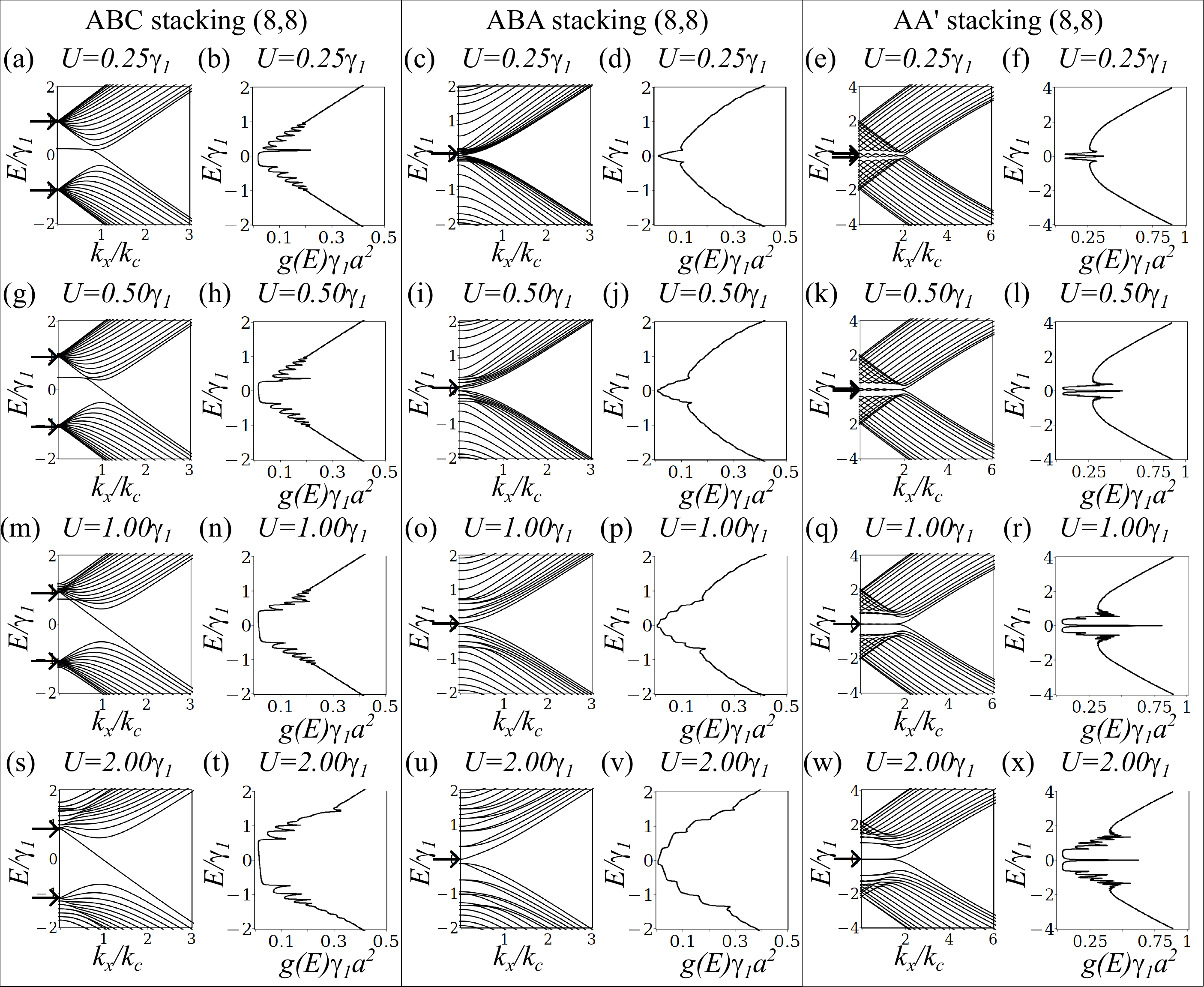}
\caption{Band structures and density of states $g(E)$ for smooth solitons with width $\zeta = 8$ at the center $(8,8)$ of a system with $N=16$ layers for ABA, ABC, and AA$^{\prime}$ stacking.
In the band structure plots, horizontal arrows show the energies at $k_x=0$ of the states localized on the soliton.
There is a different axis scale for AA$^{\prime}$ stacking.
In all plots, parameter values are $\gamma_0 = 3.16\,$eV, $\gamma_1 = 0.381\,$eV~\cite{kuzmenko09}, $a = 2.46$\AA~\cite{saito98}. For the band structures, $k_y=0$, and, for the density of states, $\delta = 0.01 \gamma_1$.
}\label{figsmooth1}
\end{figure*}

Atomically sharp solitons support localized states within the bulk band gap for ABC, ABA and AA$^{\prime}$ stacking, Fig.~\ref{fig:summary}.
Now we will describe the properties of smooth solitons with a finite width for these three stacking types.
Band structures and density of states are determined as described in the methodology, Section~\ref{s:methodology}, by numerically diagonalizing Hamiltonian~(\ref{h1}) after replacing the diagonal elements with a smooth soliton texture. Specifically, for ABC and ABA stacking, we model the onsite energy of an A site on the $j$th layer, where $j = 1,2,\ldots,N$, as
\begin{eqnarray}
\epsilon_{A,j} = - U \tanh \bigg( \frac{j-m-1/2}{\zeta} \bigg) , \label{smooth1}
\end{eqnarray}
where $U$ is the magnitude of the texture at infinity, $m$ is the number of layers below the fault, and $\zeta$ is the soliton width in dimensionless units, i.e., measured in units of the interlayer spacing. 
For AA$^{\prime}$ stacking,
\begin{eqnarray}
\epsilon_{A,j} = (-1)^{j} U \tanh \bigg( \frac{j-m-1/2}{\zeta} \bigg) .
\end{eqnarray}
For all solitons, we assume charge neutrality within each layer so that, for the onsite energy of a B site on the $j$th layer, $\epsilon_{B,j} = - \epsilon_{A,j}$.
Band structures and density of states for the three stacking types are shown in Figure~\ref{figsmooth1} for a smooth soliton of width $\zeta = 8$ at the center $(8,8)$ of a system with $N=16$ layers, and different values of $U/\gamma_1$.

\begin{figure}[t]
\includegraphics[scale=0.5]{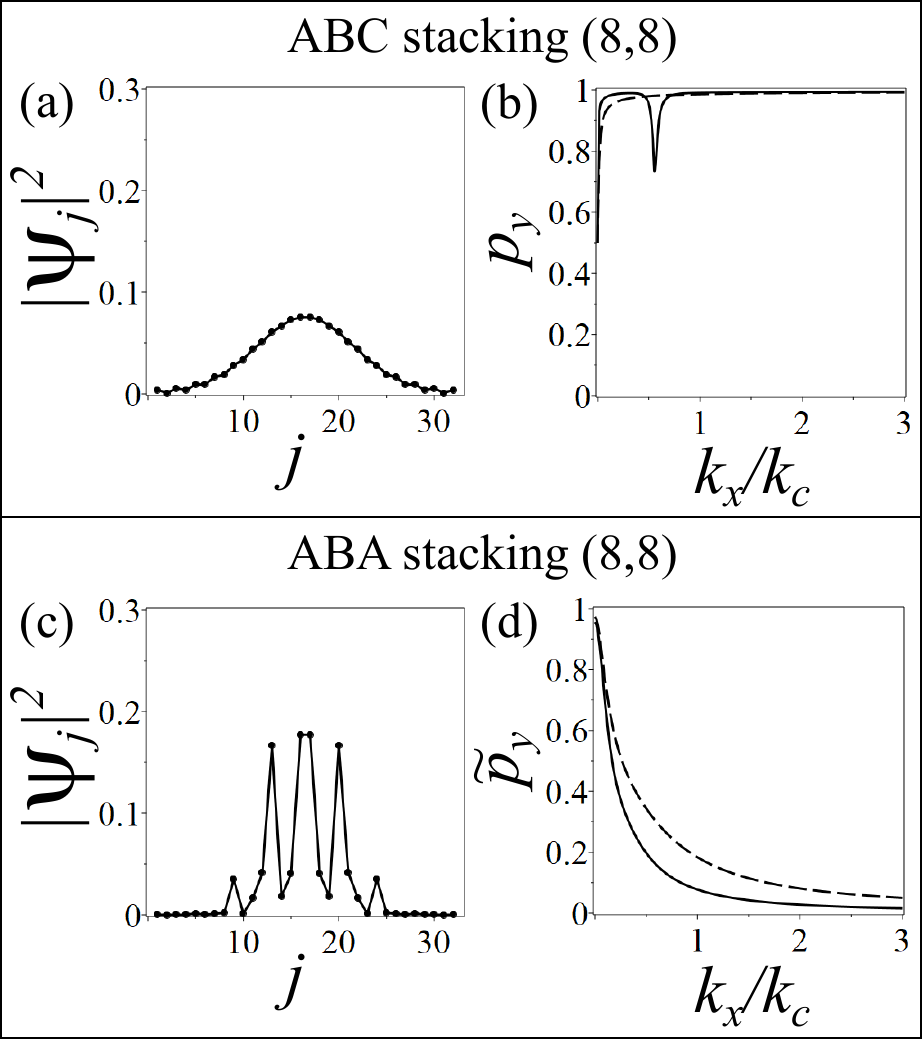}
\caption{A smooth soliton of width $\zeta = 8$ at the center of the system $(8,8)$, $N=16$ layers, for ABC stacking (top) and ABA stacking (bottom).
(a) The probability density $|\psi_j|^2$ per site $j = 1, 2, \ldots , 32$ for the energy level localized on the soliton at $k_x = k_c$, $k_y = 0$ and $U/\gamma_1 = 0.6$. (b) The polarization $p_y = \langle \psi | T_{a/2}S_z | \psi \rangle$, Eq.~(\ref{py1}), of the soliton state as a function of $k_x$ with $k_y = 0$ for $U/\gamma_1 = 0.6$ (solid) and $U/\gamma_1 = 1.8$ (dashed).
(c) The probability density $|\psi_j|^2$ per site $j = 1, 2, \ldots , 32$ for the energy level localized on the soliton at $k_x = k_c$, $k_y = 0$ and $U/\gamma_1 = 0.6$. (d) The polarization $\tilde{p}_y = \langle \psi | \tilde{T}_{a/2}S_z | \psi \rangle$, Eq.~(\ref{altpy1}), of the soliton state as a function of $k_x$ with $k_y = 0$ for $U/\gamma_1 = 0.6$ (solid) and $U/\gamma_1 = 1.8$ (dashed).
In all plots, $\gamma_0 = 3.16\,$eV, $\gamma_1 = 0.381\,$eV~\cite{kuzmenko09}, $a = 2.46$\AA~\cite{saito98}.
}\label{figABC2smooth}
\end{figure}

\subsection{ABC stacking}

For ABC stacking, Fig.~\ref{figsmooth1} shows that the soliton level remains near the same energy $E_{\mathrm{sol}} \approx \gamma_1$ at $k=0$ for a range of $U$ values, in contrast to a sharp soliton, Eq.~(\ref{abcesol1}).
Since the soliton level moves into the valence band for large $k$, this means that the soliton state crosses the bulk band gap, leading to a nonzero density of states.

The behavior at $k=0$ may be understand because the system separates into dimers and isolated atoms there, and the soliton state is localized on a dimer consisting of site $B_m$ and $A_{m+1}$ connected by interlayer hopping $\gamma_1$, as in Fig.~\ref{fig:summary}(g).
The onsite energy of these sites is identical,
$\epsilon_{A,m+1} = \epsilon_{B,m} = - U \tanh ( 1 / 2\zeta )$ according to Eq.~(\ref{smooth1}).
The dimer yields two energy levels and the soliton level is the highest energy of the pair,
\begin{eqnarray}
E_{\mathrm{sol}} = \gamma_1 - U \tanh \bigg( \frac{1}{2\zeta} \bigg)  \quad \mathrm{for} \, k = 0 . \label{smooth3}
\end{eqnarray}
This is in agreement with the sharp soliton energy~(\ref{abcesol1}), $E_{\mathrm{sol}} =  \gamma_1 - U$, in the limit $\zeta \rightarrow 0$, as expected.
For $\zeta \gg 1/2$, we expand the hyperbolic tangent to give $E_{\mathrm{sol}} \approx \gamma_1 - U/(2\zeta)$ which shows that $E_{\mathrm{sol}} \approx \gamma_1$ for a range of $U$ values if the width $\zeta$ is large enough.

As in the case of a sharp soliton, the two surface states, on $A_1$ and $B_N$, are completely disconnected at $k=0$ and they give two degenerate states at energy $E \approx U$ according to Eq.~(\ref{smooth1}) (they are at $E=U$ in a sufficiently large system $N \gg \zeta$).
For $N \gg \zeta \gg 1$, the soliton level moves from $E_{\mathrm{sol}} \approx \gamma_1$ at $k=0$ to the valence band at large $k$, crossing the surface states at $E \approx U$ provided that $U < \gamma_1$.

At $k=k_c$, the soliton level is at $E_{\mathrm{sol}} \approx 0$ for a wide range of $U$ values. This is because the system is approximately equivalent to the CDW model at this point and, for smooth solitons $\zeta \gg 1$, the system approaches the continuum limit which supports a zero-energy soliton level, as discussed in detail previously for the CDW model~\cite{brzezicki20,allen22}.

For $U < \gamma_1$, all levels in the conduction and valence bands originate from near $E = \pm \gamma_1$ at $k=0$, as for rhombohedral graphite at $U=0$.
For $U > \gamma_1$, the levels no longer coalesce at $k=0$, but bifurcations are visible in the conduction and valence bands whereby levels that are doubly degenerate at $k=0$ split at nonzero $k$. This is because of spatial inversion of the texture of onsite energies~(\ref{smooth1}) about the center of the soliton. At $k=0$, there are identical dimers either side of the soliton and equidistant from it, creating a degeneracy which is broken at nonzero $k$.

As stated above, the soliton state at $k=0$ is localized on two adjacent sites near the center of the soliton as $\psi_{\mathrm{sol}} = (\psi_{B_{m}} + \psi_{A_{m+1}})/\sqrt{2}$.
For $k = k_c$, it remains localized on the soliton, but has a broader extent in position space as shown in Fig.~\ref{figABC2smooth}(a) for $U/\gamma_1 = 0.6$.
Fig.~\ref{figABC2smooth}(b) shows the polarization $p_y = \langle \psi | T_{a/2}S_z | \psi \rangle$, Eq.~(\ref{py1}), as a function of $k_x$ (with $k_y = 0$) for $U/\gamma_1 = 0.6$ (solid line) and $U/\gamma_1 = 1.8$ (dashed line).
There is a noteable dip in $p_y$ for $U/\gamma_1 = 0.6$ (solid line) at $k_x \approx 0.6 k_c$ which arises when the soliton level crosses the surface states, Fig.~\ref{figsmooth1}(g).
Otherwise, $p_y \approx 1$ for a range of $k$ values. This is related to the fact that, for $k=k_c$, the Hamiltonian~(\ref{habc}) is approximately equal to that of the CDW model. However, the polarization $p_y$ is never exactly one because the nonsymmorphic chiral symmetry is broken by the ends of the system and the finite width of the soliton~\cite{brzezicki20,allen22}.

\subsection{ABA stacking}

For ABA stacking, Fig.~\ref{figsmooth1} shows that the antisoliton level remains near the same energy $E_{\mathrm{sol}} \approx 0$ at $k=0$ for a wide range of $U$ values, and it moves into the conduction band at large $k$.
The level is near zero at $k=0$ because the central $N$-mer of the system is separated from the individual non-dimer atoms at this point and, thus, the $N$-mer is equivalent to the CDW model. For smooth solitons $\zeta \gg 1$, the $N$-mer approaches the continuum limit with a soliton level approaching zero exponentially quickly as a function of $\zeta$, as discussed in detail previously for the CDW model~\cite{brzezicki20,allen22}.

In addition to the soliton state, however, there are many levels in the conduction and valence bands, and also in the nominal bulk band gap, which bifurcate, being doubly degenerate at $k=0$ and splitting at nonzero $k$. These levels arise from the individual non-$N$-mer atoms which are separated from the rest of the system at $k=0$. Owing to spatial inversion of the texture of onsite energies~(\ref{smooth1}) about the center of the soliton, there are single atoms with identical onsite energies either side of the soliton and equidistant from it, creating a degeneracy which is broken at nonzero $k$.
The two atoms nearest to the center of the soliton have energies $E = - U \tanh ( 1 / 2\zeta )$ at $k=0$ according to Eq.~(\ref{smooth1}), so, for $\zeta \gg 1$, they are very close to zero energy as $E \approx - U/(2\zeta)$.

For $k = 0$, the soliton state is localized near the center of the soliton~\cite{brzezicki20,allen22}, and, for $k_x = k_c$ (and $k_y = 0$), the state is still localized near the soliton center as shown in Fig.~\ref{figABC2smooth}(c).
Since the system is equivalent to the CDW model at $k = 0$, the soliton state is polarized, with maximum polarization approaching one at $k = 0$, Fig.~\ref{figABC2smooth}(d).
Again, the polarization $\tilde{p}_y$ is never exactly one because the nonsymmorphic chiral symmetry is broken by the ends of the system and the finite width of the soliton~\cite{brzezicki20,allen22}.

\subsection{AA$^{\prime}$ stacking}

For AA$^{\prime}$ stacking, Fig.~\ref{figsmooth1} shows that there are a pair of bands (soliton and antisoliton) near zero energy, as for a sharp soliton, but the major difference is that these two bands do not separate as $U/\gamma_1$ increases, but remain near zero energy as flat bands with a large extent up to $k \approx 2k_c$.

At $k=0$, the system consists of two separated CDW chains, each of them has either a soliton or an antisoliton. For smooth solitons $\zeta \gg 1$, the CDW chains approach the continuum limit with a soliton (or antisoliton) level approaching zero exponentially quickly as a function of $\zeta$, as discussed previously~\cite{brzezicki20,allen22}.

For small values of $U/\gamma_1$, Fig.~\ref{figsmooth1}(e) and (k), there are also level crossings between the soliton and antisoliton bands which appear to oscillate as a function of $k$, as described for sharp solitons in Section~\ref{s:aapss}.
The oscillations, due to anticrossings with the other conduction and valence bands, are reduced as $U/\gamma_1$ increases and the other bands move away from zero energy, separating from the soliton-antisoliton pair. The flat bands due to the soliton-antisoliton pair result in a large peak in the density of states near zero energy, Figure~\ref{figsmooth1}(r) and (x). We consider the effect of disorder on this peak in Section~\ref{s:disorder}.

\section{Single layer defects}\label{s:single}

\begin{figure*}[t]
\includegraphics[scale=0.36]{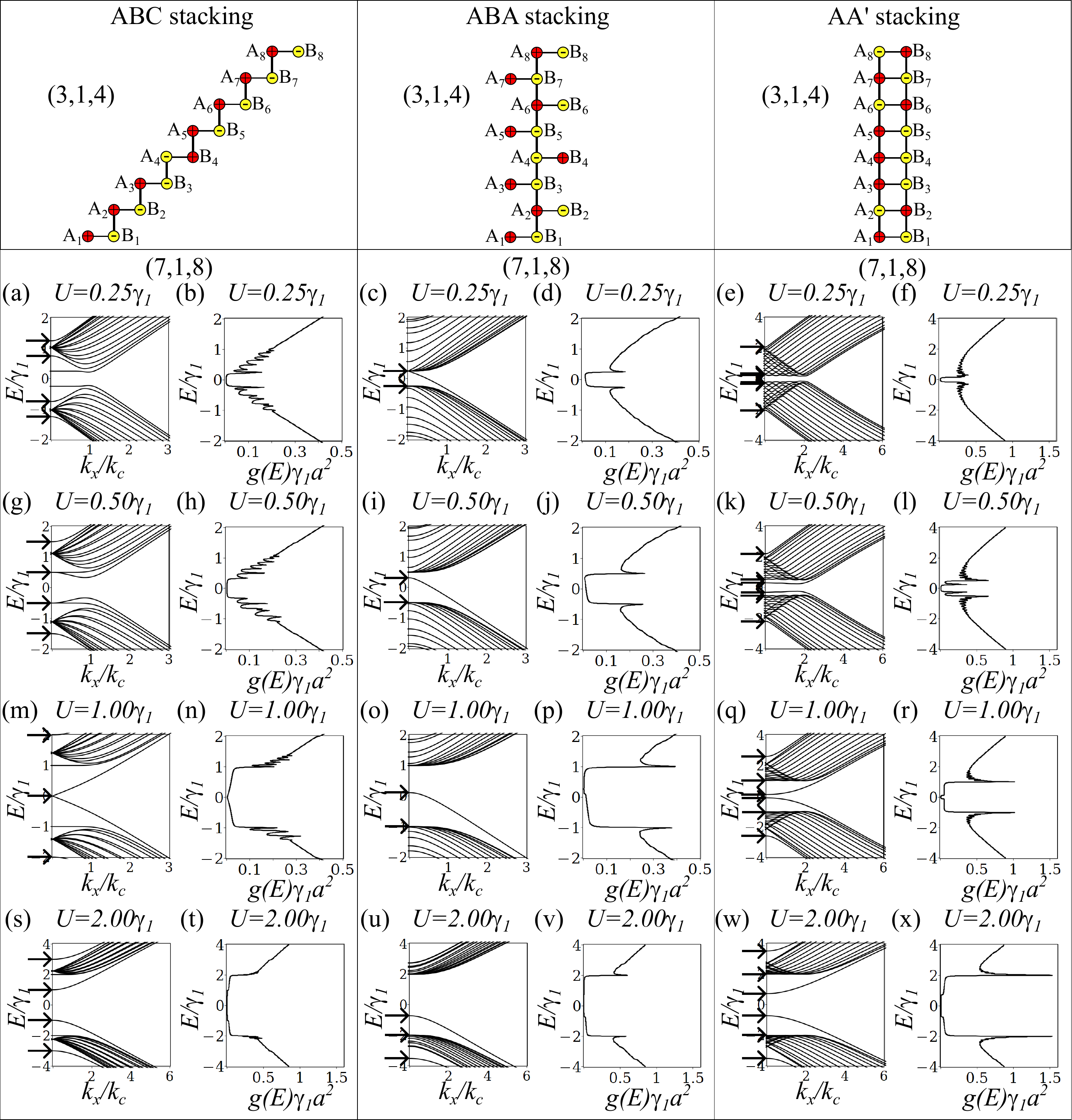}
\caption{Band structures and density of states $g(E)$ for single layer defects near the center $(7,1,8)$ of a system with $N=16$ layers for ABA, ABC, and AA$^{\prime}$ stacking.
In the band structure plots, horizontal arrows show the energies at $k_x=0$ of the states localized on the defect.
There is a different axis scale for AA$^{\prime}$ stacking, and the other stackings at large values of $U/\gamma_1$.
In all plots, parameter values are $\gamma_0 = 3.16\,$eV, $\gamma_1 = 0.381\,$eV~\cite{kuzmenko09}, $a = 2.46$\AA~\cite{saito98}. For the band structures, $k_y=0$, and, for the density of states, $\delta = 0.01 \gamma_1$.
}\label{figsingle1}
\end{figure*}

So far, we have considered solitons which consist of a change in the texture of onsite energies without changing the interatomic hopping, and they may be sharp on the atomic scale or smooth with a nonzero width $\zeta$. There are many other types of stacking faults one could consider. An example, which also involves onsite energies, is to reverse the signs of the onsite energies on a single layer only. This doesn't change the texture of the onsite energies either side of the defect, and we refer to this as a `single layer defect', not a soliton.
Such defects were modeled for AA, AA$^{\prime}$, ABA and ABA$^{\prime}$ stacking using density functional theory (DFT) in the context of hexagonal boron nitride~\cite{yin11}, and we will only discuss them briefly here, for ABC, ABA and AA$^{\prime}$ stacking.
The lattice structures, with the defects, are shown schematically in the top panels of Fig.~\ref{figsingle1} for a system we denote as $(3,1,4)$ indicating a system of $N=8$ layers consisting of $3$ layers (at the bottom) with regular stacking, followed by a single layer defect, followed by $4$ layers (at the top) with regular stacking.
Band structures and density of states for ABC, ABA, and AA$^{\prime}$ stacking, are shown in the remaining panels in Fig.~\ref{figsingle1} for a system with a larger number of layers $N=16$ and a defect near its center $(7,1,8)$.

\subsection{ABC stacking}

For ABC stacking, Fig.~\ref{figsingle1} shows that there are generally two defect states within the band gap, one merges with the conduction band for $k \gg k_c$, the other merges with the valence band. For $U < \gamma_1 / 2$, Fig.~\ref{figsingle1}(a), the two surface states are closer to zero energy than the defect states whereas, for $U > \gamma_1 / 2$, the defect states are closest to zero energy and, for $U = \gamma_1$, Fig.~\ref{figsingle1}(m), they touch at $E=0$ and there is no band gap at this point. For $U = \gamma_1 / 2$, Fig.~\ref{figsingle1}(g), the defect states hybridize strongly with the surface states.

The behavior at $k=0$ may be understand because the system separates into dimers and isolated atoms there. There are the two isolated surface states with energies $E = \pm U$, $N-3$ dimers each with energies $E = \pm \sqrt{U^2 + \gamma_1^2}$, and, near the defect, a dimer with onsite energies $-U$ giving $E = -U \pm \gamma_1$ and a dimer with onsite energies $+U$ giving $E = +U \pm \gamma_1$. These four energies are shown with the horizontal arrows in Fig.~\ref{figsingle1}.
The presence of two defect states within the band gap is in stark contrast to a single soliton which only supports one state within the gap, Fig.~\ref{figABC1}. Instead, a single layer defect has a band structure very similar to that of a soliton-antisoliton pair, Fig.~\ref{figABC4}, because the latter can also be viewed as a defect, albeit of larger spatial extent than just one layer, which doesn't change the texture of the onsite energies either side of the defect.

\subsection{ABA stacking}

For ABA stacking, Fig.~\ref{figsingle1} shows that there is a single defect state within the bulk band gap. For small $U/\gamma_1$, it is near the bottom of the conduction band for $k=0$, crossing the band gap to move into the valence band at $k \gg k_c$, whereas, for larger $U/\gamma_1$, it is always at negative energy.
The lattice structure, Fig.~\ref{figsingle1}(b), shows there are three consecutive sites in the $N$-mer with the same onsite energies $E= - U$. For $k=0$ and $U \gg \gamma_1$, this trimer gives energies $E = -U$ and $E = -U \pm \sqrt{2} \gamma_1$ (as indicated by the horizontal arrows in Fig.~\ref{figsingle1}). The level $E = -U + \sqrt{2} \gamma_1$ is the one within the band gap. A single soliton also creates a single level within the band gap, Fig.~\ref{figABA1}, and the band structure and density of states of a single soliton and a defect are similar, particularly for $U \gg \gamma_1$. For $U \ll \gamma_1$, the defect state extends across most of the band gap, creating a wide energy region with a nonzero density of states, unlike the soliton state.

\subsection{AA$^{\prime}$ stacking}

For AA$^{\prime}$ stacking, Fig.~\ref{figsingle1} shows that there are generally two defect states within the band gap, one merges with the conduction band for $k \gg k_c$, the other merges with the valence band.
The lattice structure, Fig.~\ref{figsingle1}, shows the ladder structure consisting of two coupled $N$-mers. In one of them, there is a trimer of states with onsite energy $E = +U$ near the defect, and, in the other, there is a trimer of states with onsite energy $E = -U$ near the defect. Together, they contribute six states which are, at $k=0$ and $U \gg \gamma_1$, $E = \pm U$, $E = -U \pm \sqrt{2} \gamma_1$, and $E = U \pm \sqrt{2} \gamma_1$ (as indicated by the horizontal arrows in Fig.~\ref{figsingle1}). The two levels $E = -U + \sqrt{2} \gamma_1$ and $E = U - \sqrt{2} \gamma_1$ are within the band gap. A single soliton also creates two levels within the band gap, Fig.~\ref{figAAp1}, and the band structure and density of states of a single soliton and a defect are similar. There are differences such as the hybridization of the two levels with the levels being close together and strongly hybridized for $U \ll \gamma_1$ in the soliton and for $U \approx \gamma_1$ in the defect.

\section{The role of interlayer disorder}\label{s:disorder}

We have shown that solitons, both atomically-sharp and smooth in position space, are able to support localized states with energies within the bulk band gap for ABC, ABA and AA$^{\prime}$ stacking. The soliton bands are generally dispersive as a function of the in-plane wave vector ${\bf k}$ so that they give rise to a non-zero density of states without any particularly sharp features. The exception are smooth solitons for AA$^{\prime}$ stacking which give flat bands yielding a narrow peak at zero energy in the density of states, Fig.~\ref{figsmooth1}(x).

For certain values of the in-plane wave vector ${\bf k}$, the lattice structure of these systems may be related by dimensional reduction to that of the CDW model~\cite{kivelson83,brzezicki20,cayssol21,fuchs21,allen22,mccann23} which has nonsymmorphic chiral symmetry. The influence of symmetry-breaking disorder has been considered previously for the CDW model~\cite{allen22} so here we consider it only for the most interesting case of AA$^{\prime}$ stacking, for both atomically-sharp and smooth solitons.

To focus on the topology related to the CDW model, we consider random tight-binding parameters in the perpendicular-to-layer direction while preserving translational invariance in the in-plane direction.
We consider four types of disorder: (i) ``Sharp onsite disorder'' (diagonal) which gives an additional contribution $\delta \epsilon_{A,j}$ to the onsite energy of an A site on the $j$th layer, where $j = 1,2,\ldots,N$, which is drawn randomly from a
uniform distribution $-W \leq \delta \epsilon_{A,j} \leq W$ with disorder strength $W$. We maintain charge neutrality within each layer so that $\delta \epsilon_{B,j} = - \delta \epsilon_{A,j}$ for all layers $j$.
(ii) ``Sharp hopping disorder'' (off-diagonal) where the interlayer coupling takes random values $\gamma_1 + \delta_n$ where $n = 1,2,\ldots,N-1$ indexes the $N-1$ interlayer spaces and each $\delta_n$ takes a value drawn randomly from a uniform distribution $-W \leq \delta_n \leq W$. For AA$^{\prime}$ stacking, the values of coupling between A atoms and between B atoms for any pair of adjacent layers are identical, so that there are only $N-1$ random values of $\delta_n$ in total.
(iii) ``Smooth onsite disorder'' (diagonal) with an additional contribution $\delta \epsilon_{A,j}$ to the onsite energy of an A site on the $j$th layer described by a Gaussian-correlated potential~\cite{koschny02,guo08} as given by
\begin{eqnarray}
\delta \epsilon_{A,j} = \frac{\sum_m w_m \exp ( - |j-m|^2/\eta^2 )}{\sqrt{\sum_m \exp ( - |j-m|^2/\eta^2 )}} , \label{gaussdis}
\end{eqnarray}
where $\eta$ is the correlation length in dimensionless units, i.e., measured in units of the interlayer spacing. The summation is over all layers $m = 1,2,\ldots,N$ with $w_m$ drawn randomly from a uniform distribution $-W \leq w_m \leq W$ with disorder strength $W$. We maintain charge neutrality within each layer so that $\delta \epsilon_{B,j} = - \delta \epsilon_{A,j}$ for all layers $j$.
(iv) ``Smooth hopping disorder'' (off-diagonal) where the interlayer coupling takes random values $\gamma_1 + \delta_n$ where $n = 1,2,\ldots,N-1$ indexes the $N-1$ interlayer spaces and each $\delta_n$ is described by a Gaussian-correlated weighting as in Eq.~(\ref{gaussdis}), the only difference being that now there are $N-1$ bonds and $N-1$ independent parameters $w_m$. Again, the values of coupling between A atoms and between B atoms for any pair of adjacent layers are identical, so that there are only $N-1$ random values of $\delta_n$ in total.

\begin{figure}[t]
\includegraphics[scale=0.43]{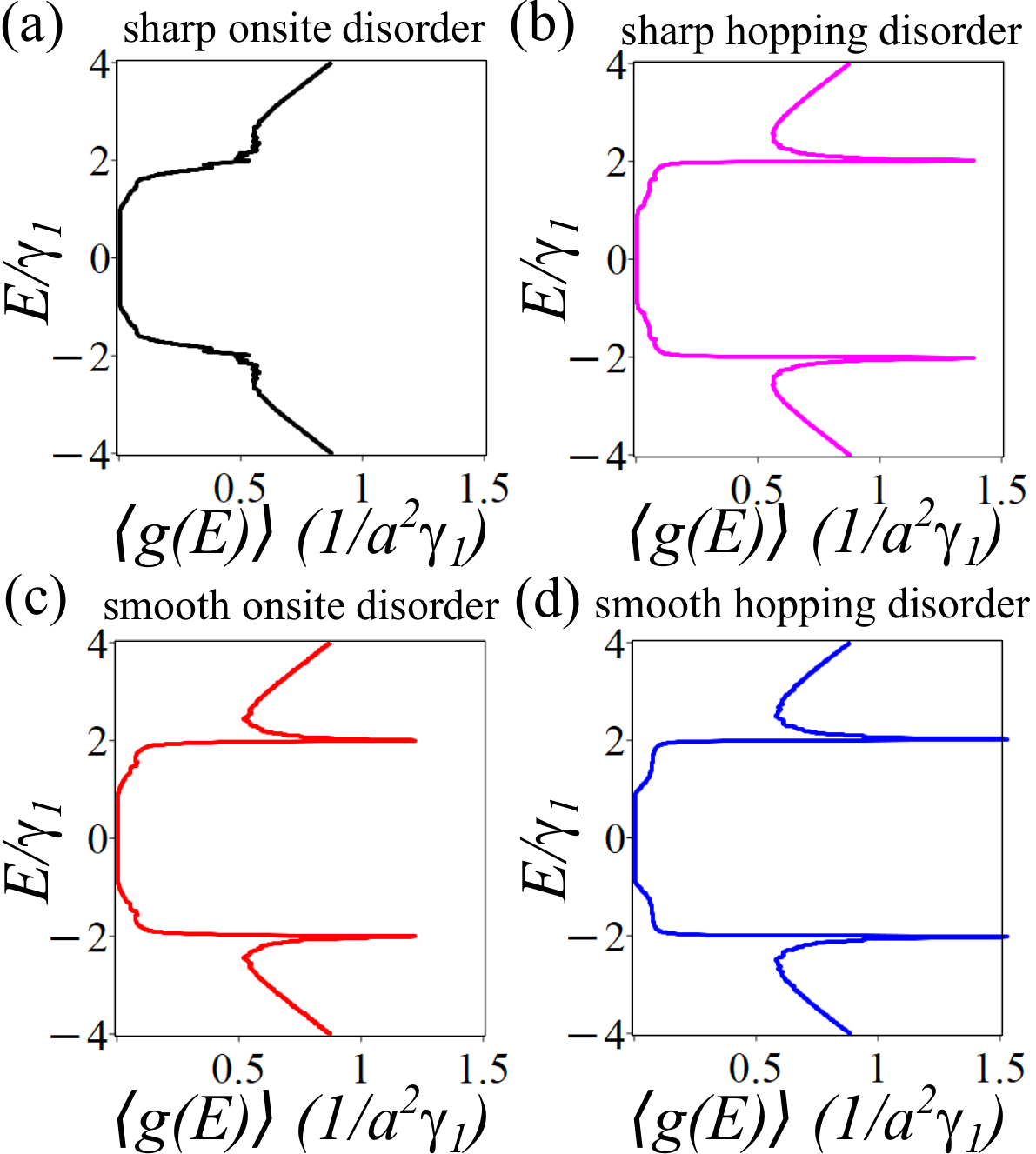}
\caption{Dependence of the disorder-averaged density of states $\langle g(E) \rangle$ on energy $E$ for an atomically-sharp soliton at the center $(8,8)$ of a system with AA$^{\prime}$ stacking and $N=16$ layers.
For all plots, the onsite energy is $U = 2.0 \gamma_1$, the disorder strength is $W = 0.5 \gamma_1$, and $\langle g(E) \rangle$ is determined using Eq.~(\ref{dos}) with broadening $\delta = 0.01 \gamma_1$ and $20$ disorder realizations. Other parameter values are $\gamma_0 = 3.16\,$eV, $\gamma_1 = 0.381\,$eV~\cite{kuzmenko09}, $a = 2.46$\AA~\cite{saito98}.
(a) is for atomically-sharp onsite disorder (black), (b) is for sharp hopping disorder (magenta), (c) is for Gaussian-correlated onsite disorder~(\ref{gaussdis}) with correlation length $\eta = 4$ (red), and (d) is for Gaussian-correlated hopping disorder with $\eta = 4$ (blue).
}\label{figdissharp1}
\end{figure}

\begin{figure}[t]
\includegraphics[scale=0.43]{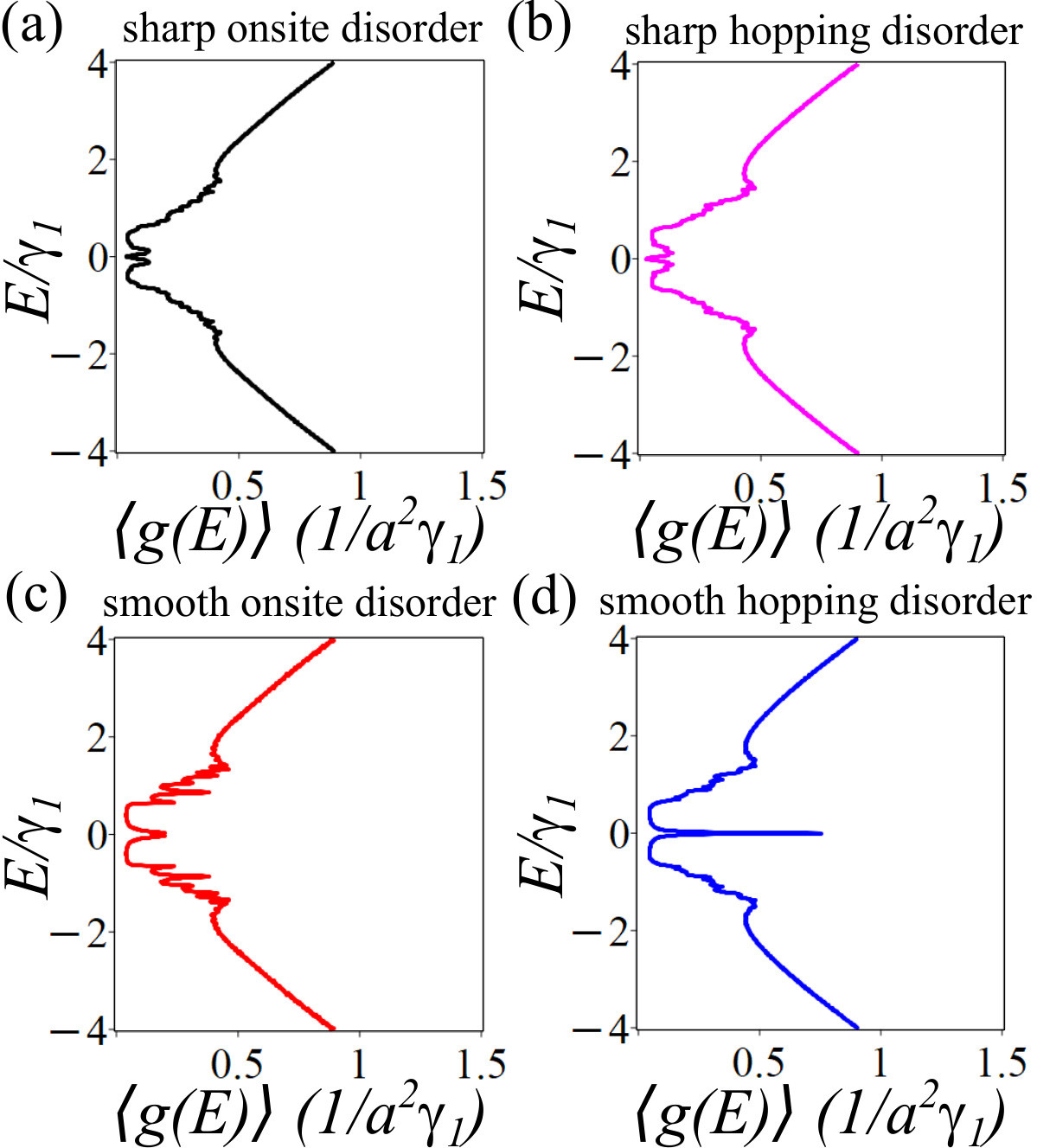}
\caption{Dependence of the disorder-averaged density of states $\langle g(E) \rangle$ on energy $E$ for a smooth soliton of width $\zeta = 8$ at the center $(8,8)$ of a system with AA$^{\prime}$ stacking and $N=16$ layers.
For all plots, the onsite energy is $U = 2.0 \gamma_1$, the disorder strength is $W = 0.5 \gamma_1$, and $\langle g(E) \rangle$ is determined using Eq.~(\ref{dos}) with broadening $\delta = 0.01 \gamma_1$ and $20$ disorder realizations. Other parameter values are $\gamma_0 = 3.16\,$eV, $\gamma_1 = 0.381\,$eV~\cite{kuzmenko09}, $a = 2.46$\AA~\cite{saito98}.
(a) is for atomically-sharp onsite disorder (black), (b) is for sharp hopping disorder (magenta), (c) is for Gaussian-correlated onsite disorder~(\ref{gaussdis}) with correlation length $\eta = 4$ (red), and (d) is for Gaussian-correlated hopping disorder with $\eta = 4$ (blue).
}\label{figdissmooth1}
\end{figure}

The normalisation with the square root factor in Eq.~(\ref{gaussdis}) is used so that smooth disorder interpolates between sharp disorder for $\eta \ll 1$ and sample-to-sample parameter variations~\cite{allen22} for $\eta \gg N$.
The latter are parameter values that are spatially uniform across a single member of the ensemble, i.e., $\delta \epsilon_{A,j} = \delta \epsilon_{A}$ for all $j$, but that differ from sample to sample with $\delta \epsilon_{A}$ drawn randomly from a uniform distribution $-W \leq \delta \epsilon_{A} \leq W$.
Smooth solitons in the CDW model are fairly robust to parameter variations~\cite{allen22} because the nonsymmorphic chiral symmetry holds in the continuum limit (of a smooth soliton with parameter variations). However, the equivalence of AA$^{\prime}$ stacking with the CDW model only holds at $k=0$, so smooth solitons for AA$^{\prime}$ stacking are not expected to be as robust.
As parameter variations were studied in detail for the CDW model~\cite{allen22}, we consider instead the case of a smooth potential with a finite range $1 \alt \eta \alt N$.

For AA$^{\prime}$ stacking and an atomically sharp soliton in the absence of disorder, Fig~\ref{figAAp1}(j), the density of states has a step-like nonzero density of states within the bulk band gap $-U \leq E \leq U$ due to the soliton bands, Fig~\ref{figAAp1}(e). 
In addition, there are also sharp peaks in the density of states at the bulk band edges, $E = \pm U$.
Figure~\ref{figdissharp1} shows the disorder-averaged density of states $\langle g(E) \rangle$ for an atomically-sharp soliton at the center $(8,8)$ of a system with AA$^{\prime}$ stacking and $N=16$ layers. We consider the example of onsite energy $U = 2.0 \gamma_1$ and disorder strength $W = 0.5 \gamma_1$.
Figure~\ref{figdissharp1} shows that all types of disorder tend to smooth out the step-like feature due to the soliton states within the bulk band gap.
The sharp peaks at $E = \pm U$ at the band edges are quite robust to disorder, as they are only smoothed out by atomically-sharp onsite disorder, Figure~\ref{figdissharp1}(a).

For AA$^{\prime}$ stacking and a smooth soliton in the absence of disorder, Fig~\ref{figsmooth1}(x), the density of states shows a narrow peak at zero energy due to the flat bands related to the soliton states.
Figure~\ref{figdissmooth1} shows the disorder-averaged density of states $\langle g(E) \rangle$ for a smooth soliton at the center $(8,8)$ of a system with AA$^{\prime}$ stacking and $N=16$ layers.
Sharp disorder, Fig.~\ref{figdissmooth1}(a) and (b), tends to destroy the narrow peak at zero energy, leaving a small but finite density of states in the bulk band gap.
The narrow peak is fairly robust to smooth disorder, however, and more robust to hopping disorder than onsite disorder, Fig.~\ref{figdissmooth1}(c) and (d). This is possibly because the solitons are textures in the onsite energies, not the hopping parameters.

\section{Electronic transport}\label{s:transport}

\begin{figure}[t]
\includegraphics[scale=0.41]{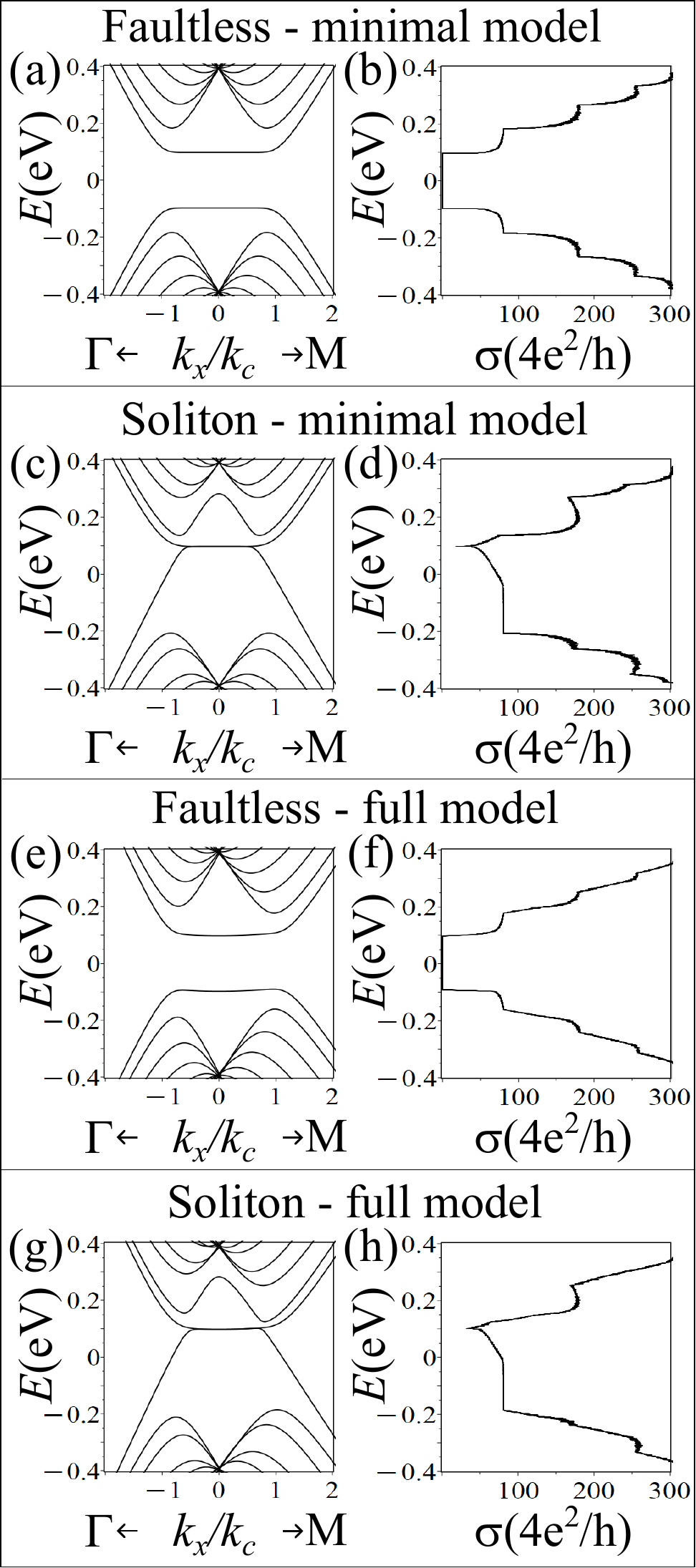}
\caption{Rhombohedral stacking with $N=10$ layers and onsite energy $U = 0.1\,$eV showing band structures (left) and electrical conductivity (right). (a), (b) show the faultless system and (c), (d) show a system with an atomically-sharp soliton at the center using the minimal model with $\gamma_2 =\gamma_3 = \gamma_4 =0$. (e), (f) show the faultless system and (g), (h) show a system with an atomically-sharp soliton at the center using the full parameter model. 
Parameter values are $\gamma_0 = 3.16\,$eV, $\gamma_1 = 0.381\,$eV, $\gamma_2 = - 0.02\,$eV, $\gamma_3 = 0.315\,$eV, $\gamma_4 = 0.044\,$eV, $a = 2.46$\AA. The conductivity calculation is performed for a system of length $L = 1754a$ and width $W = 1754a$, where $a$ is the lattice constant. Details are described in Appendix~\ref{a:transport}.
}\label{newfig1}
\end{figure}

Solitons, both atomically-sharp and smooth in position space, are able to support localized states with energies within the bulk band gap for ABC, ABA and AA$^{\prime}$ stacking. The contrasting electronic band structures and densities of states will impact transport and spectroscopic measurements~\cite{jaffrennou07,museur08,pierret14,bourrellier14,henck17,kim13,woods14,ni19}. As an example, we model coherent, ballistic, electronic transport for rhombohedral-stacking, generalizing a model developed for monolayer and bilayer graphene~\cite{tworzydlo06,snyman07} in which the central sample is connected to leads of the same material that are heavily doped in order to provide a large density of states. The conductivity is determined using the Landauer-B\"uttiker formalism~\cite{landauer88} where the transmission probability is found by wave-matching at the boundaries of the leads to the central sample, and Hamiltonians are taken in the continuum limit. Details of our numerical calculations are described in Appendix~\ref{a:transport}.

Generalizing previous results for monolayer and bilayer graphene~\cite{tworzydlo06,snyman07}, we find that the minimal conductivity for rhombohedral stacking in the absence of onsite energies ($U=0$) is given by $\sigma_{\mathrm{min}} = 4N e^2 / (\pi h)$ for $N$ layers with $W \agt L$ where $W$ and $L$ are the sample width and length, respectively. 
Figure~\ref{newfig1} shows band structures and corresponding conductivities for a rhombohedrally-stacked system with $N=10$ layers and onsite energy $U = 0.1\,$eV.
The band structures are plotted near the K point, with the direction of the $\Gamma$ and M points indicated for negative and positive $k_x$, respectively.
Fig.~\ref{newfig1}(a) and (b) are for the minimal tight-binding model (with $\gamma_0$, $\gamma_1$, and $U$ only, Eq.~(\ref{h1})) and the faultless system.
The region of the band gap, Fig.~\ref{newfig1}(a), corresponds to zero conductivity, Fig.~\ref{newfig1}(b), and, when the energy is equal to that of higher-energy bands, the presence of additional transport modes produces steps in the conductivity.
Fig.~\ref{newfig1}(c) and (d) are for the minimal tight-binding model with an atomically-sharp soliton at the center.
The state localized on the soliton creates a non-zero density of states resulting in non-zero conductivity, and the position of the strong hybridization of the soliton state with the surface states corresponds to a characteristic dip in the conductivity (at $E \approx U = 0.1\,$eV in Fig.~\ref{newfig1}(d)).

\section{Additional tight-binding parameters}\label{s:tight}

So far our calcuations have employed a minimal model with only $\gamma_0$, $\gamma_1$, and $U$, Eq.~(\ref{h1}). To assess the influence of additional tight-binding parameters, we consider three examples: Rhombohedral graphite (ABC stacking), hexagonal boron nitride~\cite{naclerio22} (AA$^{\prime}$ stacking), and hexagonal boron phosphide~\cite{wang15,wang19,hernandez21} (ABA stacking).

\subsection{ABC stacking: rhombohedral graphite}

We consider a tight-binding model of rhombohedral graphite containing skew interlayer hopping $\gamma_3$ and $\gamma_4$, and next-nearest layer hopping $\gamma_2$.
In a basis of a single orbital on each atomic site $( A_1 , B_1 , A_2 , B_2 , \ldots , A_N , B_N)$, the Hamiltonian~(\ref{h1}) is modified~\cite{koshino09} as 
\begin{eqnarray}
H = \begin{pmatrix}
D_1 & V & W & 0 & 0 & \hdots \\
V^{\dagger} & D_2 & V & W & 0 & \hdots \\
W^{\dagger} & V^{\dagger} & D_3 & V & W & \hdots \\
0 & W^{\dagger} & V^{\dagger} & D_4 & V & \hdots \\
\vdots & \vdots & \vdots & \vdots & \vdots & \ddots
\end{pmatrix} , \label{h1all}
\end{eqnarray}
where intralayer blocks $D_i$ are defined in Eq.~(\ref{di}) and we use $U_i = U$ for all $i$ (for a faultless system).
Interlayer blocks are
\begin{eqnarray}
V &=& \begin{pmatrix}
\gamma_4 f ({\bf q})  & - \gamma_3 f^{\ast} ({\bf q}) \\
\gamma_1 & \gamma_4 f ({\bf q})
\end{pmatrix} , \label{vfull}
\end{eqnarray}
and next-nearest layer blocks are
\begin{eqnarray}
W &=& \begin{pmatrix}
0 & \gamma_2 / 2 \\
0 & 0
\end{pmatrix} .
\end{eqnarray}
The band structures and corresponding conductivities for this model are shown in Fig.~\ref{newfig1}(e)-(h).
We use the following parameter values~\cite{kuzmenko09,koshino09}, $\gamma_0 = 3.16\,$eV, $\gamma_1 = 0.381\,$eV, $\gamma_2 = - 0.02\,$eV, $\gamma_3 = 0.315\,$eV, $\gamma_4 = 0.044\,$eV with $U = 0.1\,$eV.
Fig.~\ref{newfig1}(e) and (f) are for the faultless system, whereas Fig.~\ref{newfig1}(g) and (h) are for an atomically-sharp soliton at the center of the system.
In comparison to the minimal model, it can be seen that the additional parameters have a generally small effect on the band structure and conductivity, without producing any qualitative changes.

\begin{figure}[t]
\includegraphics[scale=0.42]{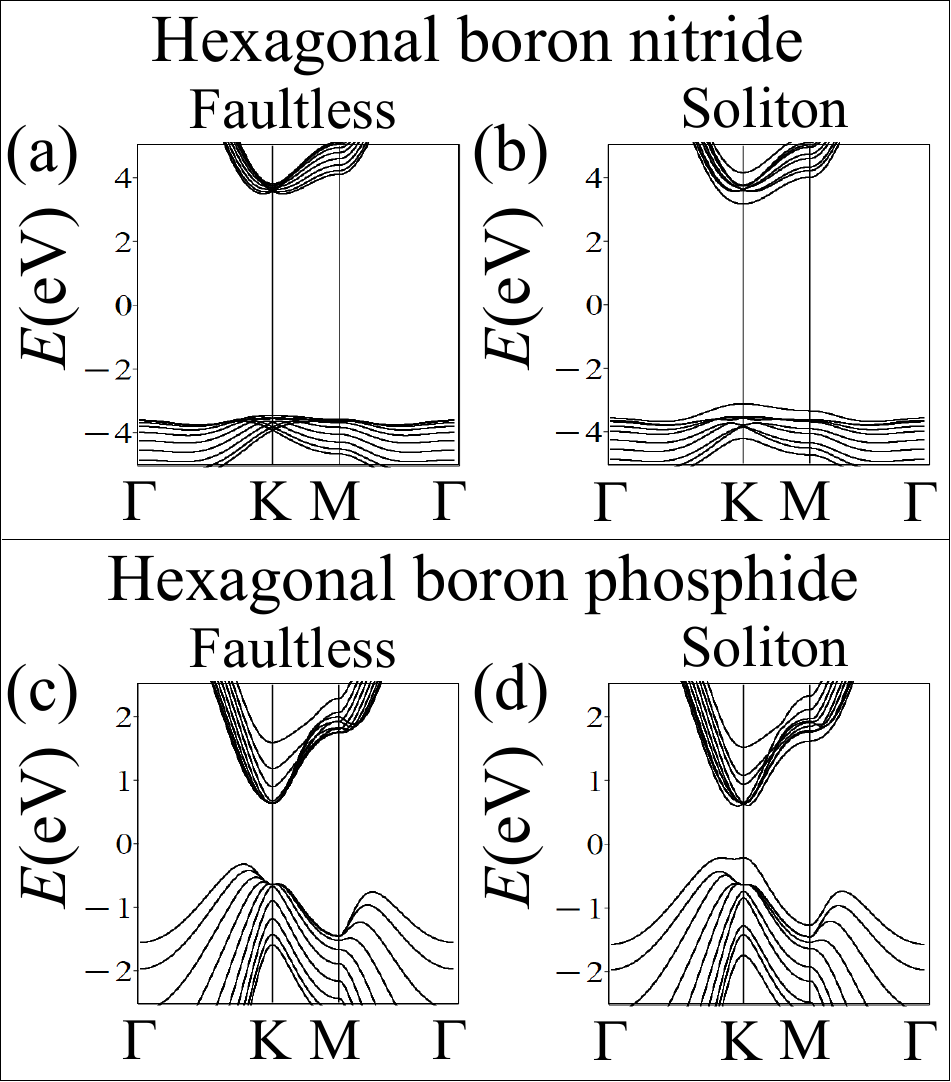}
\caption{(a), (b) band structure of hexagonal boron nitride with $N = 10$ layers and AA$^{\prime}$ stacking, plotted across the whole Brillouin zone with (a) a faultless system, and (b) a soliton at the center. Plots are obtained by diagonalizing Hamiltonian~(\ref{haap2}) with parameters $\gamma_0 = 2.33\,$eV, $\gamma_1 = 0.5\,$eV, $\gamma_n = -0.4\,$eV, $\gamma_2 = - 0.1\,$eV, and $U = 3.625\,$eV~\cite{sponza18}.
(c), (d) band structure of hexagonal boron phosphide with $N = 10$ layers and ABA stacking, plotted across the whole Brillouin zone with (c) a faultless system, and (d) a soliton at the center. Plots are obtained by diagonalizing Hamiltonian~(\ref{hbernalfull}) with parameters $\gamma_0 = 1.65\,$eV, $\gamma_1 = 0.761\,$eV, $\gamma_3 = -0.456\,$eV, $\gamma_4 = 0.255\,$eV, and $U = 0.636\,$eV~\cite{wang19}.
}\label{newfig2}
\end{figure}

\subsection{AA$^{\prime}$ stacking: hexagonal boron nitride}

As a second example of the influence of additional tight-binding parameters, we use a tight-binding model of AA$^{\prime}$-stacked hexagonal boron nitride (h-BN) with parameters fitted to calculations using density functional theory (DFT) and the GW approximation for a faultless system~\cite{sponza18}.
In this model~\cite{sponza18}, tight-binding parameters are $\gamma_0 = 2.33\,$eV, $\gamma_1 = 0.5\,$eV, and $U = 3.625\,$eV. Additional parameters, as compared to the minimal model, are in-plane next-nearest neighbor hopping $\gamma_n = -0.4\,$eV and next-nearest layer hopping $\gamma_2 = - 0.1\,$eV.

In this tight-binding model, for AA$^{\prime}$-stacking, 
the Hamiltonian~(\ref{h1}) is modified~\cite{sponza18} as 
\begin{eqnarray}
H = \begin{pmatrix}
\tilde{D}_1 & \tilde{V} & \tilde{W} & 0 & 0 & \hdots \\
\tilde{V}^{\dagger} & \tilde{D}_2 & \tilde{V} & \tilde{W} & 0 & \hdots \\
\tilde{W}^{\dagger} & \tilde{V}^{\dagger} & \tilde{D}_3 & \tilde{V} & \tilde{W} & \hdots \\
0 & \tilde{W}^{\dagger} & \tilde{V}^{\dagger} & \tilde{D}_4 & \tilde{V} & \hdots \\
\vdots & \vdots & \vdots & \vdots & \vdots & \ddots
\end{pmatrix} , \label{haap2}
\end{eqnarray}
written in terms of $2 \times 2$ blocks.
Intralayer blocks are
\begin{eqnarray}
\tilde{D}_i = \begin{pmatrix}
U_i - \gamma_n |f({\bf q})|^2 & - \gamma_0 f({\bf q}) \\
- \gamma_0 f^{\ast}\! ({\bf q}) & - U_i - \gamma_n |f({\bf q})|^2
\end{pmatrix} ,
\end{eqnarray}
where $U_i = U$ for odd $i$ and $U_i = -U$ for even $i$ (for a faultless system).
Interlayer blocks are
\begin{eqnarray}
\tilde{V} = \begin{pmatrix}
\gamma_1 & 0 \\
0 & \gamma_1
\end{pmatrix} ; \qquad
\tilde{W} = \begin{pmatrix}
\gamma_2 & 0 \\
0 & \gamma_2
\end{pmatrix} .
\end{eqnarray}
The band structure plotted across the whole Brillouin zone is shown for a faultless system in Fig.~\ref{newfig2}(a) and for an atomically-sharp soliton at the center in Fig.~\ref{newfig2}(b). For the faultless system, the band gap is near the K point but is indirect in the presence of additional parameters, Fig.~\ref{newfig2}(a). For the soliton, there are two states localized on the soliton within the bulk band gap, Fig.~\ref{newfig2}(b), and this is qualitatively similar to the minimal model, Fig.~\ref{figAAp1}(e). However, h-BN is in the wide band gap regime with $U / \gamma_1 \approx 7.25$, and, as a result, the soliton states do not extend very far into the band gap, in relative terms.

\subsection{ABA stacking: hexagonal boron phosphide}

As a third example of the influence of additional tight-binding parameters, we use a tight-binding model of ABA-stacked hexagonal boron phosphide (h-BP) with parameters fitted to calculations using DFT for a faultless system~\cite{wang19}.
As compared to the minimal model, we add skew interlayer couplings $\gamma_3$ and $\gamma_4$.
In this tight-binding model, for ABA-stacking, 
the Hamiltonian~(\ref{h1}) is modified~\cite{koshino11} as 
\begin{eqnarray}
H = \begin{pmatrix}
D_1 & V & 0 & 0 & 0 & \hdots \\
V^{\dagger} & D_2 & V^{\dagger} & 0 & 0 & \hdots \\
0 & V & D_3 & V & 0 & \hdots \\
0 & 0 & V^{\dagger} & D_4 & V^{\dagger} & \hdots \\
\vdots & \vdots & \vdots & \vdots & \vdots & \ddots
\end{pmatrix} , \label{hbernalfull}
\end{eqnarray}
where intralayer blocks $D_i$ are defined in Eq.~(\ref{di}) and we use $U_i = U$ for all $i$ (for a faultless system). 
Interlayer blocks $V$ are given by Eq.~(\ref{vfull}).
We use parameter values fit to DFT~\cite{wang19} near the K point for bilayers, neglecting small differences in the hopping between boron and phosphorous atoms, giving  
$\gamma_0 = 1.65\,$eV, $\gamma_1 = 0.761\,$eV, $\gamma_3 = -0.456\,$eV, $\gamma_4 = 0.255\,$eV, and $U = 0.636\,$eV.

The band structure plotted across the whole Brillouin zone is shown for a faultless system in Fig.~\ref{newfig2}(c) and for an atomically-sharp soliton at the center in Fig.~\ref{newfig2}(d). For the faultless system, the band gap is near the K point but is indirect in the presence of additional parameters, Fig.~\ref{newfig2}(c). For the soliton, there is a state localized on the soliton within the bulk band gap, above the bulk valence bands, Fig.~\ref{newfig2}(d). This is qualitatively similar to the minimal model, Fig.~\ref{figABA1}(c). Note that h-BP is in a regime of a relatively-small band gap with $U / \gamma_1 \approx 0.84$.

\section{Relevance to specific materials}\label{s:materials}

Although graphite occurs with both Bernal and rhombohedral stacking, our results do not apply to it because it is practically impossible to induce different onsite energies in a controlled way.
However, we note that it may also be possible to obtain isolated bands in graphite by considering rhombohedrally-stacked sections within Bernal-stacked graphite~\cite{arovas08,koshino13}, by applying a large displacement field to a stacking fault near the surface of rhombohedral graphite~\cite{shi20}, or with a large in-plane magnetic field in rhombohedral graphite~\cite{tymczyszyn23}. Localized topological states have also been considered in a different context, that of domain walls in the horizontal, in-plane direction~\cite{martin08,semenoff08}.

Our results apply to binary compounds with a honeycomb structure. A well-studied example is h-BN~\cite{naclerio22} which occurs predominantly with AA$^{\prime}$ stacking~\cite{pease50,taylor52,zunger76,blase95,han08,constantinescu13,kim13,gilbert19,hunt20,chen23}, although samples with ABA~\cite{warner10,kim13,khan16,ji17,gilbert19,rousseau22a,rousseau22b} and ABC stacking~\cite{will00,legodec00,chubarov12,moret21,olovsson22} have also been fabricated.
Density functional theory~\cite{ribeiro11,yin11,constantinescu13,gilbert19} predicts that ABA stacking has a total energy comparable to AA$^{\prime}$, but that ABC stacking is less energetically favorable.
Excitons localized on structural defects in h-BN have been discussed in the context of observed optical properties~\cite{jaffrennou07,museur08,pierret14,bourrellier14}, and
features in X-ray spectroscopy have been attributed to interlayer stacking faults~\cite{henck17}. Domain walls in the in-plane direction have been observed experimentally~\cite{kim13,woods14,ni19} and discussed theoretically~\cite{ouyang18}, and the nonlinear optical properties of multilayers with a controllable twist in the vertical (stacking) direction have been measured and analyzed~\cite{yao21}. 

The stacking types to be realized experimentally (AA$^{\prime}$, ABA and ABC) all have different atoms connected in the vertical direction, i.e., boron to nitrogen, and a fault consisting of a connection between two identical atoms, i.e., boron to boron or nitrogen to nitrogen, will be energetically expensive.
Single layer defects, as described in Section~\ref{s:single}, were modeled using DFT~\cite{yin11} for AA, AA$^{\prime}$, ABA and ABA$^{\prime}$ stacking, and it was found that the formation energy of such faults is very small. For example, an AA-type defect in AA$^{\prime}$ stacking was estimated to cost of the order of~$50\,$meV.
Even so, it is likely that such a defect is difficult to realize experimentally because a relative shift in the in-plane direction would yield the more favorable ABA stacking~\cite{warner10,gilbert19}.
Nevertheless, experiments~\cite{yasuda21,stern21} have recently demonstrated the possiblilty of engineering stacking of boron nitride (to create ferroelectric materials), and a relative twist between layers creates domains of ABA stacking separated, in the in-plane direction, by domain walls and regions of AA stacking.

Other binary compounds with a honeycomb structure in thin films and a direct band gap at the K point include boron phosphide~\cite{wang15,wang19,hernandez21} and silicon carbide~\cite{lin12,miro14,chabi16,chabi20}.
Some two-dimensional materials have a similar lattice structure, but they are not described by our model. They include compounds such as gallium nitride~\cite{sahin09,zhuang13,balushi16,qin17}, aluminium nitride~\cite{sahin09,zhuang13,tsipas13}, and zinc oxide~\cite{tusche07,topsakal09,weirum10,deng13,quang15} where the band gap doesn't lie at the K point, or materials, including transition metal dichalcogenides~\cite{mak10,splendiani10,korn11,xiao12,miro14,kormanyos15} and III-VI semiconductors~\cite{shenoy16,yu19,liu21,palepu22,lai23}, that have more than two orbitals per unit cell near the Fermi level. Nevertheless, these materials should in general support states localized on solitons analogous to the ones we discuss.

\section{Conclusions}

Using a minimal tight-binding model, we determined the electronic properties of thin films of binary compounds with two atoms per unit cell arranged as stacked two-dimensional honeycomb lattices.
The two atoms per cell are assigned different onsite energies.
We considered six different stacking orders to determine whether a fault in the texture of onsite energies in the stacking direction supports localized states.
Faults for ABC, ABA, and AA$^{\prime}$ stacking support localized states within the band gap, whereas faults for ABC$^{\prime}$, ABA$^{\prime}$, and AA stacking do not. For ABC and ABA stacking, there is a single localized state within the bulk band gap and, for ABC stacking, this state may hybridize with surface states.
For AA$^{\prime}$ stacking, there are two states within the bulk band gap which may hybridize with each other, depending on parameter values.

We consider smooth solitons where the texture of onsite energies changes over a length scale greater than the interlayer spacing, leading to different band structures as compared to atomically sharp solitons. In particular, a smooth soliton in AA$^{\prime}$ stacking results in a flat band at zero energy and a corresponding narrow peak in the density of states. We show that this feature is fairly robust to long-range correlated disorder.
Finally, we also consider the band structure due to single layer defects~\cite{yin11} where the signs of the onsite energies are reversed on a single layer only without a subsequent change of texture.
Overall, different stackings and types of fault produce a range of contrasting electronic band structures and densities of states which will manifest themselves in various transport and spectroscopic measurements~\cite{jaffrennou07,museur08,pierret14,bourrellier14,henck17,kim13,woods14,ni19}.

\begin{acknowledgments}
The authors thank V. I. Fal'ko for helpful discussions.
\end{acknowledgments}

\appendix

\section{Transport calculation}\label{a:transport}

Here we provide details of the conductivity calculation for rhombohedral stacking shown in Fig.~\ref{newfig1}.
We consider a two-contact system with a central sample at $0 < x < L$ and two semi-infinite leads on the left ($x<0$) and on the right ($x>L$). Following~\cite{tworzydlo06,snyman07}, the leads are modeled by considering the system with a very large onsite energy, creating a large density of states. Wave matching is performed to match states in the central sample to those in the leads at the same $E$ and $k_y$ values.

We consider the continuum limit near the valley at wave vector
${\mathbf{K_{+}}} = ( 4 \pi / (3a) , 0 )$ so that
$f({\mathbf{K_{+}}} + {\bf k}) \approx - \sqrt{3} a (k_x - i k_y)/2$.
Following~\cite{koshino09}, we write the Hamiltonian~(\ref{h1all}) as
\begin{eqnarray}
H = \begin{pmatrix}
D_1 & V & W & 0 & 0 & \hdots \\
V^{\dagger} & D_2 & V & W & 0 & \hdots \\
W^{\dagger} & V^{\dagger} & D_3 & V & W & \hdots \\
0 & W^{\dagger} & V^{\dagger} & D_4 & V & \hdots \\
\vdots & \vdots & \vdots & \vdots & \vdots & \ddots
\end{pmatrix} , \label{hrhom}
\end{eqnarray}
where
\begin{eqnarray}
D_j = \begin{pmatrix}
U_{j,A} & v \pi^{\dagger} \\
v \pi & U_{j,B}
\end{pmatrix} ; \quad
V = \begin{pmatrix}
- v_4 \pi^{\dagger} & v_3 \pi \\
\gamma_1 & - v_4 \pi^{\dagger}
\end{pmatrix} ,
\end{eqnarray}
with $\pi = \hbar (k_x + i k_y)$, $\pi^{\dagger} = \hbar (k_x - i k_y)$, $v_3 = \sqrt{3} a \gamma_3 / (2\hbar)$, and $v_4 = \sqrt{3} a \gamma_4 / (2\hbar)$.
Next-nearest layer hopping is described by
\begin{eqnarray}
W = \begin{pmatrix}
0 & \gamma_2 / 2 \\
0 & 0
\end{pmatrix} .
\end{eqnarray}

In the center of the system, $0<x<L$, we set $U_{A,j} = - U_{B,j} = U_{j}$ for all $j$ (for a faultless system) and numerically determine $2N$ eigenstates $\Psi_j$ of the Hamiltonian~(\ref{hrhom}) each with a value of $k_x$ denoted $k_j$ where $j=1,2,\ldots,2N$.
Then the states in the center can be written as
\begin{eqnarray}
\psi^{\ell} = \sum_j c_j^{\ell} \Psi_j e^{i k_j x + i k_y y} ,
\end{eqnarray}
where index $\ell = 1,2,\ldots,N$ denotes $N$ different incoming states.

In the contacts, we set $U_{A,j} = U_{B,j} = - U_{\infty}$ for all $j$, where $U_{\infty} \gg \gamma_1, E$.
Then we numerically determine the eigenstates of the Hamiltonian~(\ref{hrhom}).
There are $N$ right-moving solutions $\Upsilon_{\ell}$ with $k_x = Q_{\ell}^{(R)}$ where $\mathrm{Re} (Q_{\ell}^{(R)}) > 0$,
and three left-moving solutions $\Phi_{\ell}$ with $k_x = -Q_{\ell}^{(L)}$ where $\mathrm{Re} (Q_{\ell}^{(L)}) > 0$,
where $\ell = 1,2,\ldots,N$.
Each of them is normalized to carry unit flux by setting $\Upsilon_{\ell}^{\dagger} S_x \Upsilon_{\ell} = \Phi_{\ell}^{\dagger} S_x \Phi_{\ell} = 1$ where $S_x$ is proportional to the current operator,
\begin{eqnarray}
S_x =
\begin{pmatrix}
0 & 1 & 0 & 0 & 0 & 0 & \hdots \\
1 & 0 & 0 & 0 & 0 & 0 & \hdots \\
0 & 0 & 0 & 1& 0 & 0 & \hdots \\
0 & 0 & 1 & 0 & 0 & 0 & \hdots \\
0 & 0 & 0 & 0 & 0 & 1 & \hdots \\
0 & 0 & 0 & 0 & 1 & 0 & \hdots \\
\vdots & \vdots & \vdots & \vdots & \vdots & \vdots & \ddots
\end{pmatrix} ,
\end{eqnarray}
In the left contact at $x=0$, there are $N$ possible states, $\ell = 1,2,\ldots,N$, describing incoming flux,
\begin{eqnarray}
{\cal L}^{\ell} = \Upsilon_{\ell} e^{i k_y y} + \sum_j r_j^{\ell} \Phi_j e^{i k_y y} ,
\end{eqnarray}
where $r_j^{\ell}$ is a reflection amplitude from state ${\ell}$ to state $j$.
In the right contact at $x=L$, there are $N$ corresponding transmitted states:
\begin{eqnarray}
{\cal R}^{\ell} = \sum_j t_j^{\ell} \Upsilon_j e^{iQ_{j}^{(R)} L + i k_y y} ,
\end{eqnarray}
where $t_j^{\ell}$ is a transmission amplitude from state ${\ell}$ to state $j$.

We consider wave-matching $\psi^{\ell} = {\cal L}^{\ell}$ at $x=0$ and $\psi^{\ell} = {\cal R}^{\ell}$ at $x=L$ for $\ell = 1,2,\ldots,N$. For each $\ell$ value, there are $4N$ unknowns consisting of $2N$ values of $c_j^{\ell}$, $N$ values of $r_j^{\ell}$, and $N$ values of $t_j^{\ell}$, and there are $4N$ boundary conditions due to continuity of the wave function components.
Following~\cite{snyman07}, the matching conditions for all $\ell$ values may be expressed as $M C = A$ where $C$ is a $4N \times N$ matrix of unknown coefficients
\begin{eqnarray}
C = 
\begin{pmatrix}
\hat{r}^T \\
\hat{c} \\
\hat{t}^T
\end{pmatrix} ,
\end{eqnarray}
where $\hat{r}_{j\ell} = r_{\ell}^j$ is $N \times N$, $\hat{c}_{j\ell} = c_j^{\ell}$ is $2N \times N$, and $\hat{t}_{j\ell} = t_{\ell}^j$ is $N \times N$. $A$ is a $4N \times N$ matrix of incoming states,
\begin{eqnarray}
A = 
\begin{pmatrix}
- \Upsilon_1 & -\Upsilon_2 & - \Upsilon_3 & \ldots & - \Upsilon_N \\
\hat{0} & \hat{0} & \hat{0} & \ldots & \hat{0}
\end{pmatrix} ,
\end{eqnarray}
where $\hat{0}$ represents a $2N \times 1$ column vector of zeros.
$M$ is a $4N \times 4N$ matrix,
\begin{widetext}
\begin{eqnarray}
\!\!\!\!\!\!\!\!\! M = 
\begin{pmatrix}
- \Phi_1 & -\Phi_2 & - \Phi_3 & \ldots & - \Phi_N & \Psi_1 & \Psi_2 & \Psi_3 & \ldots & \Psi_{2N} & \hat{0} & \hat{0} & \hat{0} & \ldots & \hat{0} \\
\hat{0} & \hat{0} & \hat{0} & \ldots & \hat{0} & z_1 \Psi_1 & z_2 \Psi_2 & z_3 \Psi_3 & \ldots & z_{2N} \Psi_{2N} & - y_1 \Upsilon_1 & -y_2 \Upsilon_2 & - y_3 \Upsilon_3 & \ldots & - y_N \Upsilon_N
\end{pmatrix} ,
\end{eqnarray}
\end{widetext}
where $z_j = e^{i k_j L}$ for $j=1,2,\ldots,2N$, and $y_{\ell} = e^{i Q_{\ell}^{(R)} L}$ for $\ell = 1,2,\ldots,N$.
In the leads with $U_{\infty} \gg \gamma_1, E$, the dispersion is approximately linear so that $U_{\infty} \approx \hbar v Q_{\ell}^{(R)}$. For our numerical calculations, Fig.~\ref{newfig1}, we choose $U_{\infty} = 50 \gamma_1$ so that $Q_{\ell}^{(R)} \approx 50 k_c$.

The unknown coefficients are found by determining $C = M^{-1} A$.
Then, for a given energy value, the conductivity~\cite{landauer88,snyman07} is given by
\begin{eqnarray}
\sigma (E) = \frac{4e^2}{h} \frac{L}{W} \mathrm{tr} ( \hat{t} \hat{t}^{\dagger} ) ,
\end{eqnarray}
where $L$ is the length and $W$ is the width of the sample.
The trace may evaluated by summing over the eigenvalues of $\hat{t} \hat{t}^{\dagger}$ for all $k_y$ values where, for periodic boundary conditions~\cite{snyman07}, $k_y = 2 \pi n / W$ for integer $n = 0 , \pm 1 , \pm 2 , \ldots$.


\begin{thebibliography}{99}

\bibitem{pierucci15}
D. Pierucci, H. Sediri, M. Hajlaoui, J.-C. Girard, T. Brumme,
M. Calandra, E. Velez-Fort, G. Patriarche, M. G. Silly, G. Ferro,
V Souli\`ere, M. Marangolo, F. Sirotti, F. Mauri, and A. Ouerghi,
Evidence for Flat Bands near the Fermi Level in Epitaxial Rhombohedral Multilayer Graphene,
ACS Nano {\bf 9}, 5432 (2015).

\bibitem{henni16}
Y. Henni, H. P. O. Collado, K. Nogajewski, M. R. Molas, G. Usaj,
C. A. Balseiro, M. Orlita, M. Potemski, and C. Faugeras,
Rhombohedral Multilayer Graphene: A Magneto-Raman Scattering Study,
Nano Lett.\ {\bf 16}, 3710 (2016).

\bibitem{henck18}
H. Henck, J. Avila, Z. Ben Aziza, D. Pierucci, J. Baima, B. Pamuk, J. Chaste, D. Utt,
M. Bartos, K. Nogajewski, B. A. Piot, M. Orlita, M. Potemski, M. Calandra,
M. C. Asensio, F. Mauri, C. Faugeras, and A. Ouerghi,
Flat electronic bands in long sequences of rhombohedral-stacked graphene,
Phys.\ Rev.\ B {\bf 97}, 245421 (2018).

\bibitem{latychevskaia19}
T. Latychevskaia, S.-K. Son, Y. Yang, D. Chancellor, M. Brown, S. Ozdemir, I. Madan, G. Berruto, F. Carbone, A. Mishchenko and K. S. Novoselov,
Stacking transition in rhombohedral graphite,
Front.\ Phys.\ {\bf 14}, 13608 (2019).

\bibitem{yang19}
Y. Yang, Y.-C. Zou, C. R. Woods, Y. Shi, J. Yin, S. Xu, S. Ozdemir, T. Taniguchi, K. Watanabe, A. K. Geim, K. S. Novoselov, S. J. Haigh, and A. Mishchenko,
Stacking Order in Graphite Films Controlled by van der Waals Technology,
Nano Lett.\ {\bf 19}, 8526 (2019).

\bibitem{geisenhof19}
F. R. Geisenhof, F. Winterer, S. Wakolbinger, T. D. Gokus, Y. C. Durmaz, D. Priesack, J. Lenz, F. Keilmann, K. Watanabe, T. Taniguchi, R. Guerrero-Avil\'es, M. Pelc, A. Ayuela and R. Thomas Weitz,
Anisotropic Strain-Induced Soliton Movement Changes Stacking Order and Band Structure of Graphene Multilayers: Implications for Charge Transport,
ACS Appl.\ Nano Mater.\ {\bf 2}, 6067 (2019).

\bibitem{bouhafs21}
C. Bouhafs, S. Pezzini, F. R. Geisenhof, N. Mishra, V. Mi\u{s}eikis, Y. Niu, C. Struzzi, R. T. Weitz, A. A. Zakharov, S. Forti, and C. Coletti,
Synthesis of large-area rhombohedral few-layer graphene by chemical vapor deposition on copper,
Carbon {\bf 177}, 282 (2021).

\bibitem{shi20}
Y. Shi, S. Xu, Y. Yang, S. Slizovskiy, S. V. Morozov, S.-K. Son, S. Ozdemir, C. Mullan, J. Barrier, J. Yin, A. I. Berdyugin, B. A. Piot, T. Taniguchi, K. Watanabe, V. I. Fal’ko, K. S. Novoselov, A. K. Geim, and A. Mishchenko,
Electronic phase separation in multilayer rhombohedral graphite,
Nature {\bf 584}, 210 (2020).

\bibitem{kerelsky21}
A. Kerelsky, C. Rubio-Verd\'{u}, L. Xian, D. M. Kennes, D. Halbertal,
N. Finney, L. Song, S. Turkel, L. Wanga, K. Watanabe, T. Taniguchi, J. Hone,
C. Dean, D. N. Basov, A. Rubio, and A. N. Pasupathy,
Moir\'{e}less correlations in ABCA graphene,
Proc.\ Natl.\ Acad.\ Sci.\ U.S.A.\ {\bf 118}, 2017366118 (2021).

\bibitem{hagymasi22}
I. Hagym\'asi, M. S. Mohd Isa, Z. Tajkov, K. M\'arity,
L. Oroszl\'any, J. Koltai, A. Alassaf, P. Kun, K. Kandrai,
A. P\'alinkás, P. Vancs\'o, L. Tapaszt\'o, and P. Nemes-Incze,
Observation of competing, correlated ground states in the flat band of rhombohedral graphite,
Sci.\ Adv.\ {\bf 8}, eabo6879 (2022).

\bibitem{zhou21a}
H. Zhou, T. Xie, A. Ghazaryan, T. Holder, J. R. Ehrets, E. M. Spanton, T. Taniguchi, K. Watanabe, E. Berg, M. Serbyn, and A. F. Young,
Half- and quarter-metals in rhombohedral trilayer graphene,
Nature {\bf 598}, 429 (2021).

\bibitem{zhou21b}
H. Zhou, T. Xie, T. Taniguchi, K. Watanabe, and A. F. Young,
Superconductivity in rhombohedral trilayer graphene,
Nature {\bf 598}, 434 (2021).

\bibitem{su79}
W. P. Su, J. R. Schrieffer, and A. J. Heeger,
Solitons in polyacetylene,
Phys.\ Rev.\ Lett.\ {\bf 42}, 1698 (1979).

\bibitem{asboth16}
J. K. Asb\'oth, L. Oroszl\'any, and A. P\'alyi,
{\em A Short Course on Topological Insulators}
(Springer, Switzerland, 2016).

\bibitem{cayssol21}
J. Cayssol and J.-N. Fuchs,
Topological and geometrical aspects of band theory,
J.\ Phys.\ Mater.\ {\bf 4}, 034007 (2021).

\bibitem{mccann23}
E. McCann,
Catalog of noninteracting tight-binding models with two energy bands in one dimension,
Phys.\ Rev.\ B {\bf 107}, 245401 (2023).

\bibitem{ryu10}
S. Ryu, A. P. Schnyder, A. Furusaki, and A. W. W. Ludwig,
Topological insulators and superconductors: tenfold way and dimensional hierarchy,
New J.\ Phys.\ {\bf 12}, 065010 (2010).

\bibitem{heikkila11}
T. T. Heikkil\"a and G. E. Volovik,
Dimensional crossover in topological matter: Evolution of the multiple Dirac point in the layered system to the flat band on the surface,
JETP Lett.\ {\bf 93}, 59 (2011).

\bibitem{xiao11}
R. Xiao, F. Tasn\'adi, K. Koepernik, J. W. F. Venderbos, M. Richter, and M. Taut,
Density functional investigation of rhombohedral stacks of graphene: Topological surface states, nonlinear dielectric response, and bulk limit,
Phys.\ Rev.\ B {\bf 84}, 165404 (2011).

\bibitem{taut14}
M. Taut, K. Koepernik, and M. Richter,
Electronic structure of stacking faults in rhombohedral graphite,
Phys.\ Rev.\ B {\bf 90}, 085312 (2014).

\bibitem{slizovskiy19}
S. Slizovskiy, E. McCann, M. Koshino, and V. I. Fal’ko,
Films of rhombohedral graphite as two-dimensional topological semimetals,
Commun.\ Phys.\ {\bf 2}, 164 (2019).

\bibitem{garciaruiz19}
A. Garc\'ia-Ruiz, S. Slizovskiy, M. Mucha-Kruczy\'nski, and V. I. Fal’ko,
Spectroscopic Signatures of Electronic Excitations in RamanScattering in Thin Films of Rhombohedral Graphite,
Nano Lett.\ {\bf 19}, 6152 (2019).

\bibitem{muten21}
J. H. Muten , A. J. Copeland, and E. McCann,
Exchange interaction, disorder, and stacking faults in rhombohedral graphene multilayers,
Phys.\ Rev.\ B {\bf 104}, 035404 (2021).

\bibitem{garciaruiz23}
A. Garcia-Ruiz, S. Slizovskiy, and V. I. Fal'ko,
Flat Bands for Electrons in Rhombohedral Graphene Multilayers with a Twin Boundary,
Adv.\ Mater.\ Interfaces {\bf 10}, 2202221 (2023).

\bibitem{jackiw76}
R. Jackiw and C. Rebbi,
Solitons with fermion number $1/2$,
Phys.\ Rev.\ D {\bf 13}, 3398 (1976).

\bibitem{pacile08}
D. Pacil\'e, J. C. Meyer, \c{C}. \"O. Girit, and A. Zettl,
The two-dimensional phase of boron nitride: Few-atomic-layer sheets and suspended membranes,
Appl.\ Phys.\ Lett.\ {\bf 92}, 133107 (2008).

\bibitem{naclerio22}
A. E. Naclerio and P. R. Kidambi,
A Review of Scalable Hexagonal Boron Nitride (h-BN) Synthesis for Present and Future Applications,
Adv.\ Mater.\ {\bf 35}, 2207374 (2023).

\bibitem{ribeiro11}
R. M. Ribeiro and N. M. R. Peres,
Stability of boron nitride bilayers: Ground-state energies, interlayer distances, and tight-binding description,
Phys.\ Rev.\ B {\bf 83}, 235312 (2011).

\bibitem{gilbert19}
S. M. Gilbert, T. Pham, M. Dogan, S. Oh, B. Shevitski, G. Schumm, S. Liu, P. Ercius, S. Aloni, M. L. Cohen, and A. Zettl,
Alternative stacking sequences in hexagonal boron nitride,
2D Materials {\bf 6}, 021006 (2019).

\bibitem{constantinescu13}
G. Constantinescu, A. Kuc, and T. Heine,
Stacking in Bulk and Bilayer Hexagonal Boron Nitride,
Phys.\ Rev.\ Lett.\ {\bf 111}, 036104 (2013).

\bibitem{kim13}
C.-J. Kim, L. Brown, M. W. Graham, R. Hovden, R. W. Havener, P. L. McEuen, D. A. Muller, and J. Park,
Stacking Order Dependent Second Harmonic Generation and Topological Defects in h-BN Bilayers,
Nano Lett. {\bf 13}, 5660 (2013).

\bibitem{ricemele82}
M. J. Rice and E. J. Mele,
Elementary excitations of a linearly conjugated diatomic polymer,
Phys.\ Rev.\ Lett.\ {\bf 49}, 1455 (1982).

\bibitem{kivelson83}
S. Kivelson,
Solitons with adjustable charge in a commensurate Peierls insulator,
Phys.\ Rev.\ B {\bf 28}, 2653 (1983).

\bibitem{brzezicki20}
W. Brzezicki and T. Hyart,
Topological domain wall states in a nonsymmorphic chiral chain,
Phys.\ Rev.\ B {\bf 101}, 235113 (2020).

\bibitem{fuchs21}
J.-N. Fuchs and F. Pi\'{e}chon,
Orbital embedding and topology of one-dimensional two-band insulators,
Phys.\ Rev.\ B {\bf 104}, 235428 (2021).

\bibitem{allen22}
R. E. J. Allen, H. V. Gibbons, A. M. Sherlock, H. R. M. Stanfield, and E. McCann,
Nonsymmorphic chiral symmetry and solitons in the Rice-Mele model,
Phys.\ Rev.\ B {\bf 106}, 165409 (2022).

\bibitem{shiozaki15}
K. Shiozaki, M. Sato, and K. Gomi,
$\mathbb{Z}_2$ topology in nonsymmorphic crystalline insulators: M\"obius twist in surface states,
Phys.\ Rev.\ B {\bf 91}, 155120 (2015).

\bibitem{yin11}
J. L. Yin, M. L. Hu, Z. Yu, C. X. Zhang, L. Z. Sun, and J. X. Zhong,
Direct or indirect semiconductor: The role of stacking fault in h-BN,
Physica B Condens.\ Matter {\bf 406}, 2293 (2011).

\bibitem{mccann13}
E. McCann and M. Koshino,
The electronic properties of bilayer graphene,
Rep.\ Prog.\ Phys.\ {\bf 76}, 056503 (2013).

\bibitem{kuzmenko09}
A. B. Kuzmenko, I. Crassee, D. van der Marel, P. Blake, and K. S. Novoselov, 
Determination of the gate-tunable band gap and tight-binding parameters in
bilayer graphene using infrared spectroscopy,
Phys.\ Rev.\ B {\bf 80}, 165406 (2009).

\bibitem{saito98}
R. Saito, M. S. Dresselhaus, and G. Dresselhaus,
{\em Physical Properties of Carbon Nanotubes}
(Imperial College Press, London, 1998).

\bibitem{snyman07}
I. Snyman and C. W. J. Beenakker,
Ballistic transmission through a graphene bilayer,
Phys.\ Rev.\ B {\bf 75}, 045322 (2007).

\bibitem{ssh80}
W. P. Su, J. R. Schrieffer, and A. J. Heeger,
Soliton excitations in polyacetylene,
Phys.\ Rev.\ B {\bf 22}, 2099 (1980).

\bibitem{heeger88}
A. J. Heeger, S. Kivelson, J. R. Schrieffer, and W.-P. Su,
Rev.\ Mod.\ Phys.\ {\bf 60}, 781 (1988).

\bibitem{zhao16}
Y. X. Zhao and A. P. Schnyder,
Nonsymmorphic symmetry-required band crossings in topological semimetals,
Phys.\ Rev.\ B {\bf 94}, 195109 (2016).

\bibitem{han20}
S.-H. Han, S.-G. Jeong, S.-W. Kim, T.-H. Kim, and S. Cheon,
Topological features of ground states and topological solitons in generalized Su-Schrieffer-Heeger models using generalized time-reversal, particle-hole, and chiral symmetries,
Phys.\ Rev.\ B {\bf 102}, 235411 (2020).

\bibitem{conventionnote}
In this paper, we always choose the first atom ($A_1$) to have a positive onsite energy $+U$. If we were to reverse this choice (i.e., the first atom has negative energy $-U$), the band structures would be identical but inverted in energy about zero energy such that solitons would become antisolitons, and vice versa.

\bibitem{mccann06}
E. McCann and V. I. Fal'ko,
Landau-Level Degeneracy and Quantum Hall Effect in a Graphite Bilayer,
Phys.\ Rev.\ Lett.\ {\bf 96}, 086805 (2006).

\bibitem{koschny02}
T. Koschny and L. Schweitzer,
Influence of correlated disorder potentials on the levitation of current carrying states in the quantum Hall effect,
Physica E {\bf 12}, 654 (2002).

\bibitem{guo08}
Z.-Z. Guo,
Entanglement in One-Dimensional Anderson Model with Long-Range Correlated Disorder,
Chin.\ Phys.\ Lett.\ {\bf 25}, 1079 (2008).

\bibitem{jaffrennou07}
P. Jaffrennou, J. Barjon, J.-S. Lauret, B. Attal-Tr\'etout, F. Ducastelle, and A. Loiseau,
Origin of the excitonic recombinations in hexagonal boron nitride by spatially resolved cathodoluminescence spectroscopy,
J.\ Appl.\ Phys.\ {\bf 102}, 116102 (2007).

\bibitem{museur08}
L. Museur, E. Feldbach, and A. Kanaev,
Defect-related photoluminescence of hexagonal boron nitride,
Phys.\ Rev.\ B {\bf 78}, 155204 (2008).

\bibitem{pierret14}
A. Pierret, J. Loayza, B. Berini, A. Betz, B. Pla\c{c}ais, F. Ducastelle, J. Barjon, and A. Loiseau,
Excitonic recombinations in h-BN: From bulk to exfoliated layers,
Phys.\ Rev.\ B {\bf 89}, 035414 (2014).

\bibitem{bourrellier14}
R. Bourrellier, M. Amato, L. H. G. Tizei, C. Giorgetti, A. Gloter, M. I. Heggie, K. March, O. St\'ephan, L. Reining, M. Kociak, and A. Zobelli,
Nanometric Resolved Luminescence in h-BN Flakes: Excitons and Stacking Order,
ACS Photonics {\bf 1}, 857 (2014).

\bibitem{henck17}
H. Henck, D. Pierucci, Z. Ben Aziza, M. G. Silly, B. Gil, F. Sirotti, G. Cassabois, and A. Ouerghi,
Stacking fault and defects in single domain multilayered hexagonal boron nitride,
Appl.\ Phys.\ Lett.\ {\bf 110}, 023101 (2017).

\bibitem{woods14}
C. R. Woods, L. Britnell, A. Eckmann, R. S. Ma, J. C. Lu, H. M. Guo, X. Lin, G. L. Yu, Y. Cao, R. V. Gorbachev, A. V. Kretinin, J. Park, L. A. Ponomarenko, M. I. Katsnelson, Yu. N. Gornostyrev, K. Watanabe, T. Taniguchi, C. Casiraghi, H-J. Gao, A. K. Geim, and K. S. Novoselov,
Commensurate-incommensurate transition in graphene on hexagonal boron nitride,
Nat.\ Phys.\ {\bf 10}, 451 (2014).

\bibitem{ni19}
G. X. Ni, H. Wang, B.-Y. Jiang, L. X. Chen, Y. Du, Z. Y. Sun, M. D. Goldflam, A. J. Frenzel, X. M. Xie, M. M. Fogler, and D. N. Basov,
Soliton superlattices in twisted hexagonal boron nitride,
Nat.\ Commun.\ {\bf 10}, 4360 (2019).

\bibitem{tworzydlo06}
J. Tworzyd\l o, B. Trauzettel, M. Titov, A. Rycerz, and C.W. J. Beenakker,
Sub-Poissonian shot noise in graphene,
Phys.\ Rev.\ Lett.\ {\bf 96}, 246802 (2006).

\bibitem{landauer88}
R. Landauer,
Spatial variation of currents and fields due to localized scatterers in metallic conduction,
IBM J.\ Res.\ Develop.\ {\bf 32}, 306 (1988).

\bibitem{wang15}
S.-f. Wang and X.-j. Wu,
First-Principles Study on Electronic and Optical Properties of Graphene-Like Boron Phosphide Sheets,
Chin.\ J.\ Chem.\ Phys.\ {\bf 28}, 588 (2015).

\bibitem{wang19}
Y. Wang, C. Huang, D. Li, P. Li, J. Yu, Y. Zhang, and J. Xu,
Tight-binding model for electronic structure of hexagonal boron phosphide monolayer and bilayer,
J.\ Phys.: Condens.\ Matter {\bf 31}, 285501 (2019).

\bibitem{hernandez21}
O. M. Hern\'andez, J. Guerrero-S\'anchez, R. Ponce-P\'erez, R. G. D\'iaz, H. 
N. Fernandez-Escamilla, and Gregorio H. Cocoletzi,
Hexagonal boron phosphide monolayer exfoliation induced by arsenic incorporation in the BP (111) surface: A DFT study,
Appl.\ Surf.\ Sci.\ {\bf 538}, 148163 (2021).

\bibitem{koshino09}
M. Koshino and E. McCann,
Trigonal warping and Berry’s phase $N\pi$ in ABC-stacked multilayer graphene,
Phys.\ Rev. B {\bf 80}, 165409 (2009).

\bibitem{sponza18}
L. Sponza, H. Amara, C. Attaccalite, S. Latil, T. Galvani, F. Paleari, L. Wirtz, and F. Ducastelle,
Direct and indirect excitons in boron nitride polymorphs: A story of atomic configuration and electronic correlation,
Phys.\ Rev. B {\bf 98}, 125206 (2018).

\bibitem{koshino11}
M. Koshino and E. McCann,
Landau level spectra and the quantum Hall effect of multilayer graphene,
Phys.\ Rev. B {\bf 83}, 165443 (2011).

\bibitem{arovas08}
D. P. Arovas and F. Guinea,
Stacking faults, bound states, and quantum Hall plateaus in crystalline graphite,
Phys.\ Rev.\ B {\bf 78}, 245416 (2008).

\bibitem{koshino13}
M. Koshino and E. McCann,
Multilayer graphenes with mixed stacking structure: Interplay of Bernal and rhombohedral stacking,
Phys.\ Rev.\ B {\bf 87}, 045420 (2013).

\bibitem{tymczyszyn23}
M. Tymczyszyn, P. H. Cross, and E. McCann,
Solitons induced by an in-plane magnetic field in rhombohedral multilayer graphene,
Phys.\ Rev.\ B {\bf 108}, 115425 (2023).

\bibitem{martin08}
I. Martin, Y. M. Blanter, and A. F. Morpurgo,
Topological Confinement in Bilayer Graphene,
Phys.\ Rev.\ Lett.\ {\bf 100}, 036804 (2008).

\bibitem{semenoff08}
G. W. Semenoff, V. Semenoff, and F. Zhou,
Domain Walls in Gapped Graphene,
Phys.\ Rev.\ Lett.\ {\bf 101}, 087204 (2008).

\bibitem{pease50}
R. S. Pease,
Crystal Structure of Boron Nitride
Nature {\bf 165}, 722 (1950).

\bibitem{taylor52}
R. Taylor and C. A. Coulson,
Studies in Graphite and Related Compounds III: Electronic Band Structure in Boron Nitride,
Proc.\ Phys.\ Soc.\ Lond.\ A {\bf 65}, 834 (1952).

\bibitem{blase95}
X. Blase, A. Rubio, S. G. Louie, and M. L. Cohen,
Quasiparticle band structure of bulk hexagonal boron nitride and related systems,
Phys.\ Rev.\ B {\bf 51}, 6868 (1995).

\bibitem{zunger76}
A. Zunger, A. Katzir, and A. Halperin,
Optical properties of hexagonal boron nitride,
Phys.\ Rev.\ B {\bf 13}, 5560 (1976).

\bibitem{han08}
W.-Q. Han, L. Wu, Y. Zhu, K. Watanabe, and T. Taniguchi,
Structure of chemically derived mono- and few-atomic-layer boron nitride sheets,
Appl.\ Phys.\ Lett.\ {\bf 93}, 223103 (2008).

\bibitem{hunt20}
R. J. Hunt, B. Monserrat, V. Z\'olyomi, and N. D. Drummond,
Diffusion quantum Monte Carlo and GW study of the electronic properties of monolayer and bulk hexagonal boron nitride,
Phys.\ Rev.\ B {\bf 101}, 205115 (2020).

\bibitem{chen23}
X. Chen, K. Zollner, C. Moulsdale, V. I. Fal'ko, and A. Knothe,
Semimetallic and semiconducting graphene-hBN multilayers with parallel or reverse stacking,
Phys.\ Rev.\ B {\bf 107}, 125402 (2023).

\bibitem{warner10}
J. H. Warner, M. H. R\"ummeli, A. Bachmatiuk, and B. B\"uchner,
Atomic Resolution Imaging and Topography of Boron Nitride Sheets Produced by Chemical Exfoliation,
ACS Nano {\bf 4}, 1299 (2010).

\bibitem{khan16}
M. H. Khan, G. Casillas, D. R. G. Mitchell, H. K. Liu, L. Jiang, and  Z. Huang,
Carbon- and crack-free growth of hexagonal boron nitride nanosheets and their uncommon stacking order,
Nanoscale {\bf 8}, 15926 (2016).

\bibitem{ji17}
Y. Ji, B. Calderon, Y. Han, P. Cueva, N. R. Jungwirth, H. A. Alsalman, J. Hwang, G. D. Fuchs, D. A. Muller, and M. G. Spencer,
Chemical Vapor Deposition Growth of Large Single-Crystal Mono-, Bi-, Tri-Layer Hexagonal Boron Nitride and Their Interlayer Stacking,
ACS Nano {\bf 11}, 12057 (2017).

\bibitem{rousseau22a}
A. Rousseau, P. Valvin, W. Desrat, L. Xue, J. Li, J. H. Edgar,
G. Cassabois, and B. Gil,
Bernal Boron Nitride Crystals Identified by Deep-Ultraviolet Cryomicroscopy,
ACS Nano {\bf 16}, 2756 (2022).

\bibitem{rousseau22b}
A. Rousseau, P. Valvin, C. Elias, L. Xue, J. Li, J. H. Edgar, B. Gil, and G. Cassabois,
Stacking-dependent deep level emission in boron nitride,
Phys.\ Rev.\ Mater.\ {\bf 6}, 094009 (2022).

\bibitem{will00}
G. Will, G. Nover, and J. von der G\"onna,
New Experimental Results on the Phase Diagram of Boron Nitride,
J.\ Solid State Chem.\ {\bf 154}, 280 (2000).

\bibitem{legodec00}
Y. Le Godeca, D. Martinez-Garciab, V. L. Solozhenkoc, M. Mezouard, G. Syfossea,
J. M. Bessona,
Compression and thermal expansion of rhombohedral boron nitride at high pressures and temperatures,
J.\ Phys.\ Chem.\ Solids {\bf 61}, 1935 (2000).

\bibitem{chubarov12}
M. Chubarov, H. Pedersen, H. H\"ogberg, J. Jensen, and A. Henry,
Growth of High Quality Epitaxial Rhombohedral Boron Nitride,
Cryst.\ Growth Des.\ {\bf 12}, 3215 (2012).

\bibitem{moret21}
M. Moret, A. Rousseau, P. Valvin, S. Sharma, L. Souqui, H. Pederse, H. H\"ogberg, G. Cassabois, J. Li, J. H. Edgar, and B. Gil,
Rhombohedral and turbostratic boron nitride: X-ray diffraction and photoluminescence signatures,
Appl.\ Phys.\ Lett.\ {\bf 119}, 262102 (2021).

\bibitem{olovsson22}
W. Olovsson and M. Magnuson,
Rhombohedral and Turbostratic Boron Nitride Polytypes Investigated by X-ray Absorption Spectroscopy,
J.\ Phys.\ Chem.\ C {\bf 126}, 21101 (2022).

\bibitem{ouyang18}
B. Ouyang, C. Chen and J. Song,
Conjugated $\pi$ electron engineering of generalized stacking fault in graphene and h-BN,
Nanotechnology {\bf 29}, 09LT01 (2018).

\bibitem{yao21}
K. Yao, N. R. Finney, J. Zhang, S. L. Moore, L. Xian, N. Tancogne-Dejean, F. Liu, J. Ardelean, X. Xu, D. Halbertal, K. Watanabe, T. Taniguchi, H. Ochoa, A. Asenjo-Garcia, X. Zhu, D. N. Basov, A. Rubio, C. R. Dean, J. Hone, and P. J. Schuck,
Enhanced tunable second harmonic generation from twistable interfaces and vertical superlattices in boron nitride homostructures,
Sci.\ Adv.\ {\bf 7}, eabe8691 (2021).

\bibitem{yasuda21}
K. Yasuda, X. Wang, K. Watanabe, T. Taniguchi, and P. Jarillo-Herrero,
Stacking-engineered ferroelectricity in bilayer boron nitride,
Science {\bf 372}, 1458 (2021).

\bibitem{stern21}
M. Vizner Stern, Y. Waschitz, W. Cao, I. Nevo, K. Watanabe, T. Taniguchi, E. Sela, M. Urbakh, O. Hod, and M. Ben Shalom,
Interfacial ferroelectricity by van der Waals sliding,
Science {\bf 372}, 1462 (2021).

\bibitem{lin12}
S. S. Lin,
Light-Emitting Two-Dimensional Ultrathin Silicon Carbide,
J.\ Phys.\ Chem.\ C {\bf 116}, 3951 (2012).

\bibitem{miro14}
P. Mir\'o, M. Audiffred, and T. Heine,
An atlas of two-dimensional materials,
Chem.\ Soc.\ Rev.\ {\bf 43}, 6537 (2014).

\bibitem{chabi16}
S. Chabi, H. Chang, Y. Xia, and Y. Zhu,
From graphene to silicon carbide: ultrathin silicon carbide flakes,
Nanotechnology {\bf 27}, 075602 (2016).

\bibitem{chabi20}
S. Chabi and K. Kadel,
Two-Dimensional Silicon Carbide: Emerging Direct Band Gap Semiconductor,
Nanomaterials {\bf 10}, 2226 (2020).

\bibitem{balushi16}
Z. Y. Al Balushi, K. Wang, R. K. Ghosh, R. A. Vil\'a, S. M. Eichfeld, J. D. Caldwell, X. Qin, Y.-C. Lin, P. A. DeSario, G. Stone, S. Subramanian, D. F. Paul, R. M. Wallace, S. Datta4, J. M. Redwing, and J A. Robinson,
Two-dimensional gallium nitride realized via graphene encapsulation,
Nat.\ Mater.\ {bf 15}, 1166 (2016).

\bibitem{qin17}
Z. Qin, G. Qin, X. Zuo, Z. Xiong, and M. Hu,
Orbitally driven low thermal conductivity of monolayer gallium nitride (GaN) with planar honeycomb structure: a comparative study,
Nanoscale {\bf 9}, 4295 (2017).

\bibitem{sahin09}
H. \c{S}ahin, S. Cahangirov, M. Topsakal, E. Bekaroglu, E. Akturk, R. T. Senger, and S. Ciraci,
Monolayer honeycomb structures of group-IV elements and III-V binary compounds: First-principles calculations,
Phys.\ Rev.\ B {\bf 80}, 155453 (2009).

\bibitem{zhuang13}
H. L. Zhuang, A. K. Singh, and R. G. Hennig,
Computational discovery of single-layer III-V materials,
Phys.\ Rev.\ B {\bf 87}, 165415 (2013).

\bibitem{tsipas13}
P. Tsipas, S. Kassavetis, D. Tsoutsou, E. Xenogiannopoulou, E. Golias, S. A. Giamini, C. Grazianetti, D. Chiappe, A. Molle, M. Fanciulli, and A. Dimoulas,
Evidence for graphite-like hexagonal AlN nanosheets epitaxially grown on single crystal Ag(111),
Appl.\ Phys.\ Lett.\ {\bf 103}, 251605 (2013).

\bibitem{tusche07}
C. Tusche, H. L. Meyerheim, and J. Kirschner,
Observation of Depolarized ZnO(0001) Monolayers: Formation of Unreconstructed Planar Sheets,
Phys.\ Rev.\ Lett.\ {\bf 99}, 026102 (2007).

\bibitem{topsakal09}
M. Topsakal, S. Cahangirov, E. Bekaroglu, and S. Ciraci,
First-principles study of zinc oxide honeycomb structures,
Phys.\ Rev.\ B {\bf 80}, 235119 (2009).

\bibitem{weirum10}
G. Weirum, G. Barcaro, A. Fortunelli, F. Weber, R. Schennach, S. Surnev, and F. P. Netzer,
Growth and Surface Structure of Zinc Oxide Layers on a Pd(111) Surface,
J.\ Phys.\ Chem.\ C {\bf 114}, 15432 (2010).

\bibitem{deng13}
X. Deng, K. Yao, K. Sun, W.-X. Li, J. Lee, and C. Matranga,
Growth of Single- and Bilayer ZnO on Au(111) and Interaction with Copper,
J.\ Phys.\ Chem.\ C {\bf 117}, 11211 (2013).

\bibitem{quang15}
H. T. Quang, A. Bachmatiuk, A. Dianat, F. Ortmann, J. Zhao, J. H. Warner, J. Eckert, G. Cunniberti, and M. H. R\"ummeli,
In Situ Observations of Free-Standing Graphene-like Mono- and Bilayer ZnO Membranes,
ACS Nano {\bf 9}, 11408 (2015).

\bibitem{mak10}
K. F. Mak, C. Lee, J. Hone, J. Shan, and T. F. Heinz,
Atomically Thin MoS$_2$: A New Direct-Gap Semiconductor,
Phys. Rev. Lett. {\bf 105}, 136805 (2010).

\bibitem{splendiani10}
A. Splendiani, L. Sun, Y. Zhang, T. Li, J. Kim, C.-Y. Chim, G. Galli, and F. Wang,
Emerging Photoluminescence in Monolayer MoS$_2$,
Nano Lett.\ {\bf 10}, 1271 (2010).

\bibitem{korn11}
T. Korn, S. Heydrich, M. Hirmer, J. Schmutzler, and C. Sch\"uller,
Low-temperature photocarrier dynamics in monolayer MoS$_2$,
Appl.\ Phys.\ Lett.\ {\bf 99}, 102109 (2011).

\bibitem{xiao12}
D. Xiao, G.-B. Liu, W. Feng, X. Xu, and W. Yao,
Coupled Spin and Valley Physics in Monolayers of MoS$_2$ and Other Group-VI Dichalcogenides,
Phys.\ Rev.\ Lett.\ {\bf 108}, 196802 (2012).

\bibitem{kormanyos15}
A. Korm\'anyos, G. Burkard, M. Gmitra, J. Fabian, V. Z\'olyomi, N. D Drummond, and V. Fal’ko,
$k \cdot p$ theory for two-dimensional transition metal dichalcogenide semiconductors,
2D Mater.\ {\bf 2}, 022001 (2015).

\bibitem{shenoy16}
U. S. Shenoy, U. Gupta, D. S. Narang, D. J. Late, U. V. Waghmare, and C. N. R. Rao,
Electronic structure and properties of layered gallium telluride,
Chem.\ Phys.\ Lett.\ {\bf 651}, Pages 148 (2016).

\bibitem{yu19}
Y. Yu, M. Ran, S. Zhou, R. Wang, F. Zhou, H. Li, L. Gan, M. Zhu, and T. Zhai,
Phase-Engineered Synthesis of Ultrathin Hexagonal and Monoclinic GaTe Flakes and Phase Transition Study,
Adv.\ Funct.\ Mater.\ {\bf 29}, 1901012 (2019).

\bibitem{liu21}
M. Liu, S. Yang, M. Han, S. Feng, G.-G. Wang, L. Dang, B. Zou, Y. Cai, H. Sun, J. Yu, J.-C. Han, and Z. Liu,
Controlled Growth of Large-Sized and Phase-Selectivity 2D GaTe Crystals,
Small {\bf 17}, 2007909 (2021).

\bibitem{palepu22}
J. Palepu, A. Tiwari, P. Sahatiya, S. Kundu, and S. Kanungo,
Effects of artificial stacking configurations and biaxial strain on the structural, electronic and transport properties of bilayer GaSe- A first principle study,
Mater.\ Sci.\ Semicond.\ Process.\ {\bf 137}, 106236 (2022).

\bibitem{lai23}
K. Lai and J. Dai,
Stacking effect on the electronic structures of hexagonal GaTe,
J.\ Phys.\ D: Appl.\ Phys.\ {\bf 56}, 275301 (2023).

\end{thebibliography}
\end{document}